\documentclass[english,preprintnumbers,amsmath,amssymb]{revtex4}
\usepackage[T1]{fontenc}
\usepackage[latin9]{inputenc}
\usepackage{color}
\usepackage{float}
\usepackage{amsbsy}
\usepackage{amstext}
\usepackage{graphicx}
\usepackage{esint}

\usepackage{amsmath}

\makeatletter

\@ifundefined{textcolor}{}
{%
 \definecolor{BLACK}{gray}{0}
 \definecolor{WHITE}{gray}{1}
 \definecolor{RED}{rgb}{1,0,0}
 \definecolor{GREEN}{rgb}{0,1,0}
 \definecolor{BLUE}{rgb}{0,0,1}
 \definecolor{CYAN}{cmyk}{1,0,0,0}
 \definecolor{MAGENTA}{cmyk}{0,1,0,0}
 \definecolor{YELLOW}{cmyk}{0,0,1,0}
 }

\usepackage{dcolumn}% Align table columns on decimal point
\usepackage{bm}% bold math
\usepackage{array}

\def\ket{\rangle}
\def\bra{\langle}

\def\im{\mathop{\textrm{Im}}}
\def\re{\mathop{\textrm{Re}}}

\def\GamA{\bar{\Gamma}_\textrm{A}}
\def\GamB{\bar{\Gamma}_\textrm{B}}
\def\GamRIC{\Gamma_\textrm{RIC}^0}
\def\GamRICp{\Gamma_\textrm{RIC}^+}
\def\GamRICm{\Gamma_\textrm{RIC}^-}

\def\kRICm{k_\textrm{RIC}^-}

\newcommand{\PT}{\mathcal{PT}}

\newcommand{\beq}{\begin{equation}}
\newcommand{\eeq}{\end{equation}}
\newcommand{\beqa}{\begin{eqnarray}}
\newcommand{\eeqa}{\end{eqnarray}}

\makeatother

\usepackage{babel}

\begin{document}

\title{Bound states, scattering states and resonant states in $\PT$-symmetric open quantum systems}

\author{Savannah Garmon}
%\email{sgarmon@iis.u-tokyo.ac.jp}
\affiliation{Department of Physical Science,
Osaka Prefecture University,
Gakuen-cho 1-1, Sakai 599-8531, Japan}

\author{Mariagiovanna Gianfreda}
%\email{ }
%\affiliation{Washington University}
\affiliation{ Institute of Industrial Science,
University of Tokyo,
Komaba 4-6-1, Meguro, Tokyo 153-8505, Japan}

\author{Naomichi Hatano}
%\email{ }
%\affiliation{Washington University}
\affiliation{ Institute of Industrial Science,
University of Tokyo,
Komaba 4-6-1, Meguro, Tokyo 153-8505, Japan}

\begin{abstract}
We study a simple open quantum system with a $\PT$-symmetric defect potential as a prototype in order to illustrate a number of general features of $\PT$-symmetric open quantum systems; however, the potential itself could be mimicked by a number of $\PT$ systems that have been experimentally studied quite recently.
%We use our prototype model to illustrate general features of both the discrete spectrum associated with the defect region as well as the scattering states.
One key feature is the resonance in continuum (RIC), which appears in both the discrete spectrum and the scattering spectrum of such systems.
%From the discrete spectrum perspective, the RIC appears when a resonance/anti-resonance pair in the second Riemann sheet of the complex energy plane meet directly on the branch cut associated with the energy continuum.
The RIC wave function forms a standing wave extending throughout the spatial extent of the system, and in this sense represents a resonance between the open environment associated with the leads of our model and the central $\PT$-symmetric potential.  We also illustrate that as one deforms the system parameters, the RIC may exit the continuum by splitting into a bound state and a virtual bound state at the band edge, a process which should be experimentally observable.  
We also study the exceptional points appearing in the discrete spectrum at which two eigenvalues coalesce; we categorize these as either EP2As, at which two real-valued solutions coalesce before becoming complex-valued, and EP2Bs, for which the two solutions are complex on either side of the exceptional point.  The EP2As are associated with $\PT$-symmetry breaking; we argue that these are more stable against parameter perturbation than the EP2Bs.  We also study complex-valued solutions of the discrete spectrum for which the wave function is nevertheless spatially localized, something that is not allowed in traditional open quantum systems; we illustrate that these may form quasi-bound states in continuum (QBICs) under some circumstances.  We also study the scattering properties of the system, including states that support invisible propagation and some general features of perfect transmission states.  We finally use our model as a prototype for the construction of scattering states that satisfy $\PT$-symmetric boundary conditions; while these states do not conserve the traditional probability current, we introduce the $\PT$-current which is preserved.  The perfect transmission states appear as a special case of the $\PT$-symmetric scattering states.
\end{abstract}
\maketitle

%\global\long\def\ket#1{\left| #1\right\rangle }
%\global\long\def\bra#1{\left\langle #1 \right|}
%\global\long\def\kket#1{\left\Vert #1\right\rangle }
%\global\long\def\bbra#1{\left\langle #1\right\Vert }
%\global\long\def\braket#1#2{\left\langle #1\right. \left| #2 \right\rangle }
%\global\long\def\bbrakket#1#2{\left\langle #1\right. \left\Vert #2\right\rangle }
%\global\long\def\av#1{\left\langle #1 \right\rangle }
%\global\long\def\tr{\text{Tr}}
%\global\long\def\im{\text{Im}}
%\global\long\def\re{\text{Re}}
%\global\long\def\sign{\text{sgn}}
%\global\long\def\abs#1{\left|#1\right|}

\section{Introduction}
\label{sec:intro}
\label{sec1}

\subsection{Two non-Hermitian systems: open quantum systems and $\PT$-symmetric systems}

In the conventional formulation of quantum mechanics, the Hamiltonian operator $H$ describing a given physical system is generally required to satisfy the Hermitian symmetry $H = H^\dagger$, a sufficient (but \textit{not} necessary) condition to obtain a real-valued energy spectrum.  
Since the theory was originally developed, however, a number of researchers have found it useful to introduce non-Hermitian elements to the Hamiltonian, either as an extension of the original theory to accommodate certain physical situations~\cite{HNPRL96,HNPRB97,HNPRB98,Feinberg97,Goldsheid98,FukuiKawakami98,Mudry98,Ahmed02,Bagchi02,Heiss04,Swanson04,Graefe08,Longhi10NH} or as a useful reformulation in others~\cite{Feshbach58,Feshbach62,Moiseyev1980,Petrosky96,Petrosky97,Albeverio96,Rotter91,Sadreev03,Okolowicz03,Rotter_review,Fyodorov97,Dittes00,Pichugin01,Kunz06,Kunz08,Sasada08,SHO11,Klaiman11,NH_H_eff,Hatano14,Moiseyev_NHQM}.  
In the latter case, various non-Hermitian Hamiltonians have been introduced to describe open quantum systems. 

Open quantum systems generally consist of a finite system coupled with an infinite environment, and thereby give rise to an energy spectrum with both discrete and continuous eigenvalues; 
the continuum is associated with the environmental degrees of freedom, while the discrete eigenvalues are a consequence of scattering due to the finite system.
Some of the discrete eigenvalues can be complex, a signature of resonance phenomena in open systems.  
Resonances are associated with transient phenomena such as transport and exponential decay~\cite{Feshbach58,Feshbach62,Rotter91,Sadreev03,Okolowicz03,Dittes00,Kunz06,SHO11,Klaiman11,Hatano14,PPTasaki91,TGP06,HSNP08} and may be viewed as generalized solutions of the Schr\"odinger equation with complex eigenvalues~\cite{PPTasaki91,HSNP08,Hatano14} or as complex poles of the analytically continued S-matrix, among other perspectives~\cite{Moiseyev_NHQM,GP_resonance}.

The reason why open quantum systems may accommodate complex eigenvalues can be summarized as follows.
Eigenfunctions that are normalizable in open quantum systems, namely bound states and norm-preserving scattering states, lie within the Hilbert space and can only have real eigenvalues.
This corresponds to the fact that the Hamiltonian operator is Hermitian in the Hilbert space.
However, even the standard Hamiltonian operator may be non-Hermitian in a space wider than the Hilbert space~\cite{HSNP08}.
Open quantum systems indeed can harbor unnormalizable eigenfunctions, which lie outside the Hilbert space and can have complex eigenvalues depending on the boundary conditions.
(Note, however, that we can still give a probabilistic interpretation for such eigenfunctions~\cite{HSNP08,HKF09}.)

%This non-Hermiticity hidden in the boundary conditions of open quantum systems manifests itself as an explicitly non-Hermitian Hamiltonian operator once we trace out the degrees of freedom of the environment, which corresponds to the part of continuum spectrum, and come up with an effective Hamiltonian which has only the degrees of freedom of the finite system of interest, which corresponds to the discrete part of the spectrum.
While usually hidden in the boundary conditions, this non-Hermitian aspect of open quantum systems manifests itself when we trace out the continuous degrees of freedom associated with the environment;
the resulting effective Hamiltonian is then explicitly non-Hermitian.
This effective Hamiltonian has only finite degrees of freedom remaining, corresponding to the discrete portion of the open quantum system, which is usually of primary interest. 
The first and most celebrated example in the literature may be the optical potential in nuclear physics.
It was perhaps first introduced as a phenomenological potential but various researchers, Feshbach in particular, formulated it more rigorously~\cite{Feshbach58,Feshbach62,Foldy69}.

%We can translate the formulation for a tight-binding model, which leads to an energy-dependent effective potential; see Appendix C of Ref.~\cite{SHO11}.
%We then explicitly have a complex potential with a positive imaginary part at the point of the system of interest that had a coupling to the environment with the boundary condition of incoming energy, which represents an effective gain, or a source. 
%On the other hand, a complex potential with a negative imaginary part appears at the point that had a coupling to the environment with the boundary condition of outgoing energy, which represents an effective loss, or a sink.
In the case of the well-known tight-binding model this formulation leads to an energy-dependent effective potential; see Appendix C of Ref.~\cite{SHO11}.
With the boundary condition of incoming energy we then have an effective Hamiltonian with a positive imaginary complex potential at the point where the discrete system couples to the environment, which represents an effective gain (or a source). 
On the other hand, a negative imaginary complex potential appears where the discrete system couples to the environment with the boundary condition of outgoing energy, which represents an effective loss (or a sink).

As a recent development in the study of non-Hermitian physics, systems with both gain and loss have attracted a great deal of attention over the past two decades.
Bender and Boettcher in 1998 demonstrated that one may relax Hermiticity in favor of $\PT$-symmetry (parity-time) in quantum mechanics and still obtain a real-valued energy spectrum in certain regions of parameter space~\cite{BB98,BBM99}. 
This has led some researchers to consider whether quantum mechanics could be reformulated in terms of $\PT$ symmetry;
see, for example, Refs.~\cite{BQZ01,DDT01,BBJ02,Weigert03,AM_brach} and particularly the references appearing in Refs.~\cite{Bender_review,Bender_review2}.
This theoretical question in turn inspired the idea of constructing physical systems that exhibit $\PT$-symmetry in the form of balanced gain and loss components arranged in a spatially-symmetric manner.

A number of investigations have been carried out along these lines, both theoretically and experimentally, particularly in the realm of optics~\cite{MGCM,KGM08,PTOptExpt1,PTOptExpt2,Kottos_Nature,ZinPRL11,LonghiJPA11,PTOptExpt3,Uni_Nature,PTOptExpt4,PT_WGM,Longhi_PT_bound}, but also with examples in condensed matter physics~\cite{BFKS09}, simple electronic circuits~\cite{PTCircuitExpt}, coupled mechanical oscillators~\cite{BGOPY13,BBPS13,BGK14}, and mesoscopic superconducting wires \cite{CGBV12}.
A number of intriguing phenomena have been studied in the optical context, including power oscillations~\cite{MGCM,KGM08,PTOptExpt2}, double refraction~\cite{MGCM} unidirectional invisibility~\cite{ZinPRL11,LonghiJPA11,Uni_Nature,AliPRA13} and localized states with novel transient behavior~\cite{PTOptExpt4}.

One central issue in the investigation of $\PT$-symmetric systems is $\PT$ symmetry-breaking.
%If the Hamiltonian is $\PT$-symmetric, its eigenstates can be asymmetric;
%there is no contradiction here because the asymmetric eigenstates always appear as a pair of a state $|\psi\rangle$ and its partner $\PT|\psi\rangle$.
%The $\PT$-symmetric states must have real eigenvalues but the pair of asymmetric states has a pair of complex-conjugate eigenvalues.
In many $\PT$-symmetric systems, one finds a transition between a phase in which all states are $\PT$-symmetric and a phase in which at least some states are not;
the former is often referred to as the unbroken $\PT$-symmetric phase and the latter as the broken phase.
At the $\PT$-symmetry breaking point, two real eigenvalues on the unbroken side coalesce and reappear on the other side of the transition as a complex-conjugate pair;
their associated eigenfunctions are no longer $\PT$-symmetric individually, but only so as a pair (i.e. they appear as a state $|\psi\rangle$ and its partner $\PT|\psi\rangle$).
We emphasize that at the transition point the eigenstates are not merely degenerate, but coalesce into a single state with a fixed universal phase between them~\cite{HeissChirality,HeissEP3}, as verified by experiment~\cite{EPexpt1a,EPexpt1b,EPexpt1c}.  

The $\PT$-symmetry breaking point, where the eigenstates coalesce, is an example of an {\it exceptional point}~\cite{Kato}.  
Similar transitions occur even in open quantum systems described by a Hamiltonian that is Hermitian within the Hilbert space.
In this case the exceptional point is typically associated with the appearance of a resonance state (along with its anti-resonance partner)~\cite{Klaiman10,Hatano11,GRHS12} after two real eigenvalues collide.  
While the large majority of studies on exceptional points appearing in the literature focus on the case of two coalescing eigenvalues (EP2s), the standard nomenclature is to refer to an exceptional point at which $N$ eigenvalues coalescence as an EP$N$~\cite{HeissEP3,GraefeEP3}.  
In this paper we divide the EP2s into two further subcategories:
we refer to an exceptional point at which two real-valued solutions meet to form complex conjugate partners as an EP2A; 
meanwhile we refer to an exceptional point at which two complex solutions with negative (positive) imaginary part coalesce to form two new solutions with negative (positive) imaginary part as an EP2B.

\subsection{$\PT$-symmetric open quantum system}

In this paper, we combine these two non-Hermitian systems in order to analyze a $\PT$-symmetric open quantum system. 
Specifically, we incorporate a centralized $\PT$-symmetric scattering potential $\pm i\Gamma$ into an infinite tight-binding chain with otherwise real-valued site potentials.
In the perspective given above, we can interpret this model as an otherwise standard open quantum system except that two sites are equipped with a direct environmental influence, one with $+i\Gamma$ that injects energy into the chain, the other with $-i\Gamma$ that represents an energy drain.
This may be realized as an optical lattice array in which one waveguide attenuates photon propagation (the `lossy' component) and a second has a compensating amplifying character (the `gain' component). 
We observe how the $\PT$-symmetric gain and loss modify the usual open quantum system properties under two different boundary conditions: outgoing waves and scattering waves. 
For both of these we first consider the general case, including solutions that are $\PT$-asymmetric, and then further investigate the solutions for which the boundary conditions themselves satisfy $\PT$-symmetry.
%then further impose the restriction that the boundary conditions themselves should be $\PT$-symmetric.

First we consider the boundary condition consisting of purely outgoing waves (often called the Siegert boundary condition)~\cite{Gamow28,Siegert39,Peierls59,Landau77,Ostrovsky05,Kunz06,Kunz08,Sasada08,HSNP08,NH_H_eff}, which yields the discrete spectrum for the system, including all bound states and other solutions.
We also observe the location of all exceptional points and other spectral features of interest.  
Here we demonstrate that for moderately small values of the $\PT$-parameter $\Gamma$, the spectral characteristics remain typical of traditional Hermitian open quantum systems.
However, as we increase $\Gamma$ explicitly non-Hermitian spectral properties emerge.

We find a resonance state with vanishing decay width for certain specific values of $\Gamma$.  
In the context of a Hermitian open quantum system we would refer to this as a bound state in continuum (BIC) (see, e.g. Refs.~\cite{BIC_1929,BIC_1975,ONK06,LonghiEPJ07,TGOP07,BIC_opt_expt1} and references therein).
While BICs typically appear owing to geometric effects and their wave functions discontinuously vanish outside a finite support, 
the present phenomenon results in a \emph{delocalized} wave function with an eigenvalue that appears directly in the scattering continuum.  
For this reason, we refer to this state as a {\it resonance in continuum} (RIC).

\begin{figure}
\begin{center}
%\hspace*{0.05\textwidth}
\includegraphics[width=0.5\textwidth]{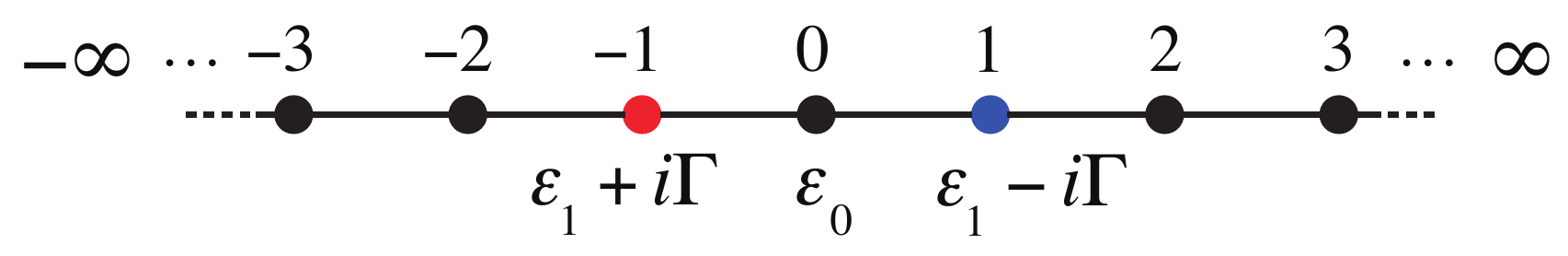}
%\hfill
% \includegraphics[width=0.4\textwidth]{inline_k_par_plot}
% \hspace*{0.05\textwidth}
\end{center}
%\\
\vspace*{-\baselineskip}
\caption{Geometry for $\PT$-symmetric optical lattice with scattering potential given in Eq.~(\ref{eq-potential})
 }
 \label{fig:PT.open quantum system.geo}
 \label{fig1}
 \end{figure}

We further demonstrate the presence of localized states with complex eigenvalues that have recently been observed in an experiment~\cite{PTOptExpt4} and have since been considered in the theoretical works Refs.~\cite{Longhi_PT_bound,BGK14}.
We note that, unlike the RIC, these complex bound states appear over a wide range of parameter values, and, as observed in Ref.~\cite{PTOptExpt4}, the real part of the eigenvalue for these states may appear in the scattering continuum.  
Here we clarify that these localized states have complex conjugate values that sit in the first Riemann sheet in the complex energy plane, something that is not allowed in Hermitian open quantum systems.  
We also emphasize that while these states are indeed localized, they are not stationary states of the Hamiltonian.  
Instead, in an experiment they demonstrate either an amplifying or an absorbing characteristic~\cite{PTOptExpt4}. 
However, given that the real part of these eigenvalues may reside within the continuum, in Ref.~\cite{Longhi_PT_bound} the author classifies these states as a type of generalized BIC.  
By contrast, in this paper we emphasize that since these solutions are localized but non-stationary they would generally behave in a manner that is quite distinct from the usual concept of a BIC.  
That having been acknowledged, we further point out that there are some parameter ranges for which the imaginary part of the eigenvalues for these states will be very small, and hence they should take on a quasi-bound state behavior for these parameter values, similar to the quasi-bound state in continuum appearing in Ref.~\cite{NHGP07,GNHP09}. Specifically, these states should behave as bound states on time-scales $t < \Gamma^2 / 4$, where the gain-loss defect parameter $\Gamma$ exceeds the energy scale of the embedding optical bandwidth; we propose that these states might be detectable, for example, in a $\PT$-symmetric optical fiber loop array with a defect region \cite{PTOptExpt4} that is modified to imitate our potential introduced in Sec. \ref{sec:PT.model} below (see Fig. \ref{fig1}). 

We then focus our attention on the ordinary bound state solutions appearing in our system and demonstrate that the wave function for these states satisfies $\PT$-symmetric boundary conditions.  
Further, we clarify that the wave function for virtual bound states (with real eigenvalue) is also $\PT$-symmetric, despite the fact that these states do not appear in the usual diagonalization scheme.  
%This emphasizes that one should associate the $\PT$-symmetric breaking point with the \emph{appearance} of a complex conjugate pair of eigenvalues rather than the mere disappearance of a bound state.

We then consider the case of scattering wave boundary conditions.  
In the general case ($\PT$-asymmetric scattering waves) we observe that the parameter choices associated with the RIC result in a divergence in the reflection and transmission coefficients.  
This phenomenon has previously appeared in the literature in which it is referred to as a spectral singularity~\cite{Bender&Wu,Shanley,AliMPRL09,AliMPRA09,AliMPRA11,Longhi09SS,Longhi10SS} and physically can be associated with both lasing and coherent perfect absorption~\cite{Chong10,Wan11,LonghiCPA11}.  
We then demonstrate that a subset of the scattering wave solutions yield perfect transmission through the scattering region.  In the special case in which the scattering potential is pure imaginary, we show that one can obtain perfect transmission for any continuum scattering states by appropriately choosing the value of $\Gamma$; this property approximately holds when small real-valued defects are introduced.  We further demonstrate in this case that invisibility (perfect transmission with no scattering phase shift) can be obtained at discrete values within the continuum.

In Sec.~\ref{sec:PT.model} below we present our prototype model for an open quantum system with a 
$\PT$-symmetric defect potential.
%, which may be realized as an optical lattice array in which one waveguide attenuates photon propagation (the `lossy' component) and a second has a compensating amplifying character (the `gain' component).  
Then in Sec.~\ref{sec:PT.outgoing} we study the model under the boundary condition of outgoing waves, which yields the discrete spectrum associated with the defect potential.  
For the simplest case of a purely complex defect potential, we locate all exceptional points in the spectrum and characterize the properties of the spectrum in their vicinity; 
we further locate the RIC eigenvalues and write the associated wave function as an outgoing plane wave from the defect region.  
We also identify the parameter ranges that give rise to the localized states with complex eigenvalues and point out the situation in which some of these solutions might behave as quasi-bound states.  
In Sec.~\ref{sec:PT.outgoing.spec.ep1} we generalize this picture by considering a potential with both real and imaginary defects.  
Here we demonstrate that as one deforms the system parameters, the RIC may exit the continuum by splitting into a bound state and a virtual bound state at the band edge; 
we believe that this point has not previously appeared in the literature.  
We note that traditional real-valued bound states also may appear for this more general potential.
We study in closer detail the formal properties of the bound states in Sec.~\ref{sec:PT.bound}, demonstrating that they satisfy $\PT$-symmetric boundary conditions as expected.
We also consider the $\mathcal{CPT}$ norm for these states, which we believe has only previously been investigated in closed $\PT$ systems.

We then turn to the scattering boundary conditions in Sec.~\ref{sec:PT.scattering}, which we use to characterize the RIC in greater detail.  
We also show that a subset of the scattering wave solutions give rise to perfect transmission through the scattering region, and in the case of a purely imaginary defect potential, there are two scattering solutions that support invisible signal propagation.  
We further demonstrate a connection between the localization transition in the discrete spectrum and the perfect transmission states that might be useful from the perspective of designing systems with predictable transport properties.
We also point out a possible application in the form of a `switch' that is sensitive to invisible transmission originating from the left (right), but ignores such transmission from the right (left).
In Sec.~\ref{sec:PT.scattering.2} we demonstrate that a scattering wave solution can be obtained that itself satisfies $\PT$-symmetric boundary conditions.  We also introduce the $\PT$-current, which is conserved for the (general) scattering wave solutions in our system, and which experiences a divergence associated with the perfect transmission states.
We summarize our work and make concluding remarks in Sec.~\ref{sec:conclusion}.   
We also present some details of the calculations from the main text in two appendices.

%%%%%%%%%%%%%%%%%%%%%
%%%%%%%%%SECTION: PT-symmetric scattering model
%%%%%%%%%%%%%%%%%%

\section{$\PT$-symmetric optical lattice model}
\label{sec:PT.model}
\label{sec2}

%Optical lattices can be experimentally realized by a variety of methods, including... 
%[\textcolor{red}{cite}]
%Optical lattices exhibiting some form of $\PT$-symmetry have also been studied~\cite{PTOptExpt4}[\textcolor{red}{more}].

In the present paper, we study a tight-binding model with a $\PT$-symmetric scattering defect potential, which can be realized as an optical lattice array or could be approximated by a modified version of the $\PT$-symmetric optical fiber loop array with a defect studied in Ref.~\cite{PTOptExpt4} or other systems appearing in the literature~\cite{PT_WGM,RDM05}.
Our tight-binding model takes the form
%Here we consider an optical array with nearest neighbor coupling given by
\begin{align}\label{eq-model}
H=-\sum_{x=-\infty}^\infty
\left(|x+1\ket \bra x|+|x\ket \bra x+1|\right)
+\sum_x V(x)|x \ket \bra x|,
\end{align}
in which the defect potential is specified as
\begin{align}\label{eq-potential}
V(x)=
\begin{cases}
\varepsilon_1+i\Gamma & \quad\mbox{for $x=-1$},
\\
\varepsilon_0 & \quad\mbox{for $x=0$},
\\
\varepsilon_1-i\Gamma & \quad\mbox{for $x=-1$},
%\\
%0&\quad\mbox{otherwise},
\end{cases}
\end{align}
where $\varepsilon_0$, $\varepsilon_1$, and $\Gamma$ are all real, with $V(x) = 0$ otherwise, such that our scattering potential is confined to the central sites $|x| \le 1$.
The positive imaginary part of the complex potential contributes a factor $\exp[-i(i|\Gamma|)t]=\exp(|\Gamma|t)$ to the time evolution, and hence is interpreted as being influenced by a particle bath that constantly injects energy as $+i|\Gamma|$.
That with a negative imaginary part is similarly interpreted as a particle bath that constantly drains energy as $-i|\Gamma|$.

The off-diagonal part of the Hamiltonian~\eqref{eq-model} is Hermitian (real symmetric) while the diagonal potential is not.
It nonetheless satisfies the condition $V(x)^\ast = V(-x)$, which guarantees the system is $\PT$-symmetric~\cite{Ganainy07,Makris08,KGM08}.  
Stated explicitly, the parity transformation $\mathcal{P}$ swaps the potentials at $x=-1$ and $x=+1$ while the time reversal operator $\mathcal{T}$ (which is complex conjugation) flips them back to the original configuration.  We note that several studies on $\PT$-symmetric tight-binding models may be found in the literature, some of which are related to our model above~\cite{Longhi_PT_bound,JSBS10,JB11,DVL13,VHIC14}.

%%%%%%%%%
%%%%%%%%%
%%%%%%%%%
%%%%%%%%%

The Schr\"{o}dinger equation $H|\psi\rangle=E|\psi\rangle$ for the Hamiltonian~\eqref{eq-model} can be written explicitly in the following way.
First, let us consider the projection $\langle x|H|\psi\rangle=E\langle x|\psi\rangle$
for the system component outside of the scattering potential $|x|\geq 2$, in which case $V(x)\equiv0$.
We thus obtain
\begin{align}\label{eq-Sch.TB.gen}
-\psi(x-1)-\psi(x+1)=E\psi(x),
\end{align}
where $\psi(x)=\langle x|\psi\rangle$.
The solution is given by $\psi(x)=e^{\pm ikx}$ with the eigenvalue
\begin{align}\label{eq-dispersion-in-k}
E(k)=-2\cos k,
\end{align}
which defines the scattering continuum for our system in the range $|E(k)| \le 2$ with $k \in [ -\pi, \pi ]$.
To solve the eigenvalue problem in the scattering region, we hold the continuum dispersion $E(k)$ and evaluate the Schr\"{o}dinger equation for $x=0$ and $\pm 1$, by which we obtain
\begin{align}
\label{eq-Sch1}
-\psi(-2)-\psi(0)+(\varepsilon_1+i\Gamma)\psi(-1)&=E(k)\psi(-1),
\\
\label{eq-Sch2}
-\psi(-1)-\psi(1)+\varepsilon_0\psi(0)&=E(k)\psi(0),
\\
\label{eq-Sch3}
-\psi(2)-\psi(0)+(\varepsilon_1-i\Gamma)\psi(1)&=E(k)\psi(1).
\end{align}
A given solution $\psi(x)$ must satisfy these equations, subject to a specific choice for the boundary conditions.  In Sec.~\ref{sec:PT.outgoing} below we consider the boundary condition for resonant states that consist of purely outgoing waves, while in Sec.~\ref{sec:PT.scattering} we consider the boundary conditions for scattering states.

\begin{figure}
\centering
\includegraphics[width=0.7\textwidth]{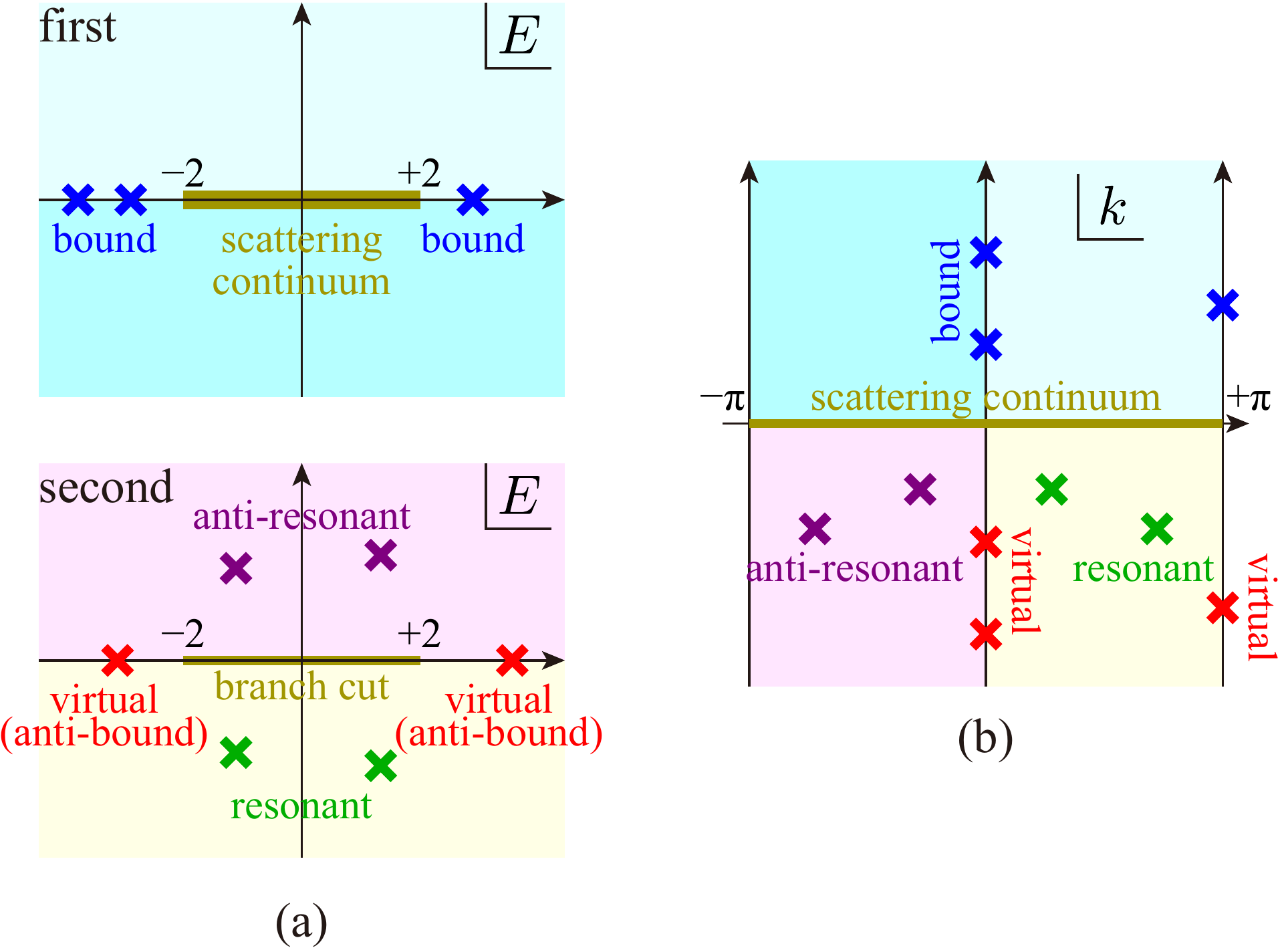}
\caption{A typical distribution of the eigenvalues of the Hermitian tight-binding model on (a) a complex energy plane and (b) a complex $k$ plane.}
\label{fig2}
\end{figure}

For later reference, let us present in Fig.~\ref{fig2} a typical distribution of the eigenvalues of the Hermitian tight-binding model, that is, for $\Gamma=0$.
We will heavily make use of the terms in the figure.

The complex $E$ plane in Fig.~\ref{fig2}(a) consists of two Riemann sheets; 
they are connected by a branch cut that extends over the range $-2\leq E\leq +2$.
The first sheet corresponds to the upper half of the complex $k$ plane in Fig.~\ref{fig2}(b), while the second sheet to the lower half.
More specifically, the upper half ($\im E > 0$) of the first Riemann sheet in the complex $E$ plane corresponds to the first quadrant of the complex $k$ plane, the lower half ($\im E < 0$) to the second quadrant, the upper half of the second sheet to the third quadrant, and the lower half to the fourth quadrant.
Notice that we can go, for example, from the upper half of the first sheet over to the lower half of the second sheet continuously through the branch cut, which corresponds to moving from the first quadrant to the fourth quadrant in the complex $k$ plane.

The scattering states continuously surround the branch cut on the real axis of the complex $E$ plane.
We hereafter refer to the scattering continuum as the energy band and to the end points of the continuum as the band edges, following the custom of condensed-matter physics.
In the complex $k$ plane, the scattering continuum is on the real axis, which is restricted to the first Brillouin zone $-\pi<k\leq +\pi$;
note that the line $\re k=-\pi$ is identified with the line $\re k=+\pi$ as a result of the lattice periodicity.

Bound states can exist on the first Riemann sheet below and above the energy band, that is, to the left ($E < -2$) and to the right ($E > 2$) of the scattering continuum.
Those below the band lie on the positive imaginary axis of the complex $k$ plane, while those above the band lie on the positive part of the line $\re k=+\pi$.
Notice that the bound states here have purely real eigenvalues;
in other words, they never exist on the first and second quadrants of the complex $k$ plane except on the lines $\re k=0$ and $\re k=+\pi$.
We will show below that once we introduce the non-Hermiticity, complex eigenvalues can appear in the first Riemann sheet (and therefore on the upper half of the complex $k$ plane);
this is one critical difference between Hermitian and non-Hermitian open systems.

The resonant states appear in the lower half of the second Riemann sheet, which is the fourth quadrant of the complex $k$ plane, while their anti-resonant partners reside on the upper half of the second sheet, which is the third quadrant of the complex $k$ plane.
These are related to one another through time-reversal symmetry~\cite{Hatano14}.
Virtual (or anti-bound) states can also appear to the left and right of the branch cut on the second Riemann sheet, which respectively correspond to the negative ($\im k < 0$) parts of the lines $\re k=0$ and $\re k=+\pi$.

%%%%%%%%%%%%%%%%%%%%%
%%%%%%%%%SECTION: outgoing waves B.C.
%%%%%%%%%%%%%%%%%%

\section{Outgoing waves boundary condition and discrete spectrum}
\label{sec:PT.outgoing}
\label{sec3}

In the present section our first consideration is the resonant states; there are several ways of computing these states~\cite{NH_H_eff}.
We here use the Siegert boundary condition~\cite{Gamow28,Siegert39,Peierls59,Landau77,Ostrovsky05,Kunz06,Kunz08,Sasada08,HSNP08,NH_H_eff}, which dictates that the system has outgoing waves only;
this is equivalent to looking for all poles of the $S$ matrix.
The solutions of the resulting polynomial equation give the discrete eigenvalues associated with the scattering region, as shown below.  

Our purely outgoing wave function takes the form
\begin{equation}
  \psi (x) =
	\begin{cases}
		B e^{-i k x} 			& \mbox{for $x \le -1$,}    	\\
		\psi(0)				& \mbox{for $x = 0$,}   	\\
		C e^{  i k x}			& \mbox{for $x \ge 1$.}
	\end{cases}
\label{outgoing.wave.fcn}
\end{equation}
This boundary condition gives $\psi(\pm2)=e^{ik}\psi(\pm1)$, which brings Eqs.~\eqref{eq-Sch1}--\eqref{eq-Sch3} into a closed form~\cite{Sasada08}.
%%
%To solve for the eigenvalues, we aim to apply this generic $\psi(x)$ and rewrite Eqs.(\ref{eq-Sch1}a-c) as  a matrix equation for the wave function elements in the scattering region $\psi (-1), \psi(0), \psi(1)$; the appropriate $k$ (and, by extension, $E(k))$, values will then be determined by the characteristic equation.  However, first we must eliminate two additional unknowns $\psi (\mp 2)$, or, equivalently, the wave function coefficients $A$ and $B$.
%We can re-write $A$ and $B$ in terms of the other unknowns by evaluating the Schr\"odinger equation at sites 
%$\bra \mp 2 | H | \psi \ket 
%= - \left[ \bra \mp 3 | \psi \ket + \bra \mp 1 | \psi \ket  \right] 
%= E(k) \bra \mp 2 | \psi \ket$ 
%%and applying Eq.~(\ref{eq-Sch.TB.gen})
% and apply Eq.~(\ref{outgoing.wave.fcn}) and $E(k) = - 2 \cos k$;
%after re-arranging we obtain
%%
%\begin{equation}
%  - \frac{1}{2} \left[ \bra \mp 3 | \psi \ket + \bra \mp 1 | \psi \ket  \right]
 %	= z \bra -2 | \psi \ket ,
%\label{PT.1.outgoing.1}
%\end{equation}
%%
%which then gives
%%
%\begin{equation}
%  \left\{ \begin{array}{c}
%  		A	\\
%		B
%  	\end{array} \right\}
%	%%
%		= \bra \mp 1 | \psi \ket e^{-i k}
%		,
%\label{outgoing.A.B}
%\end{equation}
%%%
%which is boundary condition-dependent.
%Now we apply Eqs.~(\ref{outgoing.wave.fcn}) and~(\ref{outgoing.A.B}) to Eqs.~(\ref{eq-Sch1}a-c) to write the matrix equation
%%
We thereby obtain
\begin{equation}
  \begin{pmatrix}
  	- \lambda + \varepsilon_1 + i \Gamma 	& -1			& 0    \\
	-1							& \varepsilon_0& -1   \\
	0							& -1		& - \lambda + \varepsilon_1 - i \Gamma  \\
	\end{pmatrix}
		\begin{pmatrix}
		\psi(-1) \\
		\psi(0)\\
		\psi(1)
		\end{pmatrix}
%  \left[ \begin{array}{c}
%  	\bra -1 | \phi \ket	\\
%	%%
%	\bra 0 | \phi \ket		\\
%	%%
%	\bra 1 | \phi \ket	\\
%  	%%
%	\end{array} \right]
		%%
		%%
	= E(\lambda)
		\begin{pmatrix}
		\psi(-1) \\
		\psi(0)\\
		\psi(1)
		\end{pmatrix},
%  \left[ \begin{array}{c}
%  	\bra -1 | \phi \ket	\\
%	%%
%	\bra 0 | \phi \ket		\\
%	%%
%	\bra 1 | \phi \ket	\\
%  	%%
%	\end{array} \right]
\label{outgoing.matrix0}
\end{equation}
in which we have introduced $\lambda \equiv e^{ik}$ for convenience.  In this notation the continuum dispersion Eq.~\eqref{eq-dispersion-in-k} takes the form
\begin{align}\label{eq-dispersion-in-lambda}
E(\lambda) = - (\lambda + \lambda^{-1}).
\end{align}

%The characteristic equation resulting from Eq.~(\ref{PT.outgoing.matrix}) fixes the discrete eigenvalues associated with the scattering region as the solutions of $P(\lambda_i) = 0$, where
We obtain non-trivial solutions for the discrete eigenvalues $\lambda_j$ when the determinant of the matrix in Eq.~(\ref{outgoing.matrix0}) vanishes.  
This is equivalent to the solutions of the quartic equation 
%$P(\lambda_i; \epsilon_0, \epsilon_1, \Gamma) = 0$ with
$P(\lambda_j) = 0$ with
\begin{equation}
  P(\lambda)
  	\equiv \left( \epsilon_1^2 + \Gamma^2 \right) \lambda^4
		+ \epsilon_0 \left( \epsilon_1^2 + \Gamma^2 \right) \lambda^3
		- \left( 1 - \epsilon_1^2 - 2 \epsilon_0 \epsilon_1 - \Gamma^2 \right) \lambda^2
		+ \left( \epsilon_0 + 2 \epsilon_1 \right) \lambda 
		+ 1
	.
\label{P.lambda}
\end{equation}
For a given solution $\lambda_j$, the physical energy eigenvalue is determined from $E(\lambda_j)$ and the associated wave number is given as $k_j = - i \log \lambda_j$.  
We emphasize that the number of solutions (four), is greater than the matrix dimension (three);
this is because the matrix itself depends on the energy eigenvalue through the variable $\lambda$.
% solution space associated with the matrix equation given in Eq.~(\ref{outgoing.matrix}) is larger than the dimension of the matrix itself.  This is a result of the fact that the eigenvalue $E(\lambda)$ is itself 
%$\lambda$-dependent, as discussed further in Refs.~\cite{NH_H_eff,GRHS12}.

In the remainder of this Section we investigate the four discrete eigenvalue solutions of the quartic equation $P(\lambda_j) = 0$ in detail; 
first, we study the case of a purely imaginary defect potential with $\varepsilon_0 = \varepsilon_1 = 0$ in Sec.~\ref{sec:PT.outgoing.spec}.  
Here we locate all EPs and characterize the behavior of the spectrum in the vicinity of these points.  
We further identify the RIC and write the wave function of this state as a plane wave originating from the impurity sites.
%We note that the spectral properties in this case are quite different among four different regions in the $\Gamma$ parameter space, which we describe in detail.  
%From Region I to IV the transition from one region to the next is marked by the appearance a pair of EP2As, a pair of RICs, and finally a pair of EP2Bs.
%We then discuss the complex-valued localized states that appear in two of these regions in detail.
We also discuss the complex-valued localized states and their asymptotic localization properties as well as drawing attention to the parameter ranges in which some of these states will behave as quasi-bound states.

In Sec.~\ref{sec:PT.outgoing.spec.ep1} we generalize this picture to consider the case 
$\varepsilon_1 \neq 0$ to illustrate two points (we keep $\varepsilon_0 = 0$ for now).  First we note that the EP2As still appear in the spectrum for the $\varepsilon_1 \neq 0$ case, while the EP2Bs vanish.  This suggests that the EP2As may be more robust against parameter perturbations than the EP2Bs, on which we comment in relation to experimental results.  Second, we demonstrate that as we increase the value of $\varepsilon_1$, one of the RICs approaches the band edge and eventually exits the continuum by splitting into a bound state and a virtual bound state.

%%%%%%%%%%%%%%%%%%%%%
%%%%%%%%%SUB-SECTION: discrete spectrum
%%%%%%%%%%%%%%%%%%

\subsection{Discrete spectrum for $\varepsilon_0 = \varepsilon_1 = 0$: exceptional points (EPs), resonant states in continuum (RICs) and quasi-bound states in continuum (QBICs)}\label{sec:PT.outgoing.spec}

%%%%%%%%%
%%%%%%%%%
\begin{figure}
\hspace*{0.05\textwidth}
 \includegraphics[width=0.4\textwidth]{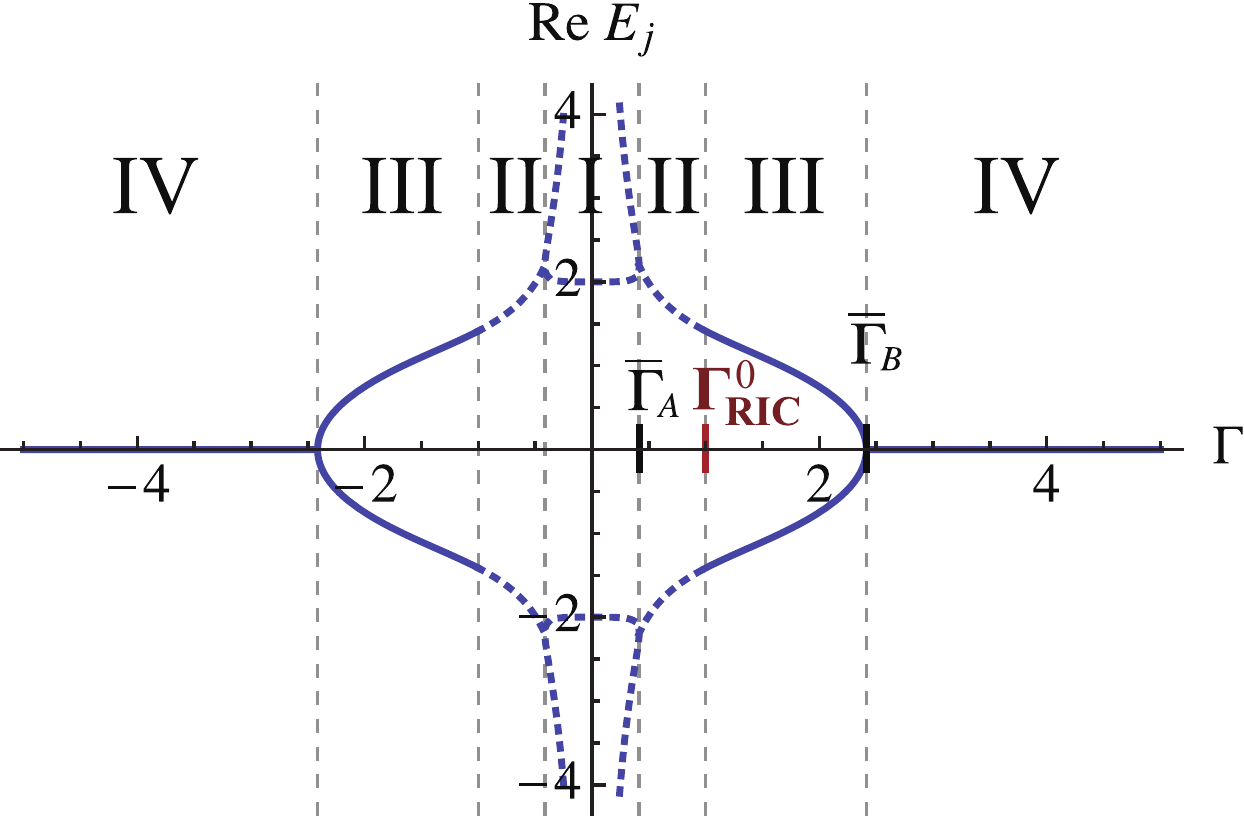}
\hfill
 \includegraphics[width=0.4\textwidth]{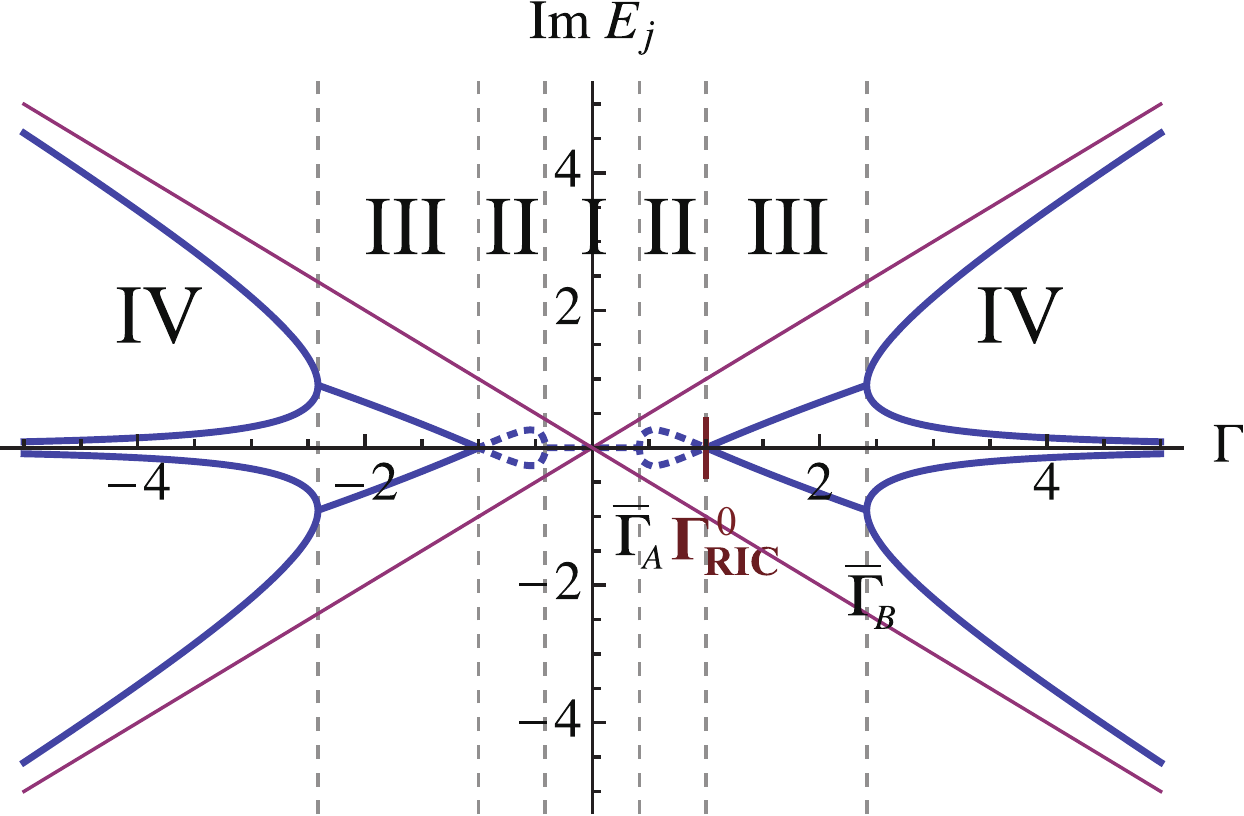}
 \hspace*{0.05\textwidth}
\\
\vspace*{\baselineskip}
\hspace*{0.05\textwidth}(a)\hspace*{0.440\textwidth}(b)\hspace*{0.4\textwidth}
\\
\vspace*{\baselineskip}
\hspace*{0.05\textwidth}
 \includegraphics[width=0.4\textwidth]{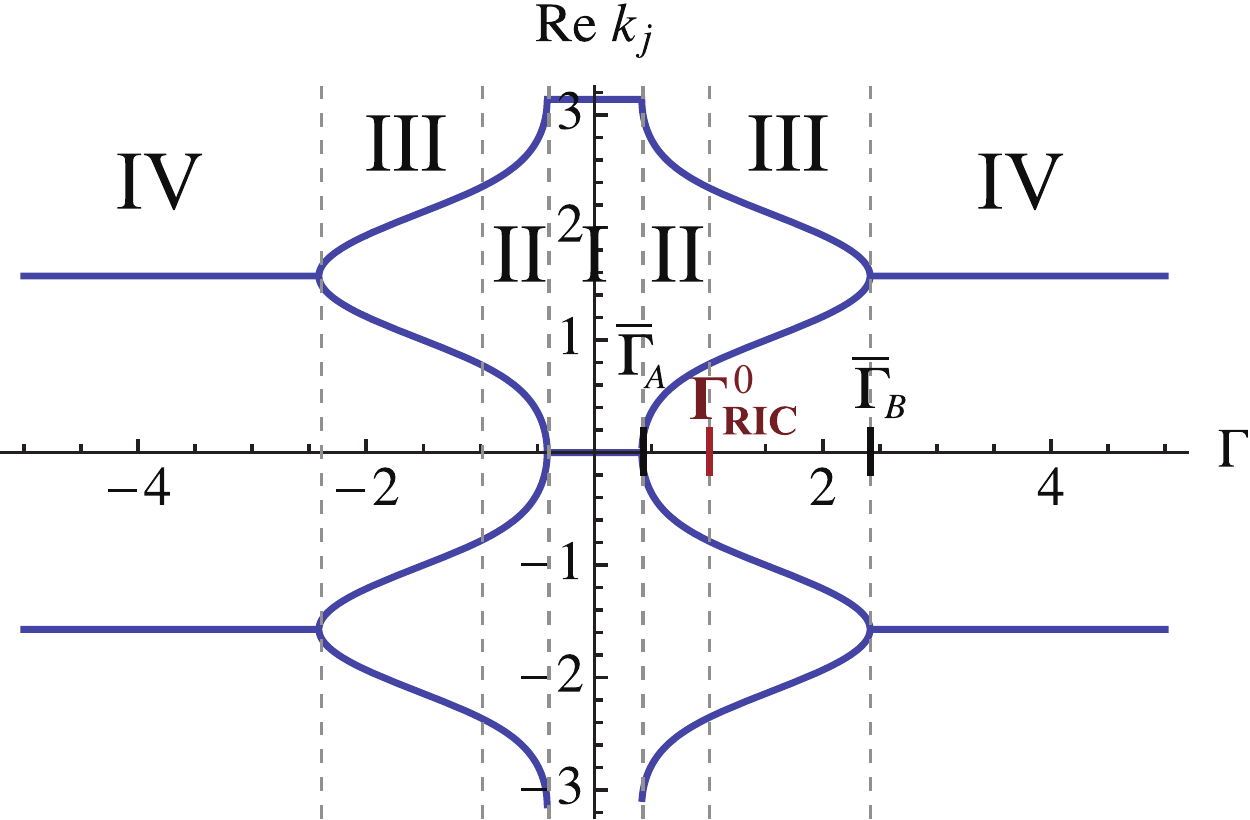}
\hfill
 \includegraphics[width=0.4\textwidth]{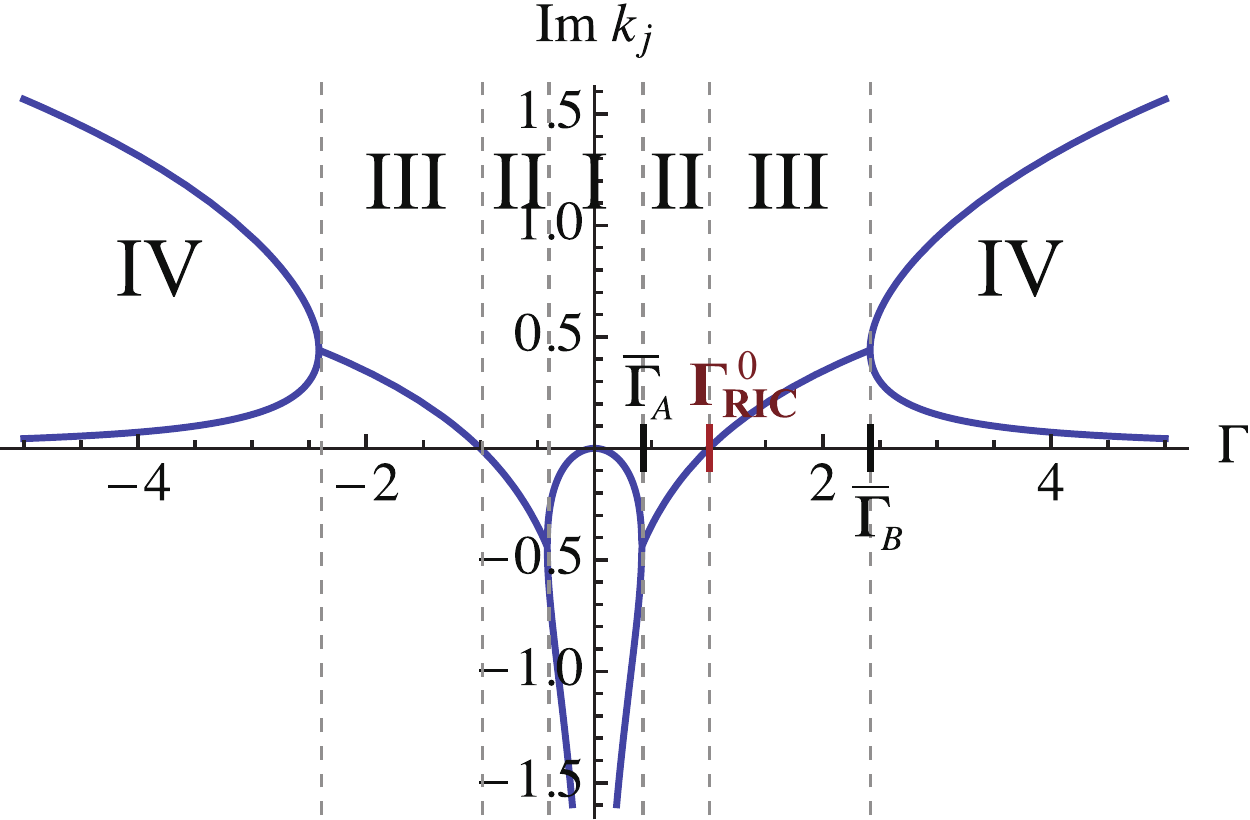}
 \hspace*{0.05\textwidth}
\\
\vspace*{\baselineskip}
\hspace*{0.05\textwidth}(c)\hspace*{0.440\textwidth}(d)\hspace*{0.4\textwidth}
\\
\vspace*{\baselineskip}
\hspace*{0.05\textwidth}
\includegraphics[width=0.4\textwidth]{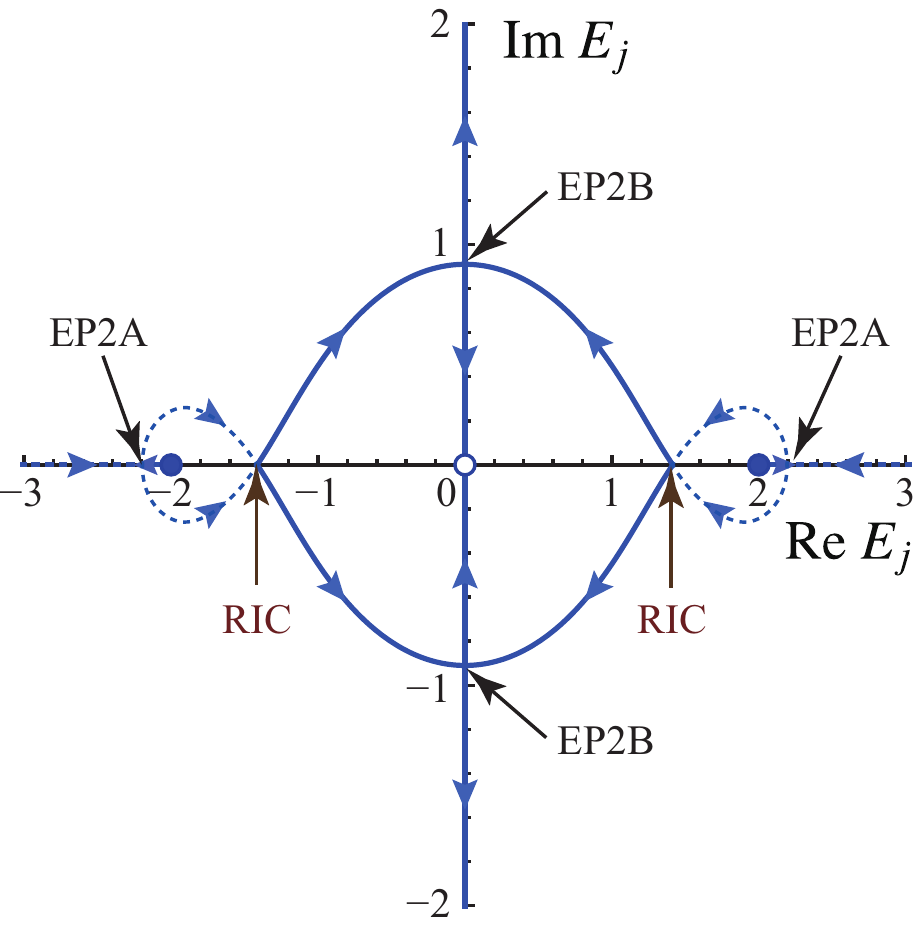}
\hfill
\includegraphics[width=0.4\textwidth]{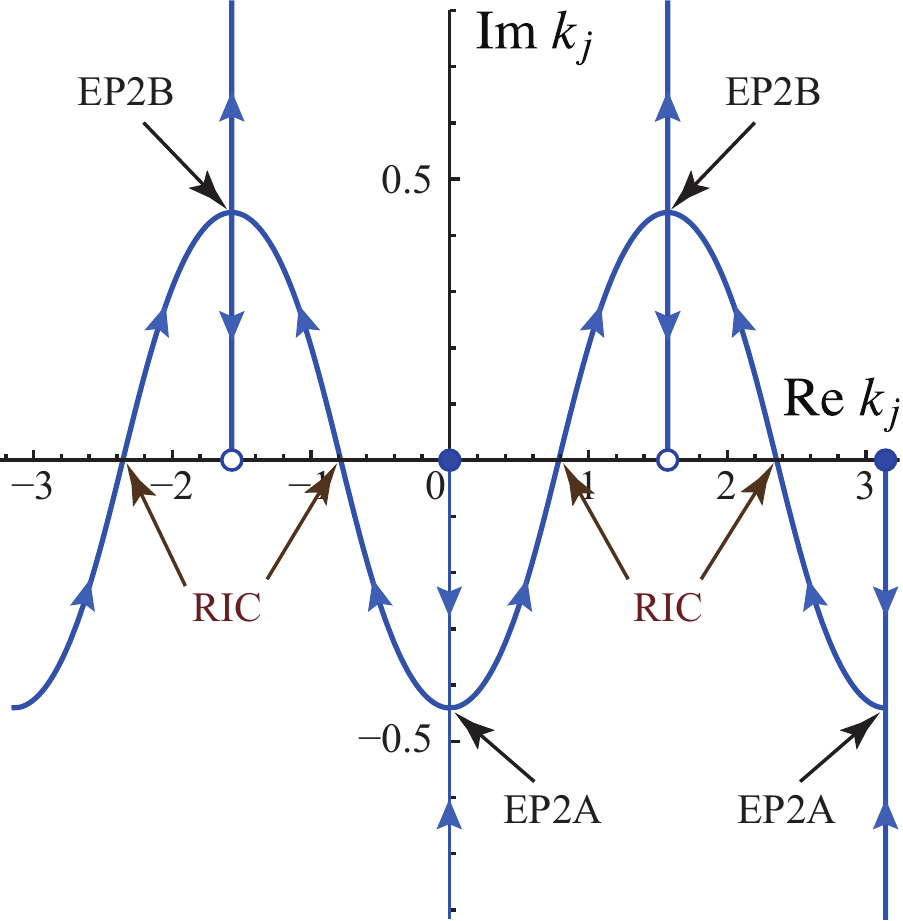}
 \hspace*{0.05\textwidth}
\\
\vspace*{\baselineskip}
\hspace*{0.05\textwidth}(e)\hspace*{0.440\textwidth}(f)\hspace*{0.4\textwidth}
\\
\vspace*{\baselineskip}
\caption{Discrete eigenvalue spectrum for the simplest case $\varepsilon_0 = \varepsilon_1 = 0$: 
(a) $\re E_j$ and (b) $\im E_j$ against the $\PT$-parameter $\Gamma$;  (c)  $\re k_j$ and (b) $\im k_j$ against $\Gamma$; parametric plots of (e) $(\re E_j(\Gamma),\im E_j(\Gamma))$ and (f) $(\re k_j(\Gamma),\im k_j(\Gamma))$ in the complex plane. 
In (e) and (f), the solid circles indicate some of the eigenvalues at $\Gamma=0$, while the open circles indicate those in the limit $\Gamma\to\infty$; 
the arrows indicate how the eigenvalues evolve as $\Gamma$ is increased from $0$ to $\infty$.
%In plots (a) and (b) full curves represents solutions in the first Riemann sheet while dotted curves represent solutions in the second Riemann sheet.  The points $\pm \Gamma_\textrm{RIC}$ denote the transition points between the two sheets.  
%[(c) and (d) are temporary place holders, annotated versions to be added]
}
 \label{fig:PT.0.spec}
 \label{fig3}
 \end{figure}
%%%%%%%%%
%%%%%%%%%

We first consider the discrete eigenvalue spectrum for the simplest case 
of our Hamiltonian $\varepsilon_0 = \varepsilon_1 = 0$, for which the only non-homogeneous element remaining in the system is the gain/loss pair governed by the $\PT$ parameter $\Gamma$.  In this case, the quartic polynomial $P(\Gamma)$ given in Eq.~(\ref{P.lambda}) simplifies to a quadratic in 
$\lambda^2$, yielding the four solutions
\begin{eqnarray}
  \lambda_{1,4}
  &	= & \pm \frac{1}{\sqrt{2} \Gamma}
		\sqrt{ 1 - \Gamma^2 
			+ \sqrt{1 - 6 \Gamma^2 + \Gamma^4} }
		,						\nonumber  \\
  \lambda_{2,3}
  &	= & \pm \frac{1}{\sqrt{2} \Gamma}
		\sqrt{ 1 - \Gamma^2 
			- \sqrt{1 - 6 \Gamma^2 + \Gamma^4} }
		.
%  \lambda_{1,4}
 % &	= & \pm \frac{1}{\sqrt{2} \Gamma}
%		\sqrt{ 1 + 4 \Gamma^2- \Gamma^4 
%			+ \left( 1 - \Gamma^2 \right) \sqrt{1 - 6 \Gamma^2 + \Gamma^4} }
%		,						\nonumber  \\
 % \lambda_{2,3}
 % &  	= & \pm \frac{1}{\sqrt{2} \Gamma}
%		\sqrt{ 1 + 4 \Gamma^2- \Gamma^4 
%			- \left( 1 - \Gamma^2 \right) \sqrt{1 - 6 \Gamma^2 + \Gamma^4} }
%		.
\label{P.lambda.0.solns}
\end{eqnarray}
We plot the real and imaginary parts of the resulting energy eigenvalues $E_j = - \lambda_j - \lambda_j^{-1}$ as a function of $\Gamma$ in Fig.~\ref{fig:PT.0.spec} (a) and (b), as well as the real and imaginary parts of the associated wave number $k_j = - i \log \lambda_j$ in Fig.~\ref{fig:PT.0.spec} (c) and (d).
Figure~\ref{fig:PT.0.spec} (e) and (f) are parametric plots in the complex $E$ plane and the complex $k$ plane in the range of $\Gamma\geq 0$.
In Fig.~\ref{fig:PT.0.spec} (a), (b) and (e), solutions plotted with full curves appear in the first Riemann sheet of the complex $E$ plane while those plotted with a dotted curve appear in the second sheet;
the former are the solutions with positive imaginary parts of $k_j$ and the latter are those with negative parts in Fig.~\ref{fig:PT.0.spec} (d) and (f).
The first and second Riemann sheets of the complex $E$ plane respectively corresponds to the upper- and lower halves of the complex $k$ plane;
a branch cut running from $E=-2$ to $E=2$ connects the two Riemann sheets.
We realize from the Siegert boundary condition~\eqref{outgoing.wave.fcn} that every solution on the first Riemann sheet has a positive imaginary part of the wave number and hence its wave function is bounded in $x$ space, while
every solution on the second Riemann sheet has a wave function that diverges along the leads of the optical array.

We immediately observe one critical difference between Hermitian and $\PT$-symmetric open quantum systems: 
in the $\PT$-symmetric case, solutions with complex eigenvalues are allowed to appear in the first Riemann sheet, with localized wave functions.
This is in stark contrast to the Hermitian case, in which complex-valued solutions are allowed to appear only in the second sheet, where they give rise to delocalized resonance and anti-resonance states.

Let us summarize the evolution of the discrete eigenvalues from $\Gamma=0$ to $+\infty$ along the lines of Fig.~\ref{fig:PT.0.spec} (e); 
the change for negative $\Gamma$ is symmetric as this just amounts to swapping the gain and loss elements.
At $\Gamma=0$, one eigenvalue is at the lower edge of the continuum $E=-2$ and another at the upper edge $E=+2$ (solid circles in Fig.~\ref{fig:PT.0.spec} (e) and (f)).
There are also two eigenvalues at $E=-\infty$ and at $E=+\infty$ both on the real axis of the second Riemann sheet.

As we increase $\Gamma$ from $0$, the eigenvalues at $E=\pm2$ separate off from the band edges and move outward, while the eigenvalues at $E=\pm\infty$ move inwards, all four along the real axis of the second Riemann sheet of the complex energy plane.
These eigenstates are referred to as virtual bound states or anti-bound states in the sense that they are real-valued solutions that are spatially delocalized~\cite{Hatano14,HSNP08,GPSS13}.  
The positive pair of solutions and the negative pair each coalesce at a point on the real axis of the second Riemann sheet at $\Gamma=\GamA = \sqrt{2} - 1$, which is a second-order exceptional point.
We label this point $\Gamma=\GamA$ as an EP2A and the region up until this point $0<\Gamma<\GamA$ as Region I.
After passing the EP2A, all four eigenvalues become complex on the second sheet, forming two resonance/anti-resonance pairs symmetrically on the positive and negative sides.
In the vicinity of the EP2A, the eigenvalues can be expanded in the characteristic form~\cite{Kato,GRHS12}
\begin{equation}
  E_\textrm{A}^- (\Gamma)
	= - \sqrt{2 \left(1 + \sqrt{2}\right)} \pm \frac{1}{2^{1/4} \sqrt{-1 + \sqrt{2}}} \sqrt{\GamA^2 - \Gamma^2}
\label{z.A.m.exp}
\end{equation}
for the resonance/anti-resonance pair with negative real part and
\begin{equation}
  E_\textrm{A}^+ (\Gamma)
	= \sqrt{2 \left(1 + \sqrt{2} \right)} \pm \frac{1}{2^{1/4} \sqrt{-1 + \sqrt{2}}} \sqrt{\GamA^2 - \Gamma^2}
\label{z.A.p.exp}
\end{equation}
for the resonance/anti-resonance pair with positive real part.  
The derivation of these expressions is detailed in App.~\ref{app:EP.calcs}.

We can regard Region I as the $\PT$-unbroken phase and the EP2A at $\Gamma=\GamA$ as the $\PT$-symmetry breaking point.
As we see in Fig.~\ref{fig:PT.0.spec}, Region I is the only continuous parameter region in which all discrete energy eigenvalues are real.

As we continue to increase $\Gamma$, the complex eigenvalues eventually turn around and then return to the real energy axis at $\Gamma=\GamRIC =1$.
Although each pair of the energy eigenvalues are degenerate when they reach the real axis, their wave numbers are all distinct as can be seen in Fig.~\ref{fig:PT.0.spec} (c) and (f), and therefore this point represents a degeneracy in the standard sense, \textit{not} a coalescence in the sense of the exceptional point.
We refer to these states as resonances in continuum (RICs) for reasons described below in Sec.~\ref{sec3B} (also see Sec. \ref{sec:PT.scattering.RIC}), and we refer to the region $\GamA<\Gamma<\GamRIC$ as Region II.

As we further increase $\Gamma$ such that $\Gamma > \GamRIC$, the four solutions pass through the branch cut running from $E=-2$ to $E=2$ and emerge on the first Riemann sheet of the $E$ plane.
%Because there is a branch cut in the complex energy plane running from $E=-2$ to $E=2$, the eigenvalues come up onto the first Riemann sheet after crossing the real energy axis.
This is equivalent to the observation that these solutions now have an effective wave number $k_j$ with positive imaginary part as shown in Fig.~\ref{fig:PT.0.spec}(d) and~(f). 
This implies that the wave function for these states $\psi_j (x) \sim e^{i k_j |x|}$ is localized, although the real part of the eigenvalues lie within the range $-2<\re E<2$ (see Fig.~\ref{fig:PT.0.spec}(a)).  These type of states were recently observed in an experiment based on light transmission through an effective $\PT$-symmetric array of optical fiber loops~\cite{PTOptExpt4} in which they gave rise to a pair of exponentially growing and decaying localized states within the continuum.

As $\Gamma$ reaches the value $\Gamma=\GamB=1 + \sqrt{2}$, 
these states coalesce on the imaginary axis of the complex energy plane at $\Gamma=\GamB=1 + \sqrt{2}$, two on the positive side and the other two on the negative side, which is another second-order exceptional point (this time occurring in the first Riemann sheet).  
We refer to this point as an EP2B, because it involves a pair of complex eigenvalues coalescing before becoming another pair of complex eigenvalues; we also refer to the region $\GamRIC<\Gamma<\GamB$ as Region III.
In the vicinity of the EP2B, the eigenvalues can be expanded as
\begin{equation}
  E_\textrm{B}^- (\Gamma)
	= - i \sqrt{2 \left(-1 + \sqrt{2}\right)} \pm \frac{i}{2^{1/4} \sqrt{1 + \sqrt{2}}} \sqrt{\Gamma^2 - \GamB^2}
\label{z.B.m.exp}
\end{equation}
for the two eigenvalues with negative imaginary part, and
\begin{equation}
  E_\textrm{B}^+ (\Gamma)
	= i \sqrt{2 \left(-1 + \sqrt{2}\right)} \pm \frac{i}{2^{1/4} \sqrt{1 + \sqrt{2}}} \sqrt{\Gamma^2 - \GamB^2}
\label{z.B.p.exp}
\end{equation}
for the two with positive imaginary part, similar to the expressions near the EP2A above (see App.~\ref{app:EP.calcs}).
%Here $| \bar{z}_\textrm{B} | = \sqrt{2 (-1 + \sqrt{2})}$ represents the point on the 

After surpassing the EP2B, two eigenvalues move to the origin while the other two go off to $\pm i\infty$, all on the imaginary axis of the first Riemann sheet of the complex energy plane. 
We refer to this region $\Gamma>\GamB$ as Region IV.
Since $\Gamma \gg 1$ generally holds here, we may expand the solutions in Eqs.~(\ref{P.lambda.0.solns}) in powers of $1 / \Gamma$ to show that two of these solutions behave as $E_{1,4} \approx \pm i (\Gamma - 2 / \Gamma)$; 
note that in the limit $\Gamma \rightarrow \infty$, these two solutions asymptotically approach the simple value of the gain or loss component of the $\PT$ parameter $\Gamma$, as is indicated in Fig.~\ref{fig3}(b), where these two solutions (blue curves) approach the two diagonal (red) lines.
Indeed, as shown in Appendix~\ref{app:iv.calcs}, the solution $E_1 \sim + i \Gamma$ is localized at site $x = -1$, while the solution $E_4 \sim - i \Gamma$ is localized at the $x = 1$; 
hence, these two solutions gradually begin to mimic the original uncoupled gain/loss pair for large $\Gamma$.  We comment further on the asymptotic localization properties of these states and show that the solutions $E_{2,3}$ behave as quasi-bound states in the continuum in Sec. \ref{sec:PT.outgoing.QBIC}.

We emphasize that the physics in Regions I and II could arise in Hermitian open quantum systems as well; 
explicitly non-Hermitian properties appear in Regions III and IV with the appearance of the RIC and then the complex eigenvalues on the first Riemann sheet.

%%%%%%%%%%%%%%%%%%%%
%%%%%%%%%SUB-SECTION break: RIC
%%%%%%%%%%%%%%%%%%%%

\subsection{Resonant state in continuum (RIC)}
\label{sec3B}

Here we describe the resonant states in continuum (RICs) at the point $\Gamma = \GamRIC$ in greater detail.
As summarized above, the eigenvalues here appear on the real axis, embedded in the energy continuum that spans $-2\leq E \leq 2$.
At a glance, these states appear similar to bound states in continuum (BICs), which in Hermitian systems appear as resonances with vanishing decay width~\cite{BIC_1929,BIC_1975,ONK06,LonghiEPJ07,TGOP07,BIC_opt_expt1}.
However, closer inspection reveals that these states are fundamentally different from BICs.

For example, in the (Hermitian) double impurity open quantum system model studied in Ref.~\cite{TGOP07} it is shown that BICs appear as localized states between the two impurities; 
due to interference, the wave function for the BIC states vanishes identically outside of the impurity region.
More generally, BICs often appear for geometrical reasons and hence are strictly confined in some spatial area.

The present RICs, however, take the form
\begin{equation}
  \psi_\textrm{RIC} (x) =
	\left\{ \begin{array}{ll}
		\displaystyle \mp \frac{1}{\sqrt{6}} e^{\pm i \pi (x+1)/4} 		& \mbox{for $x \le -1$}    	\\
		\displaystyle \frac{1}{\sqrt{3}}							& \mbox{for $x = 0$}   	\\
		\displaystyle \mp \frac{1}{\sqrt{6}} e^{\mp i \pi (x-1)/4}		& \mbox{for $x \ge 1$}
	\end{array}
	\right.  
\label{psi.RIC}
\end{equation}
at $\Gamma = \GamRIC$, for the respective eigenvalues $E_\textrm{RIC} = \sqrt{2}$ (with $k_\textrm{RIC} = \pm 3\pi/4$) and $E_\textrm{RIC} = -\sqrt{2}$ (with $k_\textrm{RIC} = \pm \pi/4$).

We refer to these points as resonant states in continuum (RICs) in part because the wave function for these states is delocalized as demonstrated in Eq.~(\ref{psi.RIC}), and because these states satisfy the Siegert boundary condition for outgoing waves.
We will comment further on this naming convention in Sec.~\ref{sec:PT.scattering.RIC} from the perspective of the scattering wave boundary conditions.
We note that these are also equivalent to the spectral singularities that have previously appeared in the literature~\cite{Bender&Wu,Shanley,AliMPRL09,AliMPRA09,AliMPRA11,Longhi09SS,Longhi10SS}. 
In Sec.~\ref{sec:PT.outgoing.spec.ep1} we will also show that for the case $\epsilon_1 \neq 0$, an RIC may approach the continuum edge and split into a bound state and a virtual bound state.
However, we add one further brief comment here to emphasize that the RIC is not an exceptional point, as the eigenstates do not coalesce, having different wave numbers,
%obviously our Hamiltonian is still diagonalizable at this point 
and hence no fractional power expansion such as Eqs.~(\ref{z.A.m.exp},~\ref{z.A.p.exp}) is possible in this case.

%%%%%%%%%%%%%%%%%%%%
%%%%%%%%SUB-SECTION: QBIC%
%%%%%%%%%%%%%%%%%%%%

\subsection{Quasi-bound states in continuum (QBICs)}
\label{sec:PT.outgoing.QBIC}
\label{QBIC}

The solutions $E_{1,4}$ from Region IV (or either pair of solutions from Region III) correspond to the localized states with complex eigenvalues that were recently experimentally observed in Ref.~\cite{PTOptExpt4}, in which the authors investigated light transmission through an effective 
$\PT$-symmetric optical lattice realized by periodically switching gain and loss in two optical fiber loops~\cite{PTOptExpt3,PTOptExpt4}.  As reported in Ref.~\cite{PTOptExpt4}, when a localized defect is introduced into the effective array (both a shift in $\PT$ pairing strength as well as a phase defect), a pair of localized complex conjugate modes appear within the continuum exhibiting exponential growth and decay in the power spectrum.
Indeed, our solutions $E_{1,4} \approx \pm i (\Gamma - 2 / \Gamma)$ in Region IV appear directly in the center of the energy continuum 
(with $\re \; E_{1,4} = 0$) and would also give rise to an exponential power output (growth or loss) 
as $\int_{- \infty}^{\infty} |\psi_{1,4} (x,t)|^2 dx \sim e^{\pm 2 t / \Gamma}$.

While the author of Ref.~\cite{Longhi_PT_bound} interprets these type of localized states with complex eigenvalue as examples of an effective BIC based on the fact that the real part of each solution may reside within the continuum, we note that since these states decay or grow exponentially, they would generally behave in a manner that is quite distinct from the usual concept of the BIC.
However, the other pair of solutions $E_{2,3}$ also have the real part of the eigenvalue residing within the continuum, yet behave quite differently in Region IV.  
Indeed we can show that the eigenvalues for these two solutions behave as $E_{2,3} \approx \pm i 2 / \Gamma^2$ such that the complex part of the eigenvalue for these states becomes arbitrarily small for increasing values of $\Gamma$.
Hence, these states should behave as effective BICs on time scales 
satisfying $t < \Gamma^2 / 4$, similar in concept to the quasi-bound state in continuum (QBIC) introduced in Refs.~\cite{NHGP07,GNHP09}, which are resonance states in the continuum with extremely long lifetime (also see Refs.~\cite{QBIC_other1,QBIC_other2}).

As shown in App.~\ref{app:iv.calcs}, the respective wave functions for the solutions $E_{2,3}$ are exponentially localized around the site $x = 0$, while those for the solutions $E_{1,4}$ are localized around the PT impurities at $x= \pm 1$; we also show that the localization for the solutions $E_{1,4}$ is very narrow as it scales for $\Gamma \gg 1$ as $1/ \log \Gamma$, while that for the quasi-bound states $E_{2,3}$ is very broad, scaling as $\Gamma^2 / 2$.
We believe that these quasi-bound states should be observable, for example, in an experiment similar either to Ref.~\cite{PTOptExpt4} or Ref.~\cite{PT_WGM} in which the $\PT$-symmetric defect potential is modified to mimic our potential appearing in Fig. \ref{fig1}.

%%%%%%%%%%%%%%%%%%%%%
%%%%%%%%%SUB-SECTION: RIC splitting
%%%%%%%%%%%%%%%%%%

\subsection{Discrete spectrum for $\varepsilon_1 \neq 0$: EP stability and RIC splitting at localization threshold}\label{sec:PT.outgoing.spec.ep1}

%%%%%%%%%
%%%%%%%%%
\begin{figure}
\hspace*{0.05\textwidth}
 \includegraphics[width=0.4\textwidth]{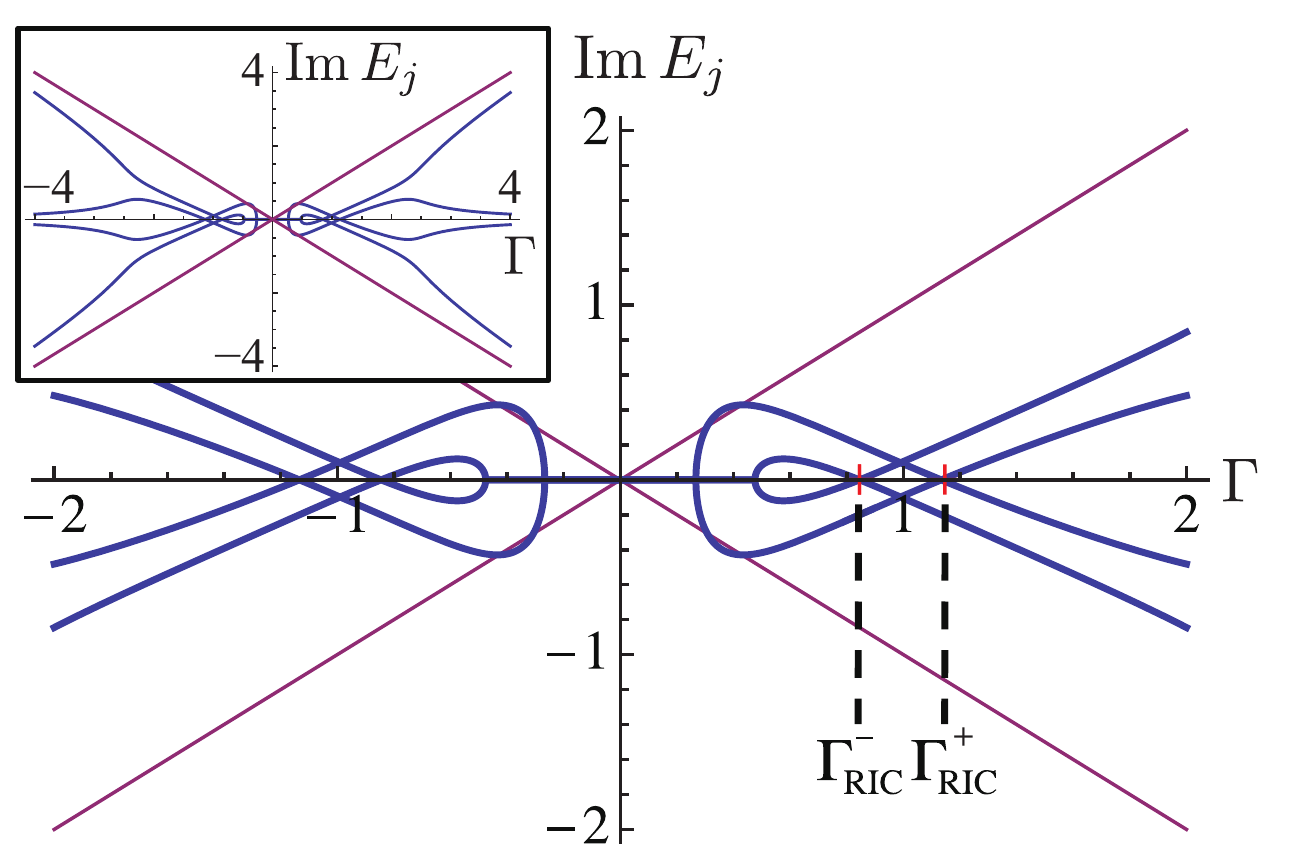}
\hfill
 \includegraphics[width=0.4\textwidth]{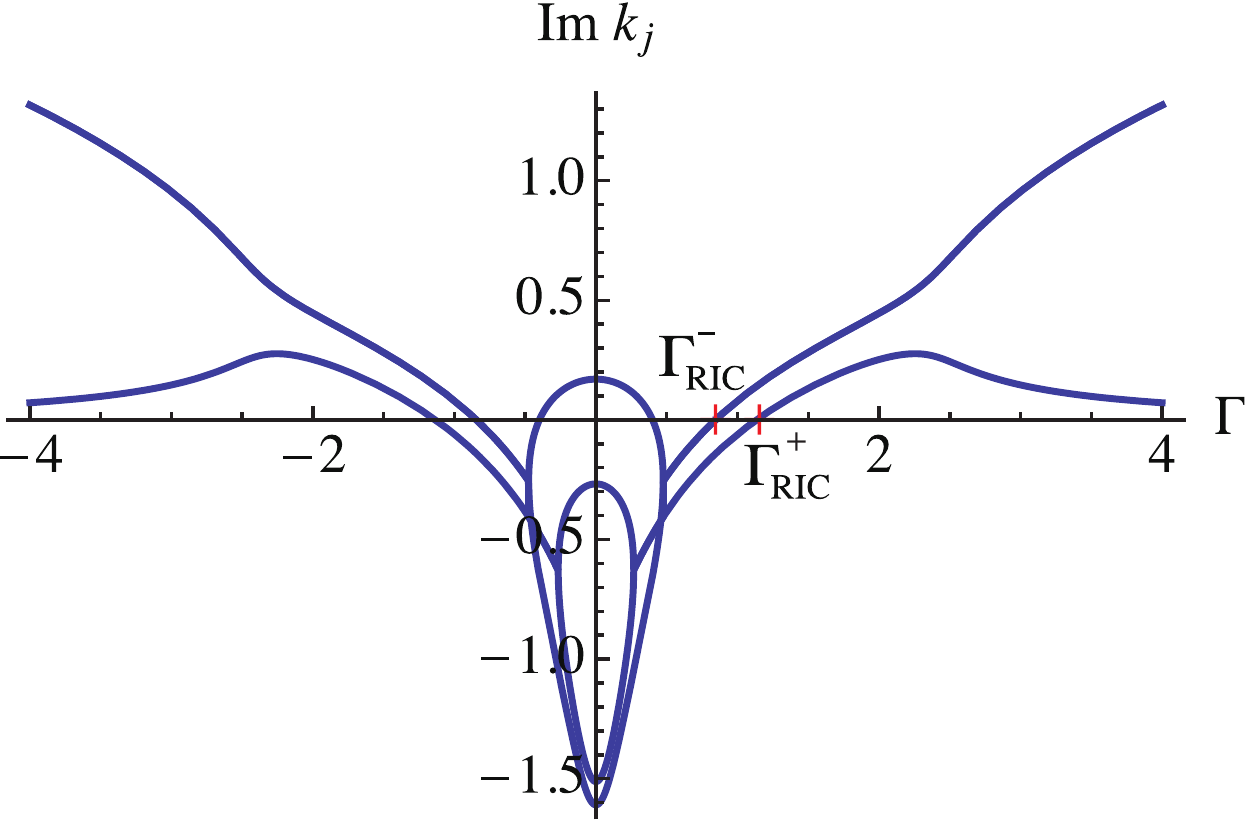}
 \hspace*{0.05\textwidth}
\\
\vspace*{\baselineskip}
\hspace*{0.05\textwidth}(a)\hspace*{0.440\textwidth}(b)\hspace*{0.4\textwidth}
\\
\vspace*{\baselineskip}
\hspace*{0.05\textwidth}
 \includegraphics[width=0.4\textwidth]{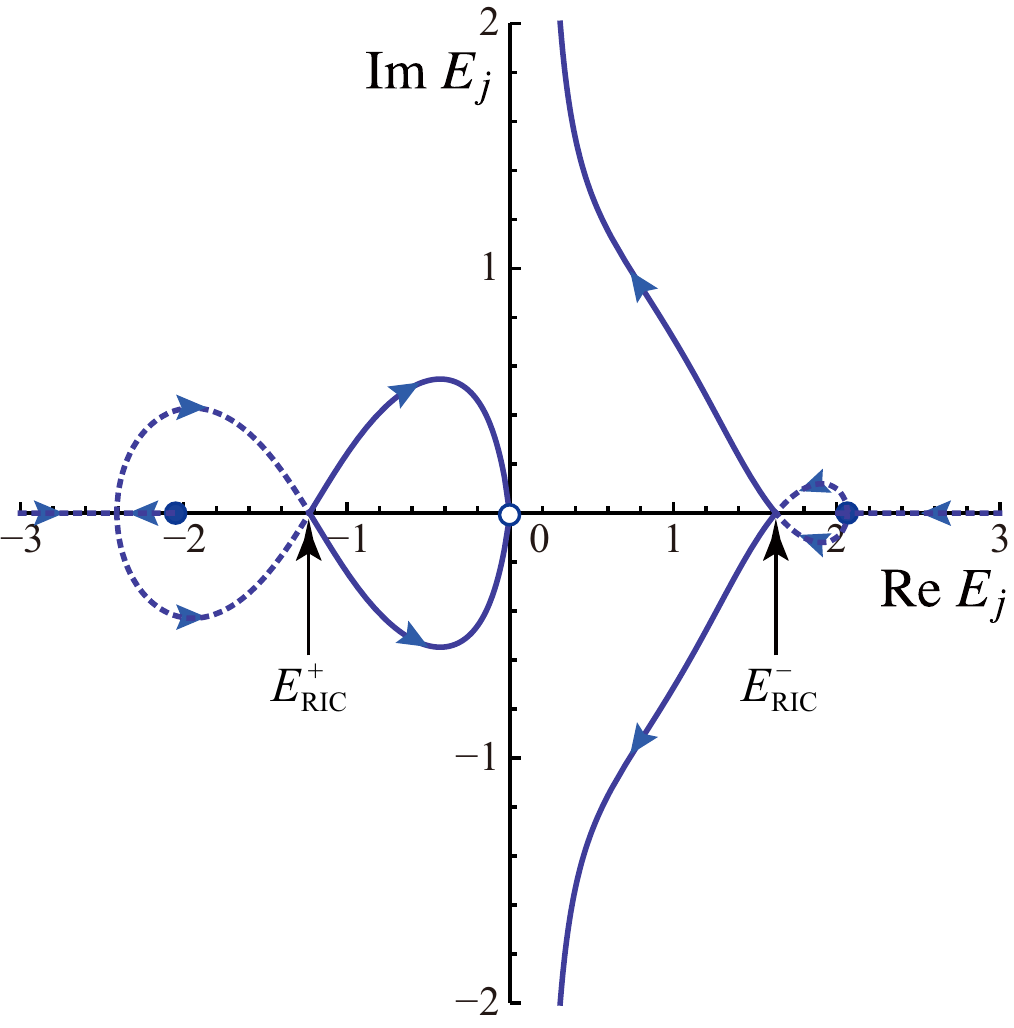}
\hfill
 \includegraphics[width=0.4\textwidth]{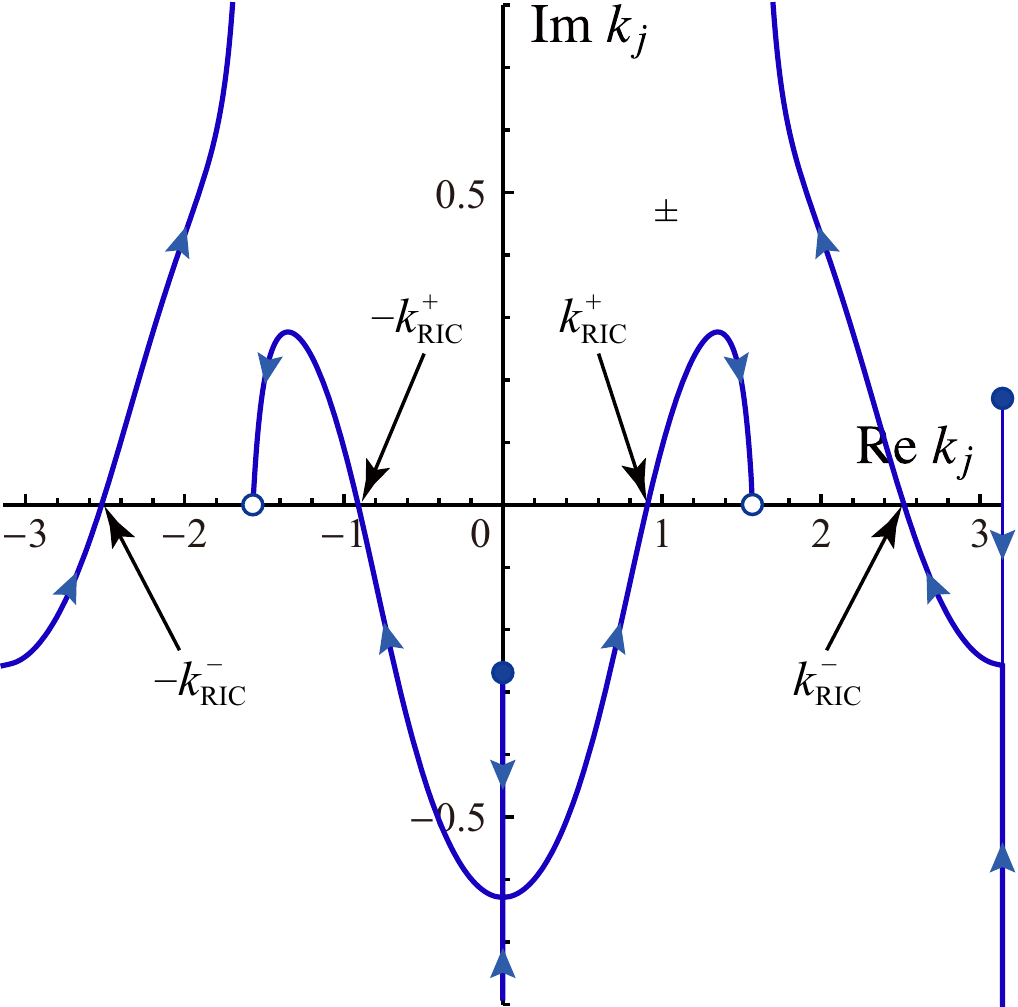}
 \hspace*{0.05\textwidth}
\\
\vspace*{\baselineskip}
\hspace*{0.05\textwidth}(c)\hspace*{0.440\textwidth}(d)\hspace*{0.4\textwidth}
\\
\vspace*{\baselineskip}
\hspace*{0.05\textwidth}
\includegraphics[width=0.4\textwidth]{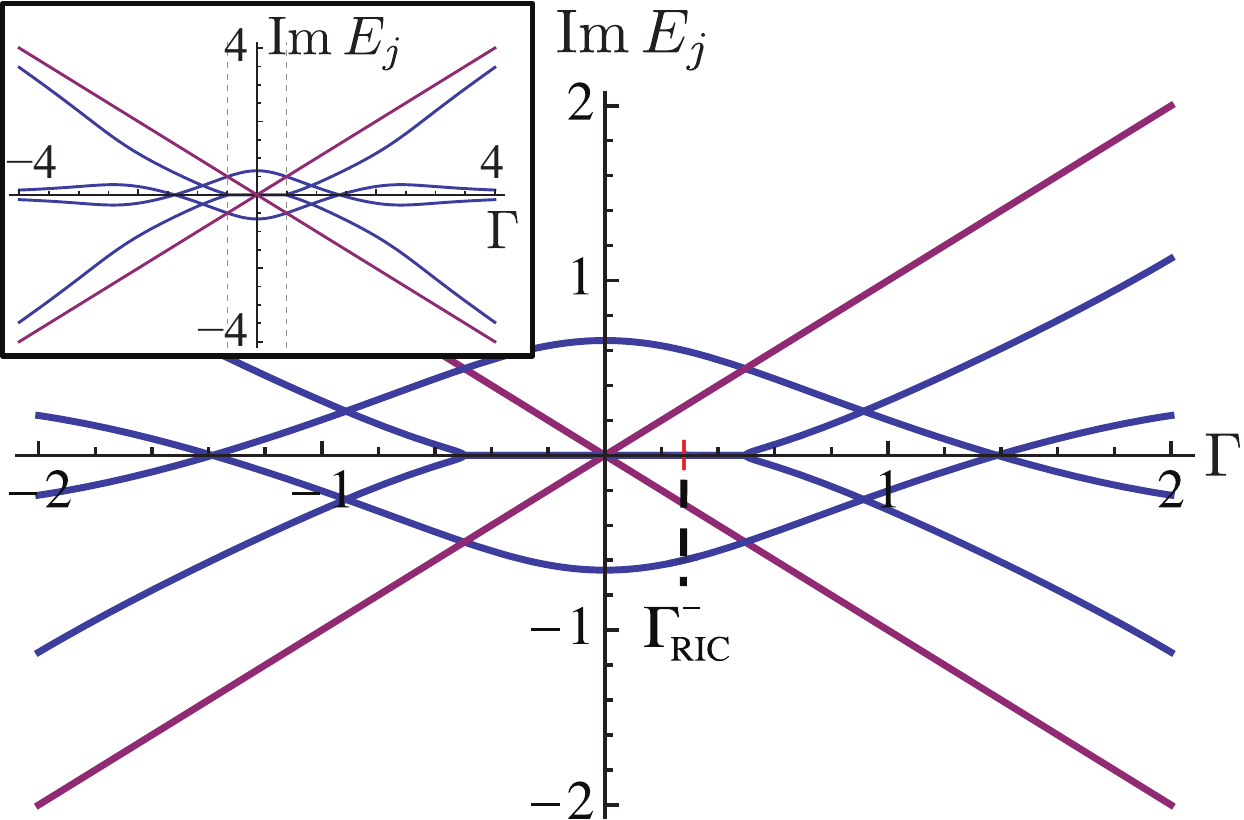}
\hfill
\includegraphics[width=0.4\textwidth]{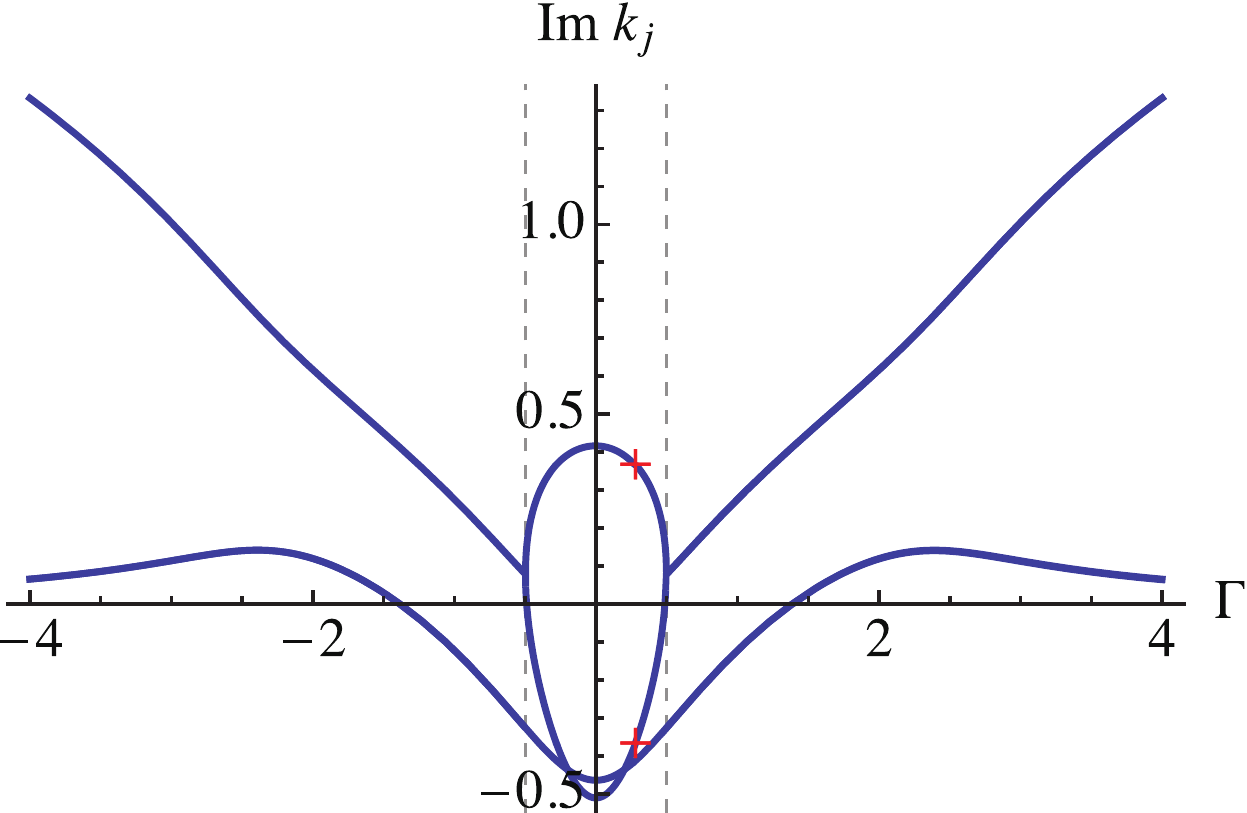}
 \hspace*{0.05\textwidth}
\\
\vspace*{\baselineskip}
\hspace*{0.05\textwidth}(e)\hspace*{0.440\textwidth}(f)\hspace*{0.4\textwidth}
\\
\vspace*{\baselineskip}
\caption{Discrete eigenvalue spectrum for the case $\varepsilon_0 = 0$.
(a) $\im E_j$ and (b) $\im k_j$ against the $\PT$-parameter $\Gamma$ for $\epsilon_1 = 0.2$; 
parametric plots of (c) $(\re E_j(\Gamma),\im E_j(\Gamma))$ and (d) $(\re k_j(\Gamma),\im k_j(\Gamma))$ in the complex plane for $\epsilon_1=0.2$;
(e) $\im E_j$ and (f) $\im k_j$ vs. the $\PT$-parameter $\Gamma$ for $\epsilon_1 = 0.6$.
In (c) and (d), the solid circles indicate some of the eigenvalues at $\Gamma=0$, while the open circles indicate those in the limit $\Gamma\to\infty$; 
the arrows indicate how the eigenvalues evolve as $\Gamma$ is increased from $0$ to $\infty$.
%In plots (a) and (b) full curves represents solutions in the first Riemann sheet while dotted curves represent solutions in the second Riemann sheet.  The points $\pm \Gamma_\textrm{RIC}$ denote the transition points between the two sheets.  [(c) and (d) are temporary place holders, annotated versions to be added]
}
\label{fig:PT.02.06.spec}
\label{fig4}
\end{figure}
%%%%%%%%%
%%%%%%%%%
As we relax the restriction $\varepsilon_1 = 0$, most of the basic features that we observed in the simplest case in Sec.~\ref{sec:PT.outgoing.spec}--C remain, although these become somewhat distorted as shown for $\varepsilon_1 = 0.2$ in Fig.~\ref{fig:PT.02.06.spec}(a)--(d).  
Here we observe that the EP2As split into two pairs, one pair of which moves outwards and away from the origin on the $\Gamma$-axis while the other pair moves inwards towards the origin (for larger values of $\varepsilon_1$, the latter pair will eventually collide at the origin, before becoming complex-valued).  
While we obtained compact analytic expressions for the eigenvalue expansions in the vicinity of the EP2As for the case $\varepsilon_1 = 0$, here those expressions become significantly more cumbersome.  
Nevertheless, following an intuitive generalization of the methods presented App.~\ref{app:EP.calcs} one may still easily obtain numerical versions of Eqs.~(\ref{z.A.p.exp}) and~(\ref{z.A.m.exp}) in the vicinity of the EP2As in the more general case.

On the other hand, the EP2Bs that we studied in Sec.~\ref{sec:PT.outgoing.spec} immediately vanish from the spectrum for $\varepsilon_1 \neq 0$, as can be seen in Fig.~\ref{fig:PT.02.06.spec}(b) and the inset of Fig.~\ref{fig:PT.02.06.spec}(a);
we can also see these coalescences vanish by comparing Fig.~\ref{fig:PT.0.spec}(e) and (f) (for 
$\varepsilon_1 = 0$) with Fig.~\ref{fig:PT.02.06.spec}(c) and (d).
Indeed, it can be shown that the EP2As also survive the generalization for $\varepsilon_0 \neq 0$, while the EP2Bs do not re-emerge.  
This seems to indicate that type EP2A exceptional points are more stable against parameter perturbations than those of type EP2B.  
We note that several experimental studies have been conducted in which an EP2A has been observed by simply passing directly through the exceptional point while varying a single parameter~\cite{PTOptExpt2,PT_WGM,PTCircuitExpt}, but experimental observation of EP2Bs have tended to rely on encircling the exceptional point~\cite{EPexpt1a,EPexpt1c} or mapping out the complex eigenvalue structure around the exceptional point in a two-dimensional parameter space~\cite{EP_Korea} (although Ref.~\cite{EPexpt1b} provides an exception where the EP2B is observed more directly).  
Theoretically, we believe that the underlying reason for this is that the EP2As seem to vanish from the real parameter space only when they collide with another EP2A (see Ref.~\cite{GRHS12} for another simple example where this occurs), while the EP2Bs do not need to collide with another EP in order to exit into the complex parameter space.

\begin{figure}
\begin{center}
%\hspace*{0.05\textwidth}
 \includegraphics[width=0.5\textwidth]{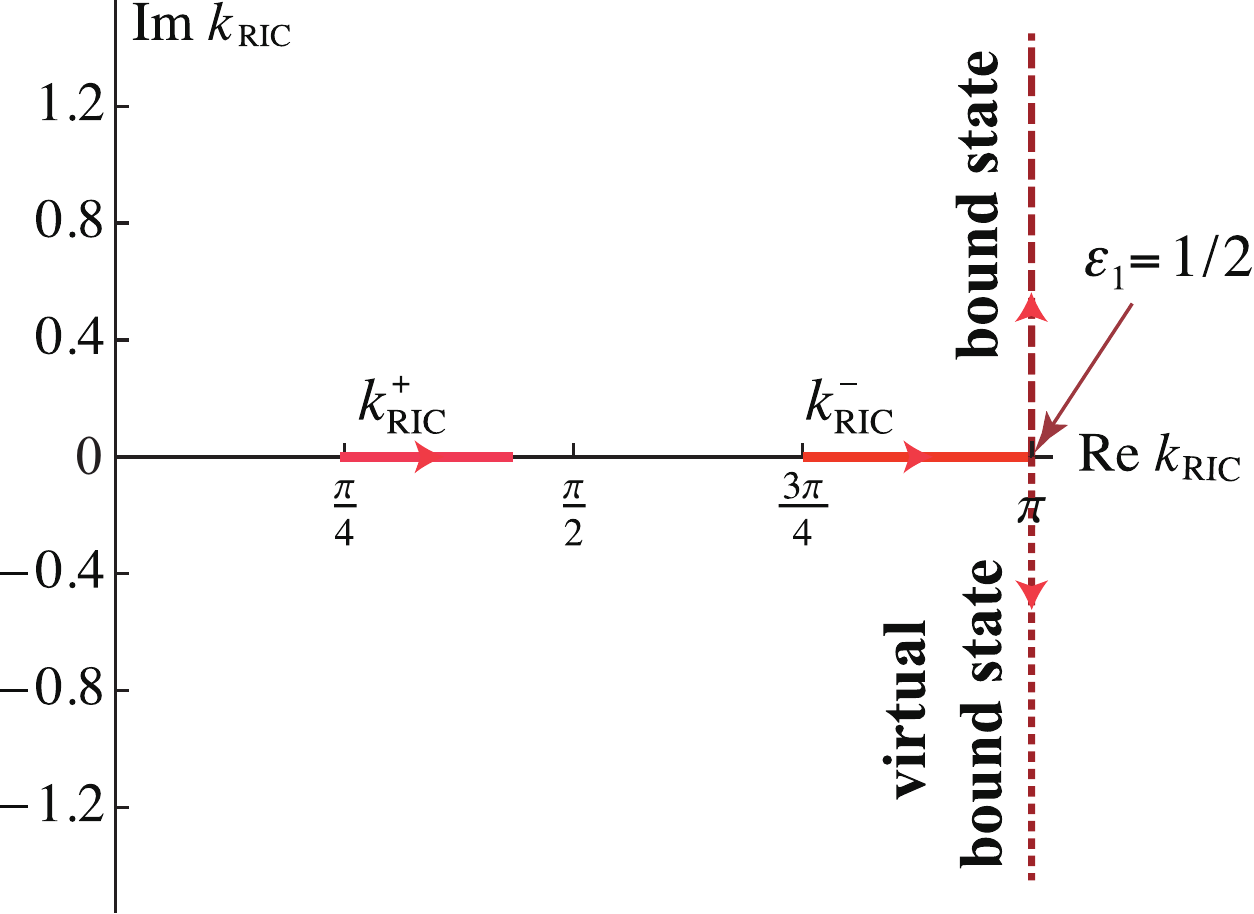}
% \includegraphics[width=0.5\textwidth]{k_RIC_plot_ann}
%\hfill
% \includegraphics[width=0.4\textwidth]{inline_k_par_plot}
% \hspace*{0.05\textwidth}
\end{center}
%\\
\vspace*{-\baselineskip}
\caption{Movement of the wave numbers of RICs, $k_\textrm{RIC}^\pm$, in the complex $k$ plane as $\varepsilon_1$ increases from 0 to 1.5.
 }
 \label{fig:PT.kRIC.evol}
 \label{fig5}
 \end{figure}

The RICs meanwhile are also split apart in the parameter space, appearing at $\pm \GamRICp$ and 
$\pm \GamRICm$, given by
\begin{equation}
   \Gamma_\textrm{RIC}^\pm (\varepsilon_1)
   	= \sqrt{1 \pm \left| \varepsilon_1 \right| \sqrt{2 + \varepsilon_1^2}}
	,
\label{Gam.RIC.pm}
\end{equation}
which we explicitly indicate by red crosses in Fig.~\ref{fig:PT.02.06.spec}(a) and (b).
%(Note that as $\epsilon_1 \rightarrow 0$ we of course have $\Gamma_\mathrm{RIC}^\pm \rightarrow \GamRIC = 1$ such that we recover the previous results from Sec.~\ref{sec:PT.outgoing.spec}).  
As we increase $\varepsilon_1$ from 0, the RIC wave numbers for $\Gamma_\mathrm{RIC}^\pm$, which we refer to as $k_\mathrm{RIC}^\pm$ in Fig.~\ref{fig:PT.02.06.spec}(d), move away from the $\varepsilon_1 = 0$ values of $\pi / 4$ and $3\pi/4$ (previously shown in Fig.~\ref{fig:PT.0.spec}(f)) and approach $\pi/2$ and the upper band edge $\pi$, respectively.  In the latter case, the RIC at $\GamRICm$ eventually reaches the upper band edge;
we can find the precise value of $\varepsilon_1$ where this occurs from the condition
\begin{equation}
   P \left( \lambda = -1; \varepsilon_1, \Gamma_\textrm{RIC}^- (\varepsilon_1) \right)
   	= 0
	,
\label{P.RIC.m.trans}
\end{equation}
which yields $\varepsilon_1 = 1/2$.  
At this precise point, one of the EP2As also touches the band edge and overlaps with the RIC.
Then for $\varepsilon_1 > 1/2$ the RIC exits the continuum and we find that it splits into a bound state and a virtual bound state as shown in Fig.~\ref{fig:PT.02.06.spec}(e) and (f) for the case $\varepsilon_1 = 0.6$.  
We also show the evolution of the wave numbers $k_\textrm{RIC}^\pm$ in the complex $k$ plane in Fig.~\ref{fig:PT.kRIC.evol} as the system evolves from $\varepsilon_1 = 0$ to $\varepsilon_1 = 1.5$.  
Here both $k_\textrm{RIC}^\pm$ move rightward on the real axis, excepting that $\kRICm$ splits into a bound state/virtual bound state pair beyond $\epsilon_1 = 1/2$.
%%%%%%%%%
%%%%%%%%%
%%%%%%%%%
%%%%%%%%%

Another key difference in the $\varepsilon_1 \neq 0$ case (as well as the $\varepsilon_0 \neq 0$ case) is the appearance of one or (at most) two bound states.  The bound state properties are discussed in greater detail in Sec. \ref{sec:PT.bound} below.

%%%%%%%%%%%%%%%%%%%%%
%%%%%%%%%SECTION: Bound states - formal details
%%%%%%%%%%%%%%%%%%

\section{Formal properties of the bound states}\label{sec:PT.bound}

We now turn to a closer investigation of the bound states that appear for various parameter ranges of our $\PT$-symmetric prototype model.  
Our focus here is on the traditional bound states with real energy eigenvalues; 
however, we will briefly comment on the virtual bound states and the localized states with complex eigenvalue at points for which they are also relevant to our discussion.

We first briefly discuss in Sec.~\ref{sec:PT.bound.param} the parameter ranges for which bound states exist in our prototype model for the general case $\varepsilon_0 \neq 0$, $\varepsilon_1 \neq 0$ and comment on the easiest method for finding bound states for a given set of parameter values.
Then in Sec.~\ref{sec:PT.bound.PT} we explore the symmetry properties of the wave function for the bound state solutions and verify that they satisfy $\PT$-symmetric boundary conditions; 
indeed, the virtual bound states (anti-bound states) are also $\PT$-symmetric.  
Finally, in Sec.~\ref{sec:PT.bound.norm} we investigate the $\mathcal{CPT}$ norm~\cite{BBJ02} for the bound states.

%%%%%%%%%%%%%%%%%%%%%
%%%%%%%%%SUB-SECTION: parameter ranges for bound states
%%%%%%%%%%%%%%%%%%

\subsection{Existence of bound states for the general case $\varepsilon_0 \neq 0$ and $\varepsilon_1 \neq 0$}
\label{sec:PT.bound.param}

As we illustrated in Fig.~\ref{fig2}, the bound states of Hermitian tight-binding models can only exist on the real axis of the first Riemann sheet, below and above the energy band (this is also true for our present model).
For our non-Hermitian system, such bound states do not appear in the particular case of $\epsilon_0=\epsilon_1=0$ as seen in Fig.~\ref{fig3}(e) and (f); however, these do appear for $\varepsilon_0 \neq 0$ or $\varepsilon_1 \neq 0$.
For example, one bound state appears above the upper band edge in the case $\varepsilon_1 > 0$ (within a specific range of $\Gamma$ values); this state is evidenced
in Fig.~\ref{fig4}(d) by the portion of the trajectory that lies on the positive side 
($\im k > 0$) of the line $\re k=+\pi$.

\begin{figure}
\hspace*{0.05\textwidth}
 \includegraphics[width=0.4\textwidth]{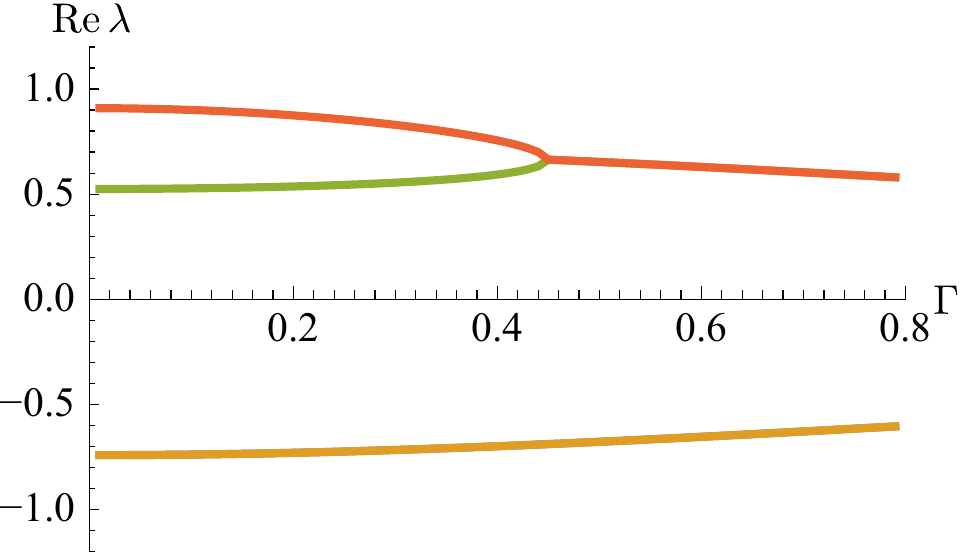}
\hfill
 \includegraphics[width=0.4\textwidth]{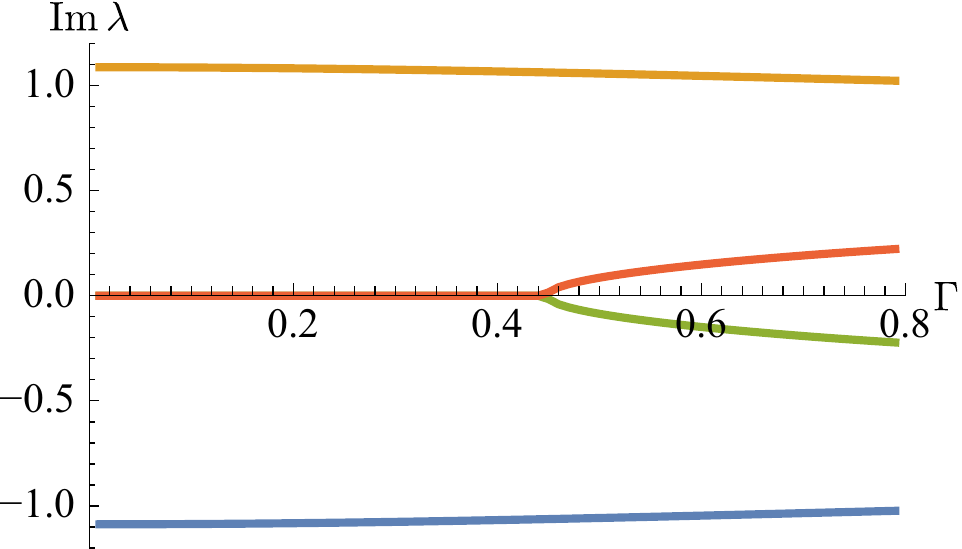}
 \hspace*{0.05\textwidth}
\\
\vspace*{\baselineskip}
\hspace*{0.05\textwidth}(a)\hspace*{0.440\textwidth}(b)\hspace*{0.4\textwidth}
\\
\vspace*{\baselineskip}
\begin{center}
 \includegraphics[width=0.4\textwidth]{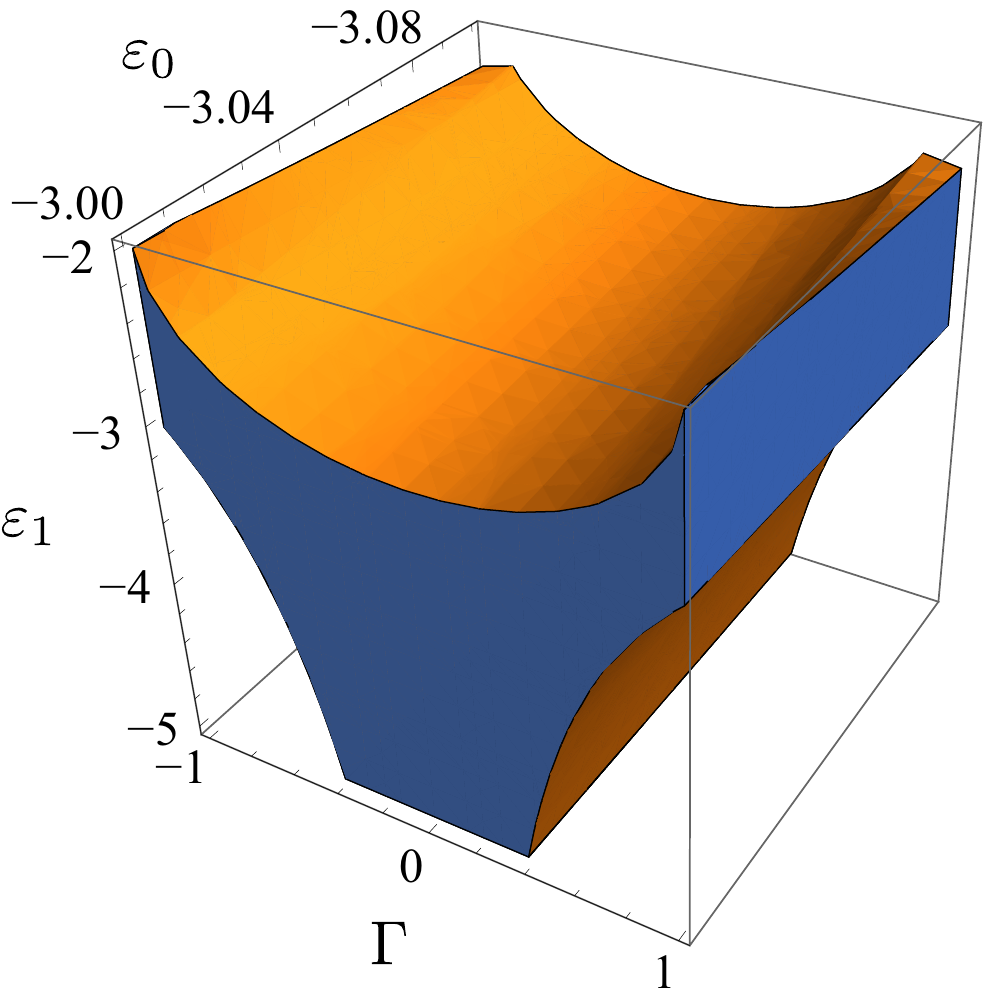}
\\
(c)
\end{center}
%\vspace*{\baselineskip}
\caption{(a) Real part and (b) imaginary part of the roots $\lambda$ of the polynomial $P(\lambda)$ in Eq.(\ref{P.lambda}) for $\epsilon_1=-1.1$ and $\epsilon_0=0.05$.
The two real roots $0<\lambda<1$ become complex at $\Gamma=0.45$.
(c)
a region of unbroken $PT$-symmetry (with all four solutions real-valued) in the parameter space 
$(\varepsilon_0, \varepsilon_1, \Gamma)$. }
\label{unbroken1}
\end{figure}

In general, we can write the wave number for the bound states as $k_j = i \kappa_j + \delta_j \pi$, with $\kappa_j > 0$ and in which $\delta_j = 0$ for bound states below the lower band edge and 
$\delta_j = 1$ for bound states above the upper band edge (the same formula also holds true for virtual bound states, except that $\kappa_j < 0$ in that case). 
Then, for a given set of parameter values, we can test for the presence of bound states within the spectrum by plugging
$\lambda_j = e^{i k_j} = \pm e^{- \kappa_j}$
into $P(\lambda_j) = 0$ from Eq.~(\ref{P.lambda}); any real solution that yields $0 < \lambda_j < 1$ represents a bound state below the lower band edge, and any solution with
$-1 < \lambda_j < 0$ represents a bound state above the upper band edge. Meanwhile, real solutions satisfying $\lambda_j > 1$ ($\lambda_j < -1$) represent virtual bound states below (above) the lower (upper) band edge.
In Fig. \ref{unbroken1}(a, b)  we plot numerical solutions of (\ref{P.lambda}) for the representative case $\varepsilon_0 = 0.01$ and $\varepsilon_1 = -1.1$.  Here we find that there exist two bound states below the lower band edge in the parameter domain $\Gamma < 0.45$. There are are also a resonance and an anti-resonance in this domain with real part of the energy eigenvalue above the upper band edge.

We define the \emph{unbroken} $\PT$-\emph{symmetry region} as any portion of the parameter space for which all of the solutions are real-valued (any combination of bound states and virtual bound states).
For example, given $\varepsilon_1 < 0$ we show in Fig. \ref{unbroken1}(c) the range of parameter values that yield real values for all four solutions of the dispersion equation.
In the following %section,
Sec. \ref{sec:PT.bound.PT}, 
we explicitly demonstrate that both bound states and virtual bound states satisfy $\PT$-symmetric boundary conditions.

%We call $\PT$-symmetric unbroken region for a bound state below the lower (above the upper) band edge the region in the parameter space $(\varepsilon_0, \varepsilon_1, \Gamma)$ where there exist a real solution $0<\lambda<1$ ($-1<\lambda<0$). 
%The unbroken $PT$-symmetric region for the first bound state $\lambda > 0$ ( $\lambda<0$) does not  coincide with the unbroken $PT$-symmetric region for the second bound state $\lambda > 0$ ( $\lambda<0$), but their intersection is not empty. 
%In Figs. \ref{unbroken1}  we represent the unbroken $PT$-symmetric region for both of the two bound states below the lower band edge, that coincides with the unbroken $PT$-symmetric region for  the two bound states above the upper band edge.

%\begin{figure}[h]

%We can obtain an expression for all localiza*tion/delocalization transitions in our system by solving 
%$P(\lambda=\pm1) = 0$ from Eq.~(\ref{P.lambda}) for $\Gamma$ in terms of the other parameters.  Doing so yields $\Gamma = \pm \Gamma_\textrm{loc,u}$ for the transitions above the upper band edge and 
%$\Gamma = \pm \Gamma_\textrm{loc,l}$ for the transition below the lower band edge;
%here $\Gamma_\textrm{loc,u} = \Gamma_\textrm{loc} (\varepsilon_0, \varepsilon_1)$ and
% $\Gamma_\textrm{loc,l} = \Gamma_\textrm{loc} (-\varepsilon_0, -\varepsilon_1)$, 
% while
%%
%\begin{equation}
 % \Gamma_\textrm{loc} (\varepsilon_0, \varepsilon_1)
  %	\equiv \sqrt{ \frac{\left( 2 \varepsilon_1 
	%				+ \varepsilon_0 \left(1 - \varepsilon_1 \right) \right) \left(1 - \varepsilon_1 \right)}
	%		{2 - \varepsilon_0} } 
%\label{Gamma.loc}
%\end{equation}
%%

%%%%%%%%%%%%%%%%%%%%%
%%%%%%%%%SUB-SECTION: PT-symmetric bound states
%%%%%%%%%%%%%%%%%%

\subsection{Verification that real-valued bound states satisfy $\PT$-symmetric boundary conditions}\label{sec:PT.bound.PT}

Here we verify that the real-valued bound states discussed in Sec.~\ref{sec:PT.bound.param} automatically satisfy $\PT$-symmetric boundary conditions.
%For bound states below the lower band edge, we can easily show that the effective wave number $k_l$ takes the pure imaginary form $k_l = i \kappa$ with $\kappa > 0$, while for those above the upper band edge we instead have $k_u = i \kappa + \pi$.  We combine these results by writing the wave number of an arbitrary bound state as $k_j = i \kappa + \delta_j \pi$, where $\delta_l = 0$ for a bound state below the lower band edge and $\delta_u = 1$ for a bound state above the upper band edge.  
To accomplish this, we again write the wave number of an arbitrary bound state in the form $k_j = i \kappa_j + \delta_j \pi$, where 
$\delta_j = 0$ for a bound state below the lower band edge and $\delta_j = 1$ for a bound state above the upper band edge.
With this formalism, the wave equation~(\ref{outgoing.wave.fcn}) for the bound states takes the form
\begin{equation}
  \psi_j (x) =
	\left\{ \begin{array}{ll}
		B e^{\kappa x - i \delta_j \pi x} 			& \mbox{for $x \le -1$,}    	\\
		\psi_0							& \mbox{for $x = 0$,}   	\\
		C e^{- \kappa x + i \delta_j \pi x} 		& \mbox{for $x \ge 1$.}
	\end{array}
	\right.
\label{outgoing.wave.bound}
\end{equation}

In order for $\psi_j$ to be a $\PT$-symmetric eigenstate of our Hamiltonian $H$, it must satisfy the condition $\PT \psi_j = e^{i \theta} \psi_j$.
Note that at any point we could introduce the state $\tilde{\psi}_j (x) = e^{i \theta / 2} \psi_j (x)$, which is then an eigenstate of $\PT$ with eigenvalue $1$.
Applying the $\PT$ operator to the bound-state wave function we obtain
\begin{equation}
  \PT \psi_j (x) =
	\left\{ \begin{array}{ll}
		C^* e^{\kappa x + i \delta_j \pi x}  
				= \left( \frac{C^*}{B} \right) B e^{\kappa x - i \delta_j \pi x}
							& \mbox{for $x \le -1$,}    	\\
		\psi_0^*				& \mbox{for $x = 0$,}   	\\
		B^* e^{- \kappa x - i \delta_j \pi x} 
				= \left( \frac{B^*}{C} \right) C e^{\kappa x + i \delta_j \pi x}							& \mbox{for $x \ge 1$,}
	\end{array}
	\right.  
\label{outgoing.wave.bound.PT}
\end{equation}
where in the last step we have taken advantage of the fact that $-\pi$ is physically equivalent to $\pi$ in the Brilliuon zone structure of our model.  If we assume $\psi_0^* = F \psi_0$, then we can write the quantity $C^*/B = B^*/C = F$ as a phase factor $F = e^{i \theta}$.

Now, for a $\PT$-symmetric solution of our Hamiltonian, we see that we must augment the outgoing boundary condition in Eq.~(\ref{outgoing.wave.fcn}) 
%(where $k_j = i \kappa_j + \delta_j \pi$) 
with an additional condition 
$B = e^{- i \theta} C^*$, %or $\bra -1 | \psi \ket = e^{-i \theta} \bra 1 | \psi \ket^*$.  
which gives $ \psi (-1) = e^{-i \theta} \psi (1)^\ast$. 
We apply this condition to re-write the matrix form of the Schr\"odinger equation in Eq.~(\ref{outgoing.matrix0}) as
\begin{equation}
  \begin{pmatrix}
  	- \lambda + \varepsilon_1 + i \Gamma 	& -e^{i \theta}		& 0    \\
	-e^{-i \theta}						& \varepsilon_0	& -1   \\
	0								& -1		& - \lambda + \varepsilon_1 - i \Gamma  \\
	\end{pmatrix}
  \begin{pmatrix}
  	\psi(1)^\ast	\\
	\psi (0)	\\
	\psi (1) 	\\
	\end{pmatrix}
	= E(\lambda)
  \begin{pmatrix}
  	\psi (1)^\ast 	\\
	\psi (0) 	\\
	\psi (1) 	\\
	\end{pmatrix}
	.
\label{outgoing.matrix.bound.PT}
\end{equation}
Taking the determinant of this modified equation yields the exact same condition for discrete eigenvalues $P(\lambda_j) = 0$ as we previously encountered at the beginning of Sec.~\ref{sec:PT.outgoing}.  Hence, any real-valued bound state of the Hamiltonian $H$ in 
Eq.~(\ref{eq-model}) is automatically an eigenstate of the $\PT$-symmetry operator with eigenvalue 
$e^{i \theta}$.

We may obtain the explicit form for the coefficient $B = e^{i \theta} C^*$ from the first and third lines of Eq.~(\ref{outgoing.matrix.bound.PT}).  For simplicity here, let us choose $\theta = 0$, such that $B = C^*$.  We then find the real and imaginary parts of $B = B_R + i B_I$ as
\begin{equation}
  B_R =
  	\frac{\lambda \left( 1+\lambda \varepsilon_1 \right)}
			{1+\Gamma^2\lambda^2 + 2 \varepsilon_1 \lambda + \varepsilon_1^2\lambda^2} \psi_0;
					\ \ \ \ \ \ \ \ \ \ \ \ \ \ \ \ 
  B_I =
  	\frac{ \Gamma \lambda}
			{1+\Gamma^2\lambda^2 + 2 \varepsilon_1 \lambda + \varepsilon_1^2 \lambda^2}  \psi_0
	,
\label{bound.PT.coeffs}
\end{equation}
with $\lambda = e^{ik_j} = e^{-\kappa + i \delta_j \pi}$.  

As a final comment, we note that according to the argument we have presented here the wave function for the virtual bound states (residing in the second Riemann sheet in the complex energy plane) must also satisfy $\PT$ symmetry.  This can immediately be seen by simply replacing the form of the wave number for the bound state $k_j = i \kappa_j + \delta_j \pi$ by that for the virtual bound states $k_j = -i \kappa_j + \delta_j \pi$ (with $\kappa_j > 0$ in either case) and proceeding with the argument as presented above.

However, we note that the bound states with complex energies are \emph{not} $\PT$-symmetric, which can be seen by writing the wave number for these states in the form $k_j = \kappa_j + i \Pi_j$ with 
$\Pi_j < | \pi |$ and noting that we can no longer make the sign replacement in the final line of Eq.~(\ref{outgoing.wave.bound.PT}) as these states reside within the Brillioun zone, rather than at the edges as do the bound states and virtual bound states.

%Taken together, these two points imply that in $\PT$-symmetric open quantum systems, we should associate the emergence of the broken $\PT$ region with the \emph{appearance} of complex-valued solutions, rather than merely the disappearance of a bound state (which in many cases will have simply become a virtual bound state).

%%%%%%%%%%%%%%%%%%%%%
%%%%%%%%%SUB-SECTION: PT-symmetric bound states
%%%%%%%%%%%%%%%%%%

\subsection{$\mathcal{CPT}$ norm of the bound states}\label{sec:PT.bound.norm}

We now set out to write the appropriate normalization condition for the bound-state wave function that we obtained in Sec.~\ref{sec:PT.bound.PT}.
For a non-symmetric Hamiltonian $H$, the  completeness relation among its eigenstates $\psi_n(x)$ assumes the form
\beq
\sum_{x=-\infty}^\infty \psi_n^L(x)\psi_m^R(x)=\delta_{n,m}
\label{e22}
\eeq 
where  $\psi_n^R(x)$ are right eigenstates and $\psi_n^L(x)$ are left eigenstates.
If $H$ is $\PT$-symmetric, we identify 
$$\psi_n^L(x)= \mathcal{CPT} \psi_n^R(x),$$
where the operator $\mathcal{C}$ \cite{BBJ02} satisfies, in the unbroken region, the three algebraic equations
\begin{equation}
  \left[ \mathcal{C}, \PT \right] = 0,
  				\ \ \ \ \ \ \ \ \ 
  	\left[ \mathcal{C}, H \right] = 0,
				\ \ \ \ \ \ \ \ \ 
	\mathcal{C}^2 = 1
	.
\label{C.relns}
\end{equation}
The completeness relation for a $\PT$-symmetric Hamiltonians then reads
\beq
\sum_{x=-\infty}^\infty[\mathcal{CPT}\psi_n(x)]\psi_m(x)=\delta_{n,m}.
\label{e21}
\eeq
Since the $\mathcal C$ operator commutes with the Hamiltonian $H$, the bound states $\psi_j (x)$ of $H$ in the unbroken region must also be  eigenstates of  $\mathcal C$ and, because $\mathcal C^2=1$, the resulting eigenvalues must be $C_j = \pm 1$ \cite{BBJ02,Weigert03}. (How the $\mathcal C$ operator might act on the complex-valued solutions in the broken region is a question presently under investigation).

In order to assign the correct eigenvalue $C_j$ to each eigenstate $\psi_j (x)$, we first evaluate the so-called $\PT$ norm as 
\begin{equation}
  \sum_{x = - \infty}^{\infty} [\PT \psi_j (x)] \psi_k (x)
  	= \sum_{x = - \infty}^{\infty} \psi_j (-x)^\ast \psi_k (x)
  	= (-1)^j \delta_{j,k}.
  	\label{PT.norm}
\end{equation}
We see that the $\PT$-norm is not positive definite \cite{BBMW02}, with alternating signs $\pm 1$ among the bound states $\psi_j (x)$; hence, we assign the eigenvalues $C_n$  to be $\pm 1$ according to the sign of  (\ref{PT.norm}) in order to obtain the positive norm introduced in Eq. (\ref{e21}).

\begin{figure}
\includegraphics[width=0.4\textwidth]{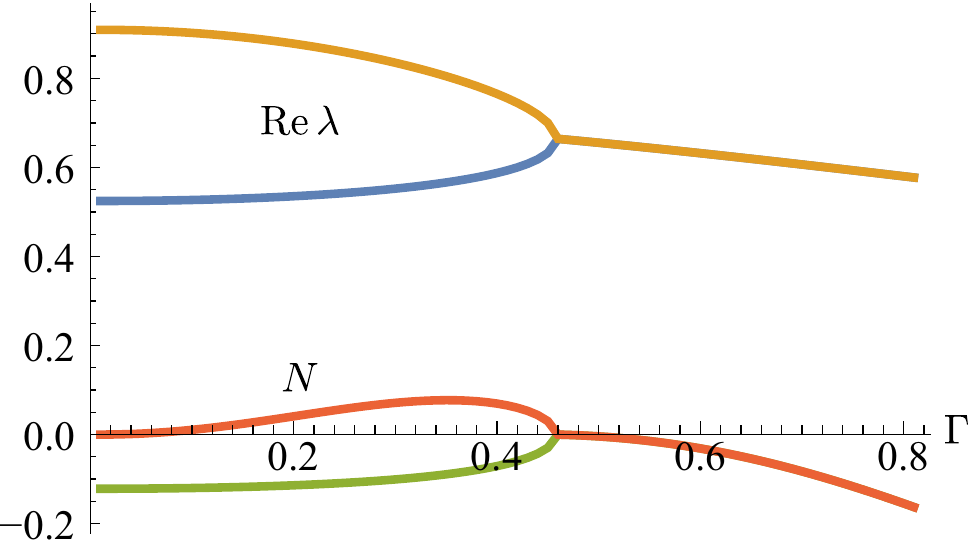}
\caption{Real part of the roots $\lambda$ of the polynomial Eq.(\ref{P.lambda}) and normalization
constant $N$ corresponding to two bound states below the lower band edge in the domain $0 < \Gamma < 0.45$ for $\varepsilon_1 = -1.1$ and $\varepsilon_0 = 0.05$.
One bound state has a positive norm and the other has a negative norm; the eigenvalues of the $\mathcal C$ operator are therefore $+1$ and $-1$, respectively. A similar picture holds for bound states appearing above the upper band edge.}
\label{norm}
\end{figure}

In either case, we may write the $\PT$-norm for our bound states given in Eq.~(\ref{outgoing.wave.bound}) as
\begin{equation}
 N_j^{PT} \equiv \sum_{x = - \infty}^{\infty} \psi_j (-x)^\ast \psi_j (x)
  	= \left( B^{*2} + B^2 \right) \sum_{x=1}^\infty e^{- 2 \kappa_j x} + \psi_0^2,
\label{PT.norm.N}
\end{equation}
where $\kappa_j$ is the imaginary component of the wave number from $k_j = i \kappa_j + \delta_j \pi$.
For convenience, we introduce $\tilde{B}=B/ \psi_0$; we then obtain 
\beq
  N_j^{PT} =\psi_0^2 \left( 2\frac{\tilde{B}_r^2-\tilde{B}_i^2}{e^{2\kappa_j}-1}+1\right)
  	.
 \label{XX1}
 \eeq
The explicit form of the coefficient $B$ %in (\ref{XX1})
 is given by Eq. (\ref{bound.PT.coeffs}), where $\lambda=e^{-\kappa}$ for the bound states  below the lower band edge and $\lambda=-e^{-\kappa}$ for the bound states above the upper band edge.
In Fig.~\ref{norm} we show the $\PT$-norm for the two bound states previously shown in Figs. \ref{unbroken1}(a,b) that appear below the lower band edge; we see that the $\PT$ norm for one of these states is positive, giving the eigenvalue of the $\mathcal C$ operator as $1$, while for the other the norm is negative, giving the eigenvalue of the $\mathcal C$ operator as $-1$.  
%A similar analysis can be applied for the bound states residing above the upper band edge.
 
%%%%%%%%%%%%%%%%%%%%%
%%%%%%%%%SECTION: outgoing waves B.C.
%%%%%%%%%%%%%%%%%%

\section{Scattering states, the resonance in continuum (RIC), and perfect transmission}
\label{sec:PT.scattering}
\label{sec5}

In this section we consider typical ($\PT$-asymmetric) scattering boundary conditions for our $\PT$-symmetric open quantum system;
we will consider $\PT$-symmetric scattering solutions later in Sec.~\ref{sec:PT.scattering.2}.  
In Sec.~\ref{sec:PT.scattering.gen} below we write the generic wave function for the states and then obtain a matrix equation for the relevant scattering coefficients.  
We then explicitly write the transmission and reflection for the simplest case with a pure imaginary potential, namely $\varepsilon_0 = \varepsilon_1 = 0$.  
We perform these calculations for both left-to-right and right-to-left scattering
and we verify that the transmission and reflectance satisfy established relations that generally hold in $\PT$ systems.  

In Sec.~\ref{sec:PT.scattering.RIC} we discuss the RIC in detail from the perspective of the scattering wave solutions; 
here we note that the RIC automatically satisfies the Siegert boundary condition for outgoing waves and argue that these states represent a resonance between the background continuum and the $\PT$-symmetric defect potential.

While the RIC is a discrete state embedded in the scattering continuum, we discuss in Sec.~\ref{sec:PT.scattering.2.perfect} outstanding scattering states, namely perfectly transmissive states, meaning that the transmission is unity while the reflectance vanishes.  
%In Sec.~\ref{sec:PT.scattering.2.perfect} we study a subset of the $\PT$-symmetric scattering wave solutions that give rise to perfect transmission
For the simplest case in which the defect potential is pure imaginary with $\varepsilon_0 = \varepsilon_1 = 0$ we demonstrate that by appropriately choosing $\Gamma$ we can obtain perfect transmission at any given value of $k$ in the spectrum; 
this property approximately holds in the case in which $\epsilon_0$ and $\epsilon_1$ are nonzero but take on small values.  
We also demonstrate that the appearance of the perfect transmission state at the band edges coincides with a delocalization transition, an observation which may be useful from an engineering perspective as one aims to construct devices with specific transmission properties.

We also examine the case in which not only the transmission is unity, but also there is no phase shift as the scattering wave passes through the defect region.  
In this case the signal is transmitted not only perfectly, but invisibly.
%Further, we demonstrate that for a precise choice of parameters, the system exhibits an interesting switching behavior in which the central scattering site at $x=0$ responds to the otherwise invisible left-to-right transmission signal but remains dormant in response to the right-to-left signal (or vice-versa).

%%%%%%%%%%%%%%%%%%%%%
%%%%%%%%%sub-SECTION: outgoing waves B.C.: assymetric
%%%%%%%%%%%%%%%%%%

\subsection{$\PT$-asymmetric scattering states}
\label{sec:PT.scattering.gen}
\label{sec5A}

Here we find scattering states of our $\PT$-symmetric model~\eqref{eq-model} under the potential~\eqref{eq-potential}.
We will limit most of the detailed calculations to the simplest case $\varepsilon_0=\varepsilon_1=0$ but the generalization to the case $\varepsilon_0\neq 0$ and $\varepsilon_1\neq 0$ is straightforward.
We first solve the Schr\"{o}dinger equations~\eqref{eq-Sch1}--\eqref{eq-Sch3} for left-to-right scattering by assuming a wave function of the form
\begin{align}
\label{eq-PTasym}
\psi(x)=
\begin{cases}
A e^{ikx} + B e^{-ikx} & \quad\mbox{for $x\leq-1$},
\\
\psi(0) & \quad\mbox{for $x=0$},
\\
C e^{ikx} &\quad\mbox{for $x\geq 1$},
\end{cases}
\end{align}
where $k$ resides within the scattering continuum $0\leq k \leq \pi$. 
The term with the coefficient $A$ gives the incoming wave, while the $B$ term is the reflected wave and the $C$ term is the transmitted wave.
Note that its eigenvalue is real: $E=-2\cos k$.

We have four parameters $A$, $B$, $C$ and $\psi(0)$ to fix under the three conditions given by the 
Schr\"odinger equations~\eqref{eq-Sch1}--\eqref{eq-Sch3}.  
Substituting the ansatz~\eqref{eq-PTasym} into them
%in the simplest case $\varepsilon_0=\varepsilon_1=0$ 
yields
\begin{equation}
  \begin{pmatrix}
  	\epsilon_1+ i \Gamma + \lambda		& -\lambda				& 0    \\
	1					& - \left( 1+ \epsilon_0 \lambda + \lambda^2 \right) 	& \lambda^2   \\
	0							& -1		& 1 + \left( \epsilon_1- i \Gamma \right) \lambda  \\
	\end{pmatrix}
  \begin{pmatrix}
  	A		\\
	\psi(0)	\\
	C		\\
	\end{pmatrix}
	= - \lambda B
  \begin{pmatrix}
  	1 + \left( \epsilon_1 + i \Gamma \right) \lambda	 \\
	\lambda								\\
	0									\\
	\end{pmatrix}
	,
\label{outgoing.matrix}
\end{equation}
%%
%%%%%%%
%%%%%%%
%%
%\begin{equation}
%  \left[ \begin{array}{ccc}
  	%%
 % 	\lambda + i \Gamma 		& -\lambda				& 0    \\
	%%
%	1					& - \left( 1 + \lambda^2 \right) 	& \lambda^2   \\
	%%
%	0					& -1						& 1 - i \lambda \Gamma  \\
	%%
%	\end{array} \right]
		%%
		%% 
 % \left[ \begin{array}{c}
  %	A		\\
	%%
  %	\psi(0)	\\
	%%
  %	C		\\
  	%%
  %	\end{array} \right]
		%%
		%%
  %	= - B  \lambda
  %\left[ \begin{array}{c}
  %	1 + i \Gamma \lambda	\\
	%%
  %	\lambda				\\
	%%
  %	0					\\
  	%%
  %	\end{array} \right]
%	,
%\label{outgoing.matrix}
%\end{equation}
%%
Let us limit ourselves from this point to the simplest case $\varepsilon_0=\varepsilon_1=0$.
Although the overall phase of the wave function~\eqref{eq-PTasym} does not affect physical quantities, %we can arbitrarily choose this phase without loss of generality.
it turns out that it is easiest to assume $B\in\mathbb{R}$.
We can represent the coefficients as
\begin{align}\label{eq300}
A&=B\frac{i\sin k-\Gamma^2e^{2ik}\cos k}%
{(\Gamma+2\sin k) \Gamma\cos k},
\\\label{eq310}
C&=B\frac{i\sin k}%
{(\Gamma+2\sin k) \Gamma\cos k},
\\\label{eq320}
\psi(0)&=B\frac{i(1-i\Gamma e^{ik})\sin k}%
{(\Gamma+2\sin k) \Gamma\cos k},
\end{align}
and thereby obtain the transmission and reflection amplitudes as
\begin{eqnarray}\label{trans.int.l1}
  t_l
  	& = & \frac{C}{A}
	= \frac{i \sin k}{i \sin k - \Gamma^2 e^{2ik} \cos k}
			,\\
  r_l 
  	& = & \frac{B}{A}
	= \frac{\left( \Gamma + 2 \sin k \right) \Gamma \cos k}{i \sin k - \Gamma^2 e^{2ik} \cos k}
		,
\label{trans.int.l2}
\end{eqnarray}
%%
%which are directly related to the transmission and reflection.
%from this matrix equation the ratios
%\begin{align}
%A&=B\frac{i\sin k-e^{2ik}\Gamma^2\cos k}%
%{(2\sin k+\Gamma) \Gamma\cos k},
%\\
%C&=B\frac{i\sin k}%
%{(2\sin k+\Gamma) \Gamma\cos k},
%\\
%\psi(0)&=B\frac{i\sin k+e^{ik}\Gamma\sin k}%
%{(2\sin k+\Gamma) \Gamma\cos k}.
%\end{align}
%Since the overall phase does not affect physical quantities, we can arbitrarily choose this phase without loss of generality.
%It turns out that it is easiest to assume $B\in\mathbb{R}$.
%%
which are followed by the transmission and reflection probabilities as
\begin{align}\label{eq-T}
T_{L\to R}&:=\left|t_l\right|^2
%T_{L\to R}&:=\left|\frac{C}{A}\right|^2
=\frac{\sin^2k}{\sin^2k+(\Gamma-2\sin k)(\Gamma+2\sin k)\Gamma^2\cos^2k},
%=\frac{2\sin^2 k}{1-\cos2k-(1-\cos4k)\Gamma^2+(1+\cos2k)\Gamma^4},
\\\label{eq-R}
R_{L\to R}&:=\left|r_l\right|^2
%R_{L\to R}&:=\left|\frac{B}{A}\right|^2
=\frac{(\Gamma+2\sin k)^2\Gamma^2\cos^2 k}{\sin^2k+(\Gamma-2\sin k)(\Gamma+2\sin k)\Gamma^2\cos^2k}.
%=\frac{2 (2\sin k+\Gamma)^2\Gamma^2\cos^2k}{1-\cos2k-(1-\cos4k)\Gamma^2+(1+\cos2k)\Gamma^4}.
\end{align}
Note that $T_{L\to R}+R_{L\to R}$ is, in general, not unity because we have a source and a sink and therefore the particle number is not conserved.  Instead, the usual probability conservation relation is replaced by a generalized rule for $\PT$-symmetric systems that relates the left-to-right and right-to-left transmission properties~\cite{Mosta14}, as shown below.

Hence, we next consider the right-to-left scattering solution given by the ansatz %instead of Eq.~\eqref{eq-PTasym}:
\begin{align}
\label{eq-PTasym1}
\psi(x)=
\begin{cases}
B e^{-ikx} & \quad\mbox{for $x\leq-1$},
\\
\psi(0) & \quad\mbox{for $x=0$},
\\
C e^{ikx} +D e^{-ikx} &\quad\mbox{for $x\geq 1$},
\end{cases}
\end{align}
in which we again have $0\leq k\leq \pi$, the $D$ term is the incoming wave, the $C$ term is the reflected wave and the $B$ term is the transmission wave.
Note again that its eigenvalue is real: $E=-2\cos k$.

Again substituting this ansatz into the Schr\"odinger equations~\eqref{eq-Sch1}--\eqref{eq-Sch3},
%for the $\varepsilon_0 = \varepsilon_1 = 0$ case, 
we obtain
\begin{equation}
  \begin{pmatrix}
  	1 + \left( \varepsilon_1 + i \Gamma\right) \lambda 	& -1						& 0    \\
	\lambda^2			& - \left( 1 + \varepsilon_0 \lambda + \lambda^2 \right) 	& 1   \\
	0					& - \lambda				& \varepsilon_1 - i \Gamma + \lambda  \\
	\end{pmatrix}
  \begin{pmatrix}
  	B		\\
	\psi(0)	\\
	D		\\
	\end{pmatrix}
	= - \lambda C
  \begin{pmatrix}
  	0					\\
	\lambda				\\
	1 + \left( \varepsilon_1 - i \Gamma \right) \lambda	\\
	\end{pmatrix}
	.
\label{outgoing.matrix1}
\end{equation}
After assuming $C\in\mathbb{R}$ this time, we obtain for the case $\varepsilon_0 = \varepsilon_1 = 0$
the coefficients as
\begin{align}
\label{eq390}
D&=C\frac{i\sin k-\Gamma^2e^{2ik}\cos k}%
{(\Gamma-2\sin k) \Gamma\cos k},
\\\label{eq400}
B&=C\frac{i\sin k}%
{(\Gamma-2\sin k) \Gamma\cos k},
\\\label{eq410}
\psi(0)&=C\frac{i(1-i\Gamma e^{ik})\sin k}%
{(\Gamma-2\sin k) \Gamma\cos k}
\end{align}
and 
the amplitudes
\begin{eqnarray}\label{trans.int.r1}
  t_r
  	& = & \frac{B}{D}
	= \frac{i \sin k}{i \sin k - \Gamma^2 e^{2ik} \cos k}
			,  \\
  r_r 
  	& = & \frac{C}{D}
	= \frac{\left( \Gamma - 2 \sin k \right) \Gamma \cos k}{i \sin k - \Gamma^2 e^{2ik} \cos k}
		,
\label{trans.int.r2}
\end{eqnarray}
which in turn lead to the
right-to-left transmission and reflection probabilities
\begin{align}\label{eq-T1}
T_{R\to L}&:=\left|t_r\right|^2
%T_{R\to L}&:=\left|\frac{B}{D}\right|^2
=\frac{\sin^2k}{\sin^2k+(\Gamma-2\sin k)(\Gamma+2\sin k)\Gamma^2\cos^2k}
\equiv T_{L\to R},
%=\frac{2\sin^2 k}{1-\cos2k-(1-\cos4k)\Gamma^2+(1+\cos2k)\Gamma^4},
\\\label{eq-R1}
R_{R\to L}&:=\left|r_r\right|^2
%R_{R\to L}&:=\left|\frac{C}{D}\right|^2
=\frac{(\Gamma-2\sin k)^2\Gamma^2\cos^2 k}{\sin^2k+(\Gamma-2\sin k)(\Gamma+2\sin k)\Gamma^2\cos^2k}
\leq R_{L\to R}.
%=\frac{2 (2\sin k+\Gamma)^2\Gamma^2\cos^2k}{1-\cos2k-(1-\cos4k)\Gamma^2+(1+\cos2k)\Gamma^4}.
\end{align}
The left-right asymmetry in \eqref{eq-R1} comes from the fact that the $\mathcal{P}$ and $\mathcal{T}$ symmetries are individually broken in our system.

We nonetheless note that we have $t_l = t_r \equiv t$, such that the transmission is equal for the left-to-right and right-to-left scattering;
this is a general property of $\PT$ systems.  
Further we note that the relations
\begin{equation}
  t (-k) = t(k)
  			\ \ \ \ \ \ \ \ \ \ \ \ \ \ \ \ \ \ \ \ 
  r_l (-k) = r_r(k)
\label{PT.t.r}
\end{equation}
are satisfied, which also hold for $\PT$ systems in general~\cite{Mosta14,Ahmed12}.  
Finally, the usual probability conservation property for Hermitian systems ($T + R = 1$) is here replaced by
\cite{Mosta14,Ahmed12,GCS12}
\begin{equation}
  \left| t (k) \right|^2 \pm \left| r_l (k) r_r (k) \right| = 1
  	,
\label{PT.t.r.cons}
\end{equation}
which we can easily verify by using Eqs.~\eqref{trans.int.l1},~\eqref{trans.int.l2},~\eqref{trans.int.r1}, and~\eqref{trans.int.r2}.
This is a result of the fact that Eqs.~\eqref{eq390}--\eqref{eq410} are obtained from the corresponding Eqs.~\eqref{eq300}--\eqref{eq320} by taking the complex conjugate and flipping the sign of $k$, which is just the $\PT$ operation in the wave-number space.
Note that the sign choice appearing in Eq.~\eqref{PT.t.r.cons} is fixed by the sign of the quantity $1 - \left| t (k) \right|^2 = 1 - T$;
it can easily be shown that for the present case with $\varepsilon_0 = \varepsilon_1 = 0$ the sign changes in this quantity generally occur at $\arcsin (\pm \Gamma / 2)$.

%%%%%%%%%%%%%%%%%%%%%
%%%%%%%%%sub-SECTION: outgoing waves B.C.: RIC
%%%%%%%%%%%%%%%%%%

\subsection{Scattering wave perspective on the resonance in continuum (RIC)}
\label{sec:PT.scattering.RIC}

We explain the resonance in continuum (RIC), which we introduced in Sec.~\ref{sec:PT.outgoing} as a discrete resonant eigenstate, here from the perspective of the scattering states presented in Sec.~\ref{sec5A}.
More specifically, we show that the RICs appear as singularities in the transmission and reflection probabilities~\eqref{eq-T}, \eqref{eq-R}, \eqref{eq-T1}, and \eqref{eq-R1} on the real axis.
In this sense, it is a discrete state embedded in the scattering continuum.

Let us recall that in Sec.~\ref{sec:PT.outgoing} the RICs appear as the points where two solutions meet on the real energy axis;
in Fig.~\ref{fig3}(e) for the simplest case $\varepsilon_0 = \varepsilon_1 = 0$, this happens at $E=\pm\sqrt{2}$ for $\Gamma=1$.
This is not an exceptional point, however, because each of them has a different (real) value of $k$;
namely, $k=\pm\pi/4$, $\pm 3\pi/4$ in the simplest case as is shown in Fig.~\ref{fig3}(c) or (f).

\begin{figure}
\begin{minipage}[t]{0.3\textwidth}
\vspace{0mm}
\begin{center}
\includegraphics[width=\textwidth]{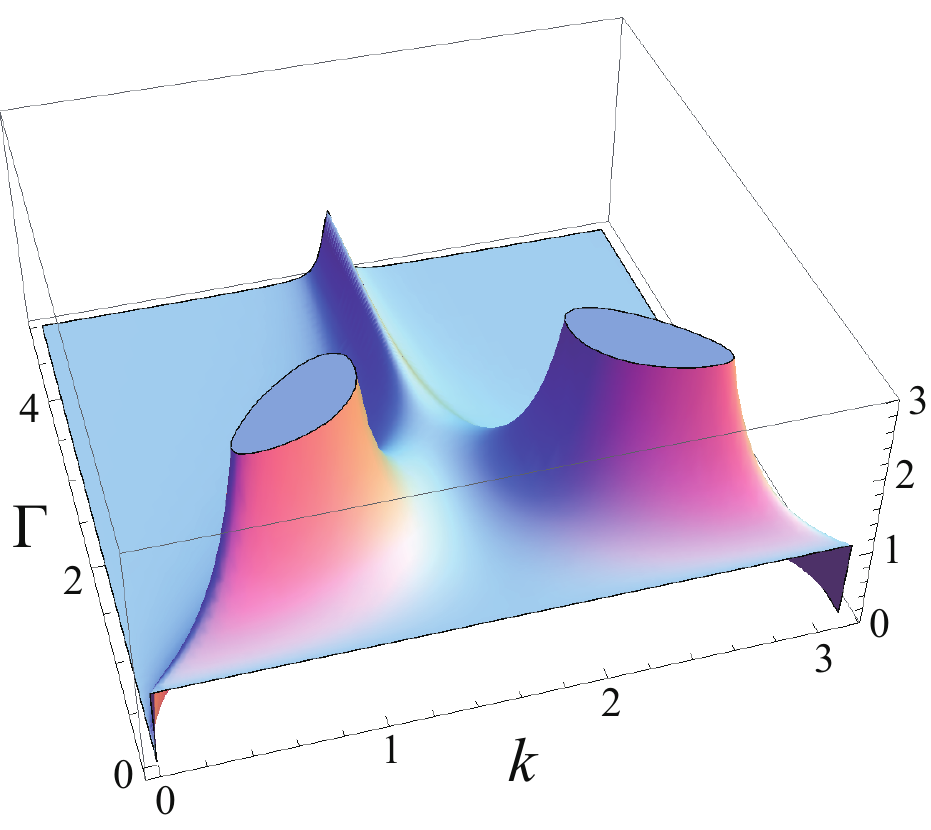}

(a) Transmission probability
\end{center}
\end{minipage}
\hfill
\begin{minipage}[t]{0.3\textwidth}
\vspace{0mm}
\begin{center}
\includegraphics[width=\textwidth]{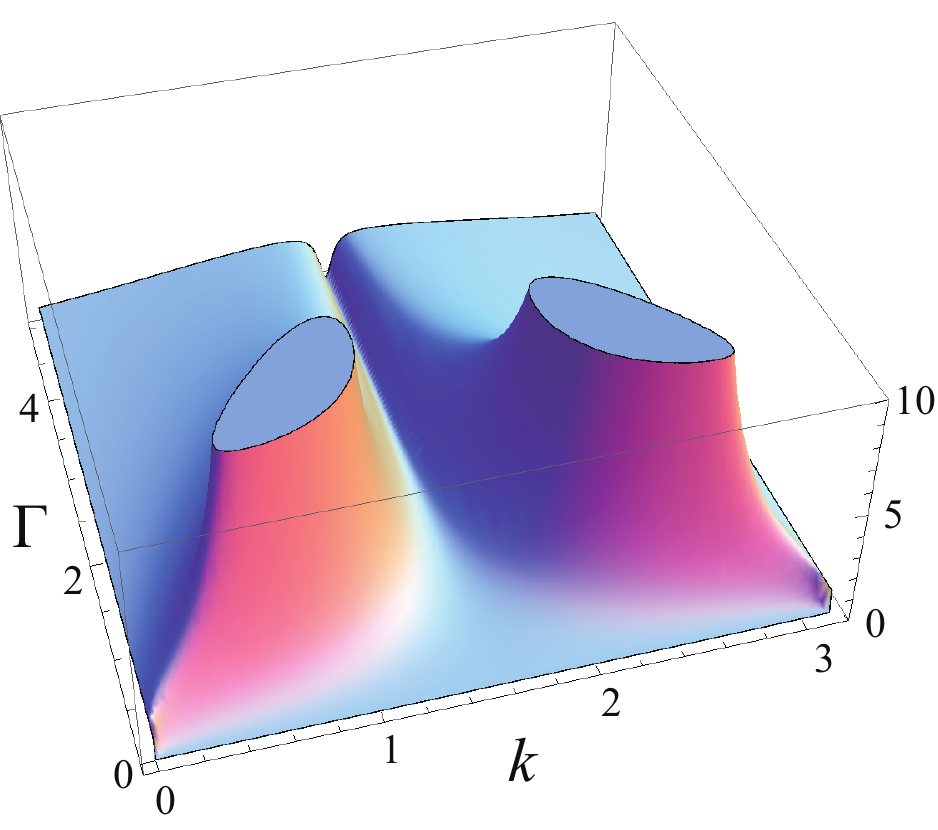}

(b) Reflection probability to the left
\end{center}
\end{minipage}
\hfill
\begin{minipage}[t]{0.3\textwidth}
\vspace{0mm}
\begin{center}
\includegraphics[width=\textwidth]{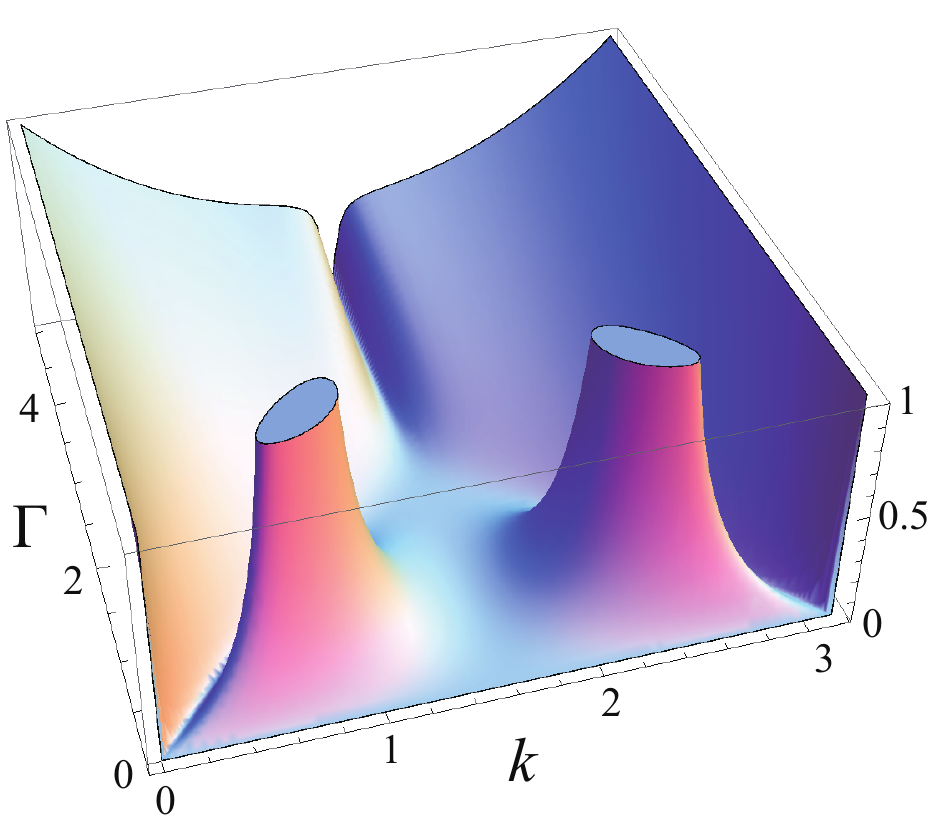}

(c) Reflection probability to the right
\end{center}
\end{minipage}
\caption{(a) The left-to-right transmission probability~\eqref{eq-T}, which is equal to the right-to-left transmission~\eqref{eq-T1},
(b) the left-to-left reflection probability~\eqref{eq-R},
and (c) the right-to-right reflection probability~\eqref{eq-R1}, 
all in the simplest case of $\varepsilon_0=\varepsilon_1=0$.
Note that the scale of the vertical axis varies from panel to panel.}
\label{140313hatanofig1}
\end{figure}

We here show that these points indeed have the properties of the resonant states in the sense that they have divergent transmission and reflection probabilities.
We plot in Fig.~\ref{140313hatanofig1} the transmission and reflection probabilities both for Eqs.~\eqref{eq-PTasym} and~\eqref{eq-PTasym1} in the case $\varepsilon_0=\varepsilon_1=0$ for positive $0 \le \Gamma \le 5$;
they are symmetric with respect to $k=0$.
%%%%%
%%%%%
%%%%%
%%%%%
%%%%%
%%%%%
All probabilities have poles at $k=\pm\pi/4$ and $k=\pm3\pi/4$ with $\Gamma=1$, namely for the RICs.
These are the only instances at which any of the probabilities diverges for real $k$.

%\begin{figure}
%\begin{minipage}[t]{0.45\textwidth}
%\vspace{0mm}
%\centering
%\includegraphics[width=\textwidth]{140313hatanofig3a}
%(a) complex $k$ plane\\
%for $\varepsilon_0=\varepsilon_1=0$
%\end{minipage}
%%
%\hfill
%%
%\begin{minipage}[t]{0.45\textwidth}
%\vspace{0mm}
%\centering
%\includegraphics[width=\textwidth]{140313hatanofig3b}
%(b) complex $E$ plane\\
%for $\varepsilon_0=\varepsilon_1=0$
%\end{minipage}
%
%\vspace{\baselineskip}
%
%\begin{minipage}[t]{0.45\textwidth}
%\vspace{0mm}
%\centering
%\includegraphics[width=\textwidth]{140313hatanofig3c}
%(c) complex $k$ plane\\
%for $\varepsilon_0=-0.2$ and $\varepsilon_1=0.3$
%\end{minipage}
%%
%\hfill
%%
%\begin{minipage}[t]{0.45\textwidth}
%\vspace{0mm}
%\centering
%\includegraphics[width=\textwidth]{140313hatanofig3d}
%(b) complex $E$ plane\\
%for $\varepsilon_0=-0.2$ and $\varepsilon_1=0.3$
%\end{minipage}
%
%\caption{The movement of the poles of the transmission and reflection probabilities,
%in the simplest case of $\varepsilon_0=\varepsilon_1=0$ and in the case of $\varepsilon_0=-0.2$ and $\varepsilon_1=0.3$, for example.
%The solid circles indicate the points for $\Gamma=0$, while the open circles indicate the points in the limit $\Gamma\to\infty$.
%The points $k=\pm \pi/4$ and $k=\pm3\pi/4$ in (a), which corresponds to the points $E=\pm\sqrt{2}$ in (b), are the incidents of the resonant states in continuum.
%The same incidents appear in (c) and (d) at slightly different points.
%Note that the continuum stretches over $-2\leq E\leq 2$.}
%\label{fig3}
%\end{figure}

From this perspective, the poles are indeed discrete states embedded in the scattering continuum.
Let us explain why they are \textit{resonant states in continuum}.
These poles are associated with the zeros of the coefficient $A$ in the wave function~\eqref{eq-PTasym} and the coefficient $D$ in the wave function~\eqref{eq-PTasym1}, as we can see in Eqs.~\eqref{trans.int.l1},~\eqref{trans.int.l2},~\eqref{trans.int.r1}, and~\eqref{trans.int.r2}.
Exactly at these zeros $A=0$ and $D=0$, the wave functions~\eqref{eq-PTasym} and~\eqref{eq-PTasym1} have only outgoing waves, which indeed matches the Siegert boundary condition~\eqref{outgoing.wave.fcn} for resonant states~\cite{Gamow28,Siegert39,Peierls59,Landau77,Ostrovsky05,Kunz06,Kunz08,Sasada08,HSNP08,NH_H_eff}.
Therefore, the poles of the transmission and reflectance shown in Fig.~\ref{140313hatanofig1} are resonance poles.
In a resonance state with $\mathop{\mathrm{Re}} k>0$, particles are ejected from the central area and flow away towards $x=\pm \infty$;
in a state with  $\mathop{\mathrm{Re}} k<0$, which is historically called an anti-resonance state, particles flow into the central area and vanish.
We indeed saw this from another perspective in Eq.~(\ref{psi.RIC}), in which the RIC took the form of an outgoing plane wave outside of the central scattering region.

In the Hermitian scattering problem, the particle number is conserved and hence the Siegert boundary condition can only be satisfied at discrete complex values of $k$ and $E$, which give the location of the resonance poles;
it can never be satisfied for real values of $k$ and $E$ in the Hermitian case.  
Under the Siegert boundary condition, %we can interpret the resonance with only outgoing waves as being associated with a process in which
%We can understand the resonant state with only outgoing waves as follows in the Hermitian case.
the particles flow away from the scattering region and hence the particle number in this area decays in time, which can be described only by a complex energy eigenvalue~\cite{HSNP08}. 
Hence, a resonance pole in the Hermitian case is strongly tied to an eigenstate with complex $E$ and $k$.

In the $\PT$-symmetric case, however, the particle number is not conserved because we have a source and a sink.
Hence it is possible for particles to emerge out of the scattering region (or vanish into it) as a stationary state without changing the particle number in this region.
This is indeed what happens with the resonant states in continuum:
because $\mathop{\mathrm{Im}}E=0$ and $\mathop{\mathrm{Im}}k=0$ for these poles, the wave function stretches over all space as a stationary state.  
%\textcolor{blue}{Indeed, we saw this from another perspective back in Eq.~(\ref{psi.RIC}), in which the RIC took the form of a standing plane wave solution outside of the central scattering region.}

In this sense, it is a remarkable characteristic of $\PT$-symmetric systems that we have resonances resting within the real energy continuum.
This is why we specifically refer to these as resonance states in continuum;
these states represent a resonance between the open environment associated with the leads and the 
$\PT$-symmetric potential.
We further note that in the special case $\varepsilon_1=0$ and $\Gamma=1$, one or two RICs appear for any values of $\varepsilon_0$, although again one or the other RIC exits the continuum at $\varepsilon_0 = \pm 1$ and splits into a bound state and a virtual bound state, as previously discussed in Sec. \ref{sec:PT.outgoing.spec.ep1}.

%%%%%%%%%%%%%%%%%%%%%
%%%%%%%%%SECTION: perfect transmission
%%%%%%%%%%%%%%%%%%

\subsection{Perfect transmission, invisibility and applications}
\label{sec:PT.scattering.2.perfect}

In the present subsection, we turn our attention from the RIC poles to the scattering continuum itself.
More specifically, we now study the system parameters that give rise to perfect transmission such that $T=1$ while the reflectance vanishes $R = 0$.  
%For the simplest case $\varepsilon_0 = \varepsilon_1 = 0$ we can obtain perfect transmission at any given $k$ value by appropriately choosing $\Gamma$, which approximately holds in the case where $\epsilon_0$ and $\epsilon_1$ are nonzero but take on small values.  
We also examine the phase associated with the perfect transmission in order to observe the case in which invisibility is obtained, and
%Invisibility occurs whenever we have perfect transmission \emph{and} there is no phase shift in the output signal.  
%If we imagine the scenario in which there perfect transmission is obtained but a phase shift is present, then an observer can still detect the impurity by performing a time-of-flight measurement on the transmitted wave~\cite{LonghiPRA10} [\textcolor{red}{other refs?}].
we comment on several points below that may be useful from an engineering perspective.

The condition to obtain perfect transmission in the left-to-right (right-to-left) case is immediately apparent %from the form of the scattering wave appearing 
in Eq.~(\ref{eq-PTasym}) (Eq.~(\ref{eq-PTasym1})), namely
%for this to be realized, we must have 
$B = 0$ ($C = 0$).  
This condition is realized whenever the determinant of the $3\times 3$ matrix on the left-hand side of Eq.~(\ref{outgoing.matrix}) (Eq.~(\ref{outgoing.matrix1})) vanishes.  
Hence, we obtain left-to-right (right-to-left) perfect transmission at a given value $k = \tilde{k}_{\textrm{X},j}$ within the continuum whenever the condition 
$M_\textrm{R} (\tilde{\lambda}_{\textrm{R},j}) = 0$ ($M_\textrm{L} (\tilde{\lambda}_{\textrm{L},j}) = 0$) is satisfied, where 
$\tilde{\lambda}_{\textrm{X},j} = e^{i \tilde{k}_{\textrm{X},j}}$ and
\begin{equation}
  M_\textrm{L,R} (\lambda)
    = \left( \varepsilon_1 \mp i \Gamma \right) \lambda^4%_\textrm{PT}^4
  	+ \left( \varepsilon_1^2 + \Gamma^2 + \varepsilon_0 \left( \epsilon_1 \mp i \Gamma \right) \right) 												                         \lambda^3%_\textrm{PT}^3
	+ \varepsilon_0 \left( 1 + \varepsilon_1^2 + \Gamma^2 \right) \lambda^2%_\textrm{PT}^2
	+ \left( \varepsilon_1^2 + \Gamma^2 + \varepsilon_0 \left( \epsilon_1 \pm i \Gamma \right) \right) 												                          \lambda%_\textrm{PT}
	+ \varepsilon_1 \pm i \Gamma
	= 0
		.
\label{perf.trans.cond}
\end{equation}
%%
%Here the upper (lower) sign is for left-to-right (right-to-left) perfect transmission.
Since $M_\textrm{X} (\lambda)$ is a quartic polynomial for either case $\textrm{X}=\textrm{L,R}$, for a given set of parameter values we generally obtain four values $\tilde{k}_{\textrm{L},j}$ for left-to-right and four values $\tilde{k}_{\textrm{R},j}$ for right-to-left perfect transmission.  However, some of these solutions might turn out to be complex-valued and hence must be discarded.

\begin{figure}
\hspace*{0.05\textwidth}
 \includegraphics[width=0.4\textwidth]{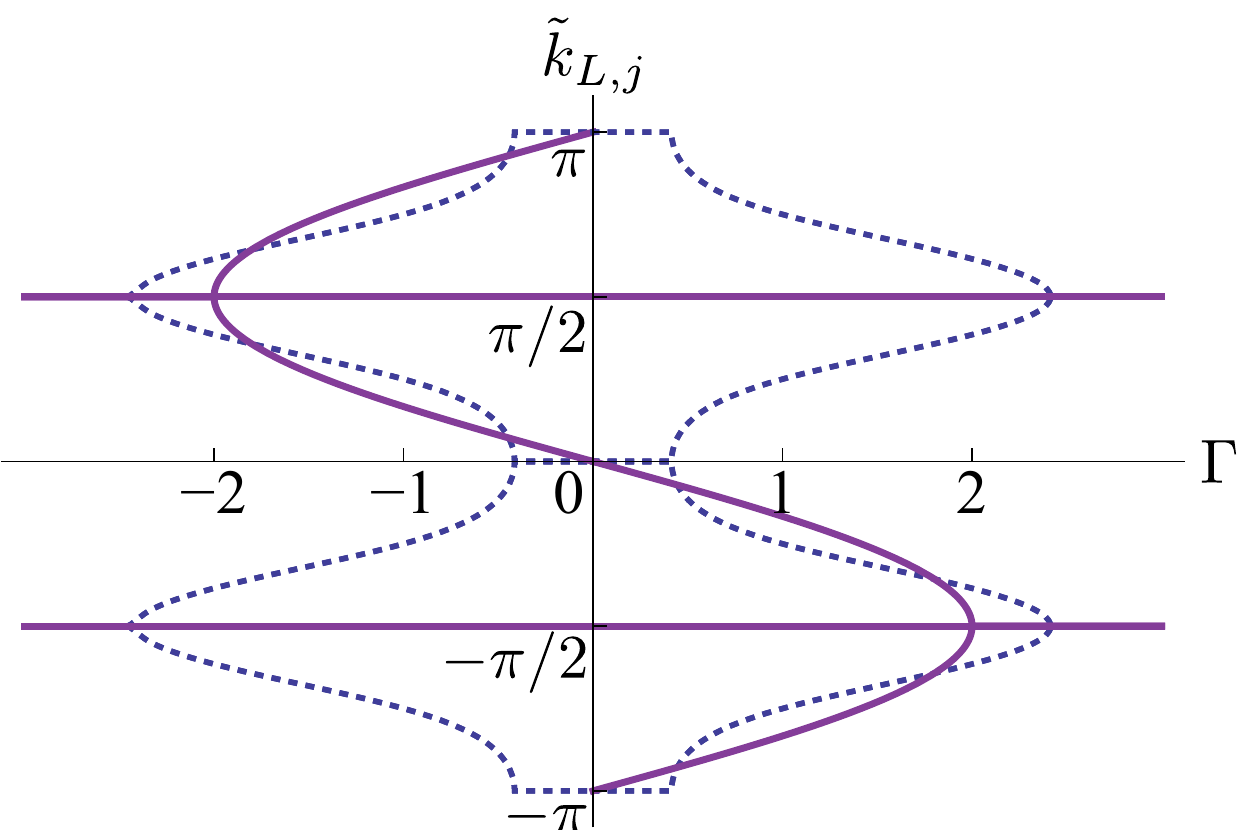}
\hfill
 \includegraphics[width=0.4\textwidth]{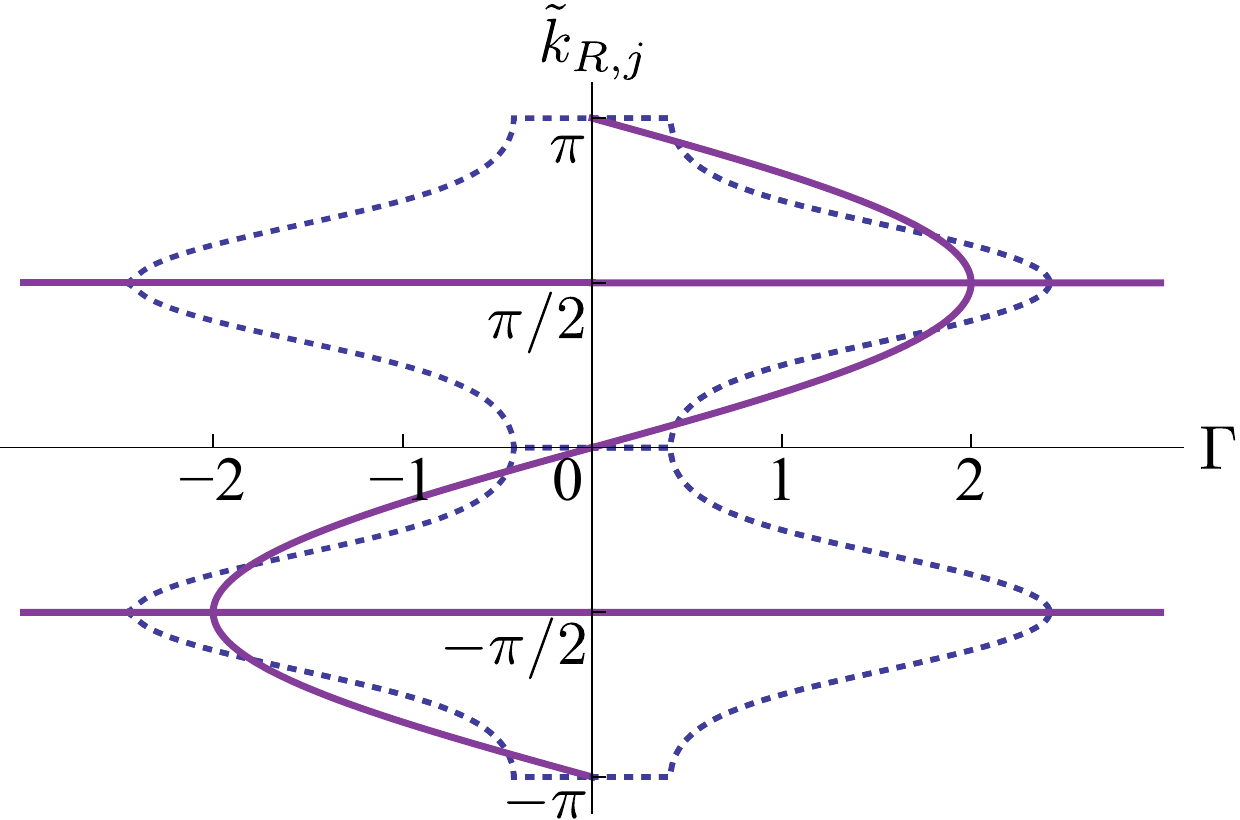}
 \hspace*{0.05\textwidth}
\\
\vspace*{\baselineskip}
\hspace*{0.05\textwidth}(a)\hspace*{0.440\textwidth}(b)\hspace*{0.4\textwidth}
\\
\vspace*{\baselineskip}
\hspace*{0.05\textwidth}
 \includegraphics[width=0.4\textwidth]{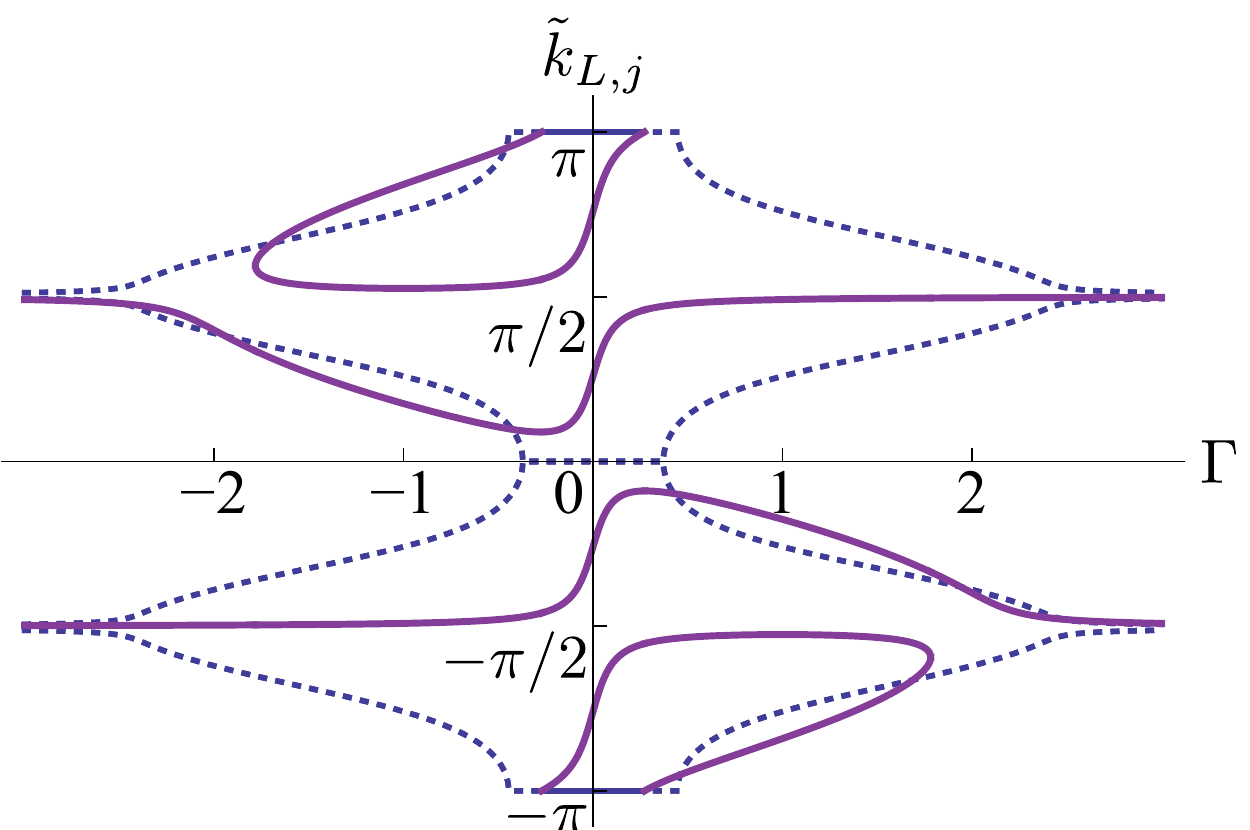}
\hfill
 \includegraphics[width=0.4\textwidth]{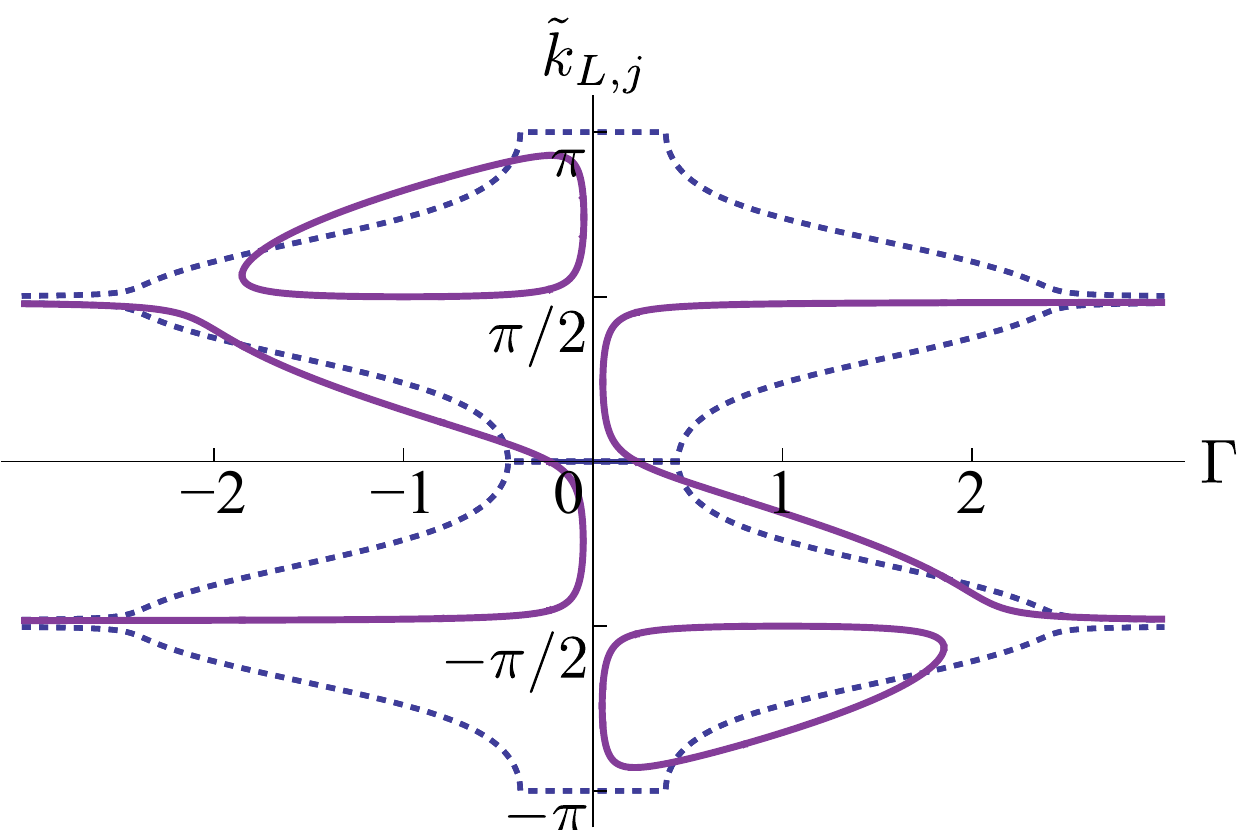}
 \hspace*{0.05\textwidth}
\\
\vspace*{\baselineskip}
\hspace*{0.05\textwidth}(c)\hspace*{0.440\textwidth}(d)\hspace*{0.4\textwidth}
\\
\vspace*{\baselineskip}
\caption{The wave numbers for perfect transmission (full curves) as a function of $\Gamma$ for fixed values of 
$\varepsilon_0$ and $\varepsilon_1$;  
the wave numbers $\re \ k_j$ for discrete eigenvalues are also shown in the background with dotted curves, except for bound states which appear as full lines at either $k = 0$ (lower band edge) or  $\pm \pi$ (upper band edge).  
We used the following parameter values:
(a) left-to-right transmission for $\varepsilon_0 = \varepsilon_1 = 0$,
(b) right-to-left transmission for $\varepsilon_0 = \varepsilon_1 = 0$,
(c) left-to-right transmission for $\varepsilon_0 = 0, \varepsilon_1 = 0.08$, and
(d) left-to-right transmission for $\varepsilon_0 = -0.1, \varepsilon_1 = 0$.
}
 \label{fig:PT.perf.trans}
 \label{fig7}
 \end{figure}

In the simplest case $\varepsilon_0 = \varepsilon_1 = 0$, Eq.~(\ref{perf.trans.cond}) for the left-to-right transmission gives the factorized form 
$M_\textrm{L} (\lambda) = i \Gamma (1 + i \lambda)(1 - i \lambda) (1 - i \Gamma \lambda - \lambda^2)$.  
The two linear factors give two $\Gamma$-independent solutions as 
$\tilde{k}_{\textrm{L},1} = \pi / 2$ and $\tilde{k}_{\textrm{L},2} = - \pi / 2$ with energy $E = 0$ appearing directly at the center of the transmission band.
Meanwhile the quadratic factor gives the solutions
\begin{equation}
  \tilde{k}_{\textrm{L},\{3,4\} } =
	\left\{ \begin{array}{ll}
		\cos^{-1} \left( \pm \sqrt{4 - \Gamma^2} / 2 \right) 	
				& \ \ \ \ \mbox{for $-2 < \Gamma <  0$,}    	\\
		\cos^{-1} \left( \pm \sqrt{4 - \Gamma^2} / 2 \right) - \pi	
				& \ \ \ \ \mbox{for $0 < \Gamma <  2$.}
	\end{array}
	\right. 
\label{perf.trans.gamma.L}
\end{equation}
These four solutions are plotted as the full curves in Fig.~\ref{fig:PT.perf.trans}(a);
%%%%%%%%%
%%%%%%%%%
%%%%%%%%%
%%%%%%%%%
note that the real parts of the wave numbers for the eigenvalues $\re k_j$ are also plotted as the dotted curves, for reasons that will be described below.  
From an applications perspective, we note for this case we can obtain perfect transmission at any given value of $k \in [-\pi, \pi]$ by choosing the appropriate value of $\Gamma$.

So far we have only considered the intensity of the transmitted signal, but not the associated phase information.  
While perfect transmission does maintain the signal intensity, if there is a phase shift between two leads an observer may still be able to detect the presence of impurities in the scattering region by performing a time-of-flight measurement~\cite{LonghiPRA10}.  
We can investigate the phase shift for the perfect transmission states by calculating the eigenvector $( A, \psi(0), C )^\textrm{T}$ on the left-hand side of the singular matrix~(\ref{outgoing.matrix}) for our four perfect transmission solutions.  
For the case of the two perfect transmission values $\tilde{k}_{\textrm{L},1} = \pi / 2$ ($\tilde{\lambda}_{\textrm{L},1} = i$) and $\tilde{k}_{\textrm{L},2} = - \pi / 2$ ($\tilde{\lambda}_{\textrm{L},2} = - i$) we have
\begin{equation}
  \begin{pmatrix}
  	i \left( \Gamma \pm 1 \right)	& \mp i		& 0    \\
	1						& 0 			& -1   \\
	0						& -1			& 1 \pm \Gamma  \\
	\end{pmatrix}
  \begin{pmatrix}
  	A		\\
	\psi(0)	\\
	C		\\
	\end{pmatrix}
	= 0
		,
\label{perf.trans.vec}
\end{equation}
which yields $A = C$ without any phase shift between the leads and
%$\psi(0) = ( 1 \pm \Gamma ) A$.
%%
\begin{equation}
  \psi(0) = ( 1 \pm \Gamma ) A
  	.
\label{perf.trans.psi.0}
\end{equation}
%%
%With no phase shift between the leads, we have thus established 
Left-to-right invisibility holds for these two cases; 
indeed, as we discuss below right-to-left invisibility will also hold for $k = \pm \pi/2$.  
As a further point, notice from Eq.~(\ref{perf.trans.psi.0}) that we can arbitrarily adjust $\psi(0)$
%$\psi(0) = ( 1 \pm \Gamma ) A$ 
by tuning the parameter $\Gamma$ and still maintain perfect transmission; 
for the special case $\Gamma = \mp 1$ we can even eliminate it entirely.  

Performing a similar calculation for the $\Gamma$-dependent perfect transmission states $\tilde{k}_{\textrm{L},3}$ and $\tilde{k}_{\textrm{L},4}$ reported in Eq.~(\ref{perf.trans.gamma.L}) reveals that a phase shift is always present for these states, apart from the special case $k = \pm \pi/2$ for a specific value of $\Gamma$.
To summarize, invisibility can only be achieved at $k = \pm \pi/2$ for arbitrary $\Gamma$.

For right-to-left transmission in the simplest case $\varepsilon_0 = \varepsilon_1 = 0$, we obtain from the lower sign in Eq.~(\ref{perf.trans.cond}) the factorized equation $M_\textrm{R} (\lambda) = - i \Gamma (1 + i \lambda)(1 - i \lambda) (1 + i \Gamma \lambda - \lambda^2)$.
Hence for the right-to-left transmission we again obtain perfect transmission at 
$\tilde{k}_{\textrm{R},1} = \pi/2$ and $\tilde{k}_{\textrm{R},2} = - \pi/2$, but for the $\Gamma$-dependent perfect transmission states we have instead
\begin{equation}
 \tilde{k}_{\textrm{R},\{3,4\} } =
	\left\{ \begin{array}{ll}
		\cos^{-1} \left( \pm \sqrt{4 - \Gamma^2} / 2 \right) - \pi	
				& \ \ \ \ \mbox{for $-2 < \Gamma <  0$,}    	\\
		\cos^{-1} \left( \pm \sqrt{4 - \Gamma^2} / 2 \right)	
				& \ \ \ \ \mbox{for $0 < \Gamma <  2$,}
	\end{array}
	\right. 
\label{perf.trans.gamma.R}
\end{equation}
as plotted in Fig.~\ref{fig:PT.perf.trans}(b).
Notice that the expressions~\eqref{perf.trans.gamma.R} are just reversed from the left-to-right case reported in Eq.~(\ref{perf.trans.gamma.L}); 
this is a natural result for a $\PT$-symmetric system since switching our scattering orientation amounts to swapping the position of the gain and loss impurities.  
Again in this case we can evaluate the scattering wave coefficients to find that $D = B$ and hence we have no phase shift for the perfect transmission states $\tilde{k}_{\textrm{R}, \{1,2\} } = \pm \pi/2$; 
this shows that these states yield bi-directional invisibility.  

Here the response at the site $0$ is given by 
\begin{equation}
  \psi(0) = (1 \mp \Gamma) D
  	.
\label{perf.trans.psi.0.R}
\end{equation}
If we choose, say $\Gamma = -1$, the invisible right-to-left wave has a finite value of $\psi(0)$ at $k = + \pi/2$, but from Eq. (\ref{perf.trans.psi.0}) the left-to-right transmission gives no response at the same frequency.  
Hence the site $0$ in this scenario can act as a kind of switch that responds to the otherwise invisible signal transmission in one direction but ignores signals from the other direction.

In Fig.~\ref{fig:PT.perf.trans}(c) and (d) we plot the left-to-right perfect transmission for more general parameter values with $\varepsilon_0 = 0, \varepsilon_1 = 0.08$ and $\varepsilon_0 = -0.1, \varepsilon_1 = 0$, respectively.  
We note that the range of the continuum that is capable of supporting perfect transmission has been reduced slightly in comparison to the `cleaner' case in Figs.~\ref{fig:PT.perf.trans}(a) and (b) for these relatively small values of the impurity energies.  
For larger impurity values the range of coverage for perfect transmission is further reduced.

We also note the following connection between the perfect transmission scattering states and the bound states of the discrete spectrum.  Notice that in the background of Fig.~\ref{fig:PT.perf.trans}(a)--(d) we have plotted the real part of the wave numbers $\re k_j$ for the discrete eigenvalues as the dotted curves.  
However, the wave numbers for the bound states are marked with full lines appearing at either $k = 0$ (lower continuum edge) or $k = \pm \pi$ (upper continuum edge).  Focusing on Fig.~\ref{fig:PT.perf.trans}(c) and (d), we note that the appearance of the perfect transmission scattering state exactly coincides with the appearance (or disappearance) of a bound state at the edges of the Brillouin zone $k=0$ or $k=\pm \pi$;
see the footnote~\footnote{Note that we can explicitly show that the appearance of the perfect transmission state and the bound state delocalization do indeed occur at the same point in parameter space in the following manner.  On  one hand, we can exactly locate the delocalization transition at the band edges $\lambda = \pm 1$ by plugging into the dispersion polynomial Eq. (\ref{P.lambda}) as $P(\lambda = \pm 1) = 0$; on the other hand we can find the appearance of the perfect transmission states at the band edges by plugging into Eq. (\ref{perf.trans.cond}) as $M_\textrm{L} (\pm 1) = M_\textrm{R} (\pm 1) = 0$, which can be analytically solved in either case.  Doing so yields the exact same points in parameter space.}.

This is a rather intuitive result if we consider the behavior of the bound-state wave function at the delocalization transition, where it brushes against one of the band edges before becoming a virtual bound state in the second Riemann sheet.  On one side of this transition we have a bound state with a wave function that is localized in the defect region, while on the other side we have a virtual bound state with a wave function that diverges into the leads (one can even view this state as being localized in the leads \cite{Hatano14}).  At precisely the transition between these two states, we have a wave function that spreads out evenly throughout the chain, which here supports perfect transmission from one lead to the other.

This explains the connection between the delocalization transition and the appearance of the perfect transmission state at either edge of the scattering continuum, which may provide an intuitive approach to designing systems with desired transport properties.  We notice a somewhat similar transition appears in Ref.~\cite{VHIC14}.

%%%%%%%%%%%%%%%%%%%%%
%%%%%%%%%SECTION: outgoing waves B.C.
%%%%%%%%%%%%%%%%%%

\section{$\PT$-symmetric scattering wave solutions}
\label{sec:PT.scattering.2}
\label{sec6}

This section is devoted to more mathematical interest.
We here investigate the $\PT$-symmetric properties of the scattering wave solutions.
%While the discrete states exhibit $\PT$-symmetry breaking as we demonstrated in Sec.~\ref{sec3} and Sec.~\ref{sec:PT.bound}, the results in Sec.~\ref{sec5} suggest that we do not see any indications of $\PT$-symmetry breaking in scattering states; the eigenvalue $E=-2\cos k$ is always real.
Previously we demonstrated in Sec.~\ref{sec3} and Sec.~\ref{sec:PT.bound} that the discrete states satisfy $\PT$-symmetric boundary conditions in certain regions of parameter space; however, we note that despite the fact that the eigenvalue $E=-2\cos k$ associated with the scattering states is always real, in Sec.~\ref{sec5} these states generically satisfied $\PT$-asymmetric boundary conditions.

This motivates us to investigate whether or not the scattering states themselves can obey
$\PT$-symmetric boundary conditions.
We will show in Secs.~\ref{sec:PT.scattering.2.sym} and~\ref{app:Jost} that we indeed always have $\PT$-symmetric scattering states.
In the former, we will $\PT$-symmetrize the scattering wave function previously obtained in Sec.~\ref{sec:PT.scattering}.
In the latter, we will present a more direct and systematic way of finding a $\PT$-symmetric scattering wave with the use of the Jost solutions.
We will then introduce the concept of the $\PT$ current in Sec.~\ref{sec:PT-current}.

\subsection{$\PT$-symmetrization of the scattering wave solutions}\label{sec:PT.scattering.2.sym}

%Here we will demonstrate how to obtain a $\PT$-symmetric scattering wave solution from our general scattering solutions appearing in Eqs.~(\ref{eq-PTasym}) and~(\ref{eq-PTasym1}).  
%We present an alternative method to obtain these solutions from the Jost solutions of the original Schr\"odinger equation in Sec.~\ref{app:Jost}.

We can construct a $\PT$-symmetric solution out of an asymmetric solution $\psi(x)$ by writing
\begin{align}\label{wave.PT.gen}
\psi_\PT(x)&=\psi(x)+\PT\psi(x).
%\\
%\psi_a(x)&=\psi(x)-\PT\psi(x).
\end{align}
%For the $\PT$ transformation, they have the eigenvalues $\pm 1$.
Let us apply this strategy to the scattering solution~\eqref{eq-PTasym}.
The $\PT$ transformation results in the changes $i\to-i$ and $x\to-x$ as well as the complex conjugation of coefficients $A$ and $C$;
we recall that $B$ is real by assumption in the relations~\eqref{eq300}--\eqref{eq320} for the simplest case $\varepsilon_0=\varepsilon_1=0$.
Note that the real-valued wave number $k$ (the momentum) is invariant under the action of $\PT$, since both $\mathcal{P}$ and $\mathcal{T}$ result in a sign flip separately for this quantity.
Substituting our result into Eq.~(\ref{wave.PT.gen}) we obtain a $\PT$-symmetric solution for $0\leq k\leq \pi$ given by
\begin{align}
\psi_\PT^{(\mathrm{L})}(x)=
\begin{cases}
(A+ C^\ast)e^{ikx}+B e^{-ikx} & \quad\mbox{for $x\leq-1$},
\\
\psi(0) + {\psi(0)}^\ast & \quad\mbox{for $x=0$},
\\
(A^\ast + C) e^{ikx} + B e^{-ikx} &\quad\mbox{for $x\geq 1$}
\end{cases}
\end{align}
%where we have used the assumption $B\in\mathbb{R}$.
%In the simplest case $\varepsilon_0=\varepsilon_1=0$, the equation similar to Eq.~\eqref{outgoing.matrix} produces
with the relations~\eqref{eq300}--\eqref{eq320} producing
\begin{align}
A+C^\ast&=-B\frac{e^{2ik}\Gamma}{2\sin k+\Gamma},
\\
\psi(0)+{\psi(0)}^\ast&=B\frac{2\sin k}{2\sin k+\Gamma}
%\\
%A-C^\ast&=B\frac{2i\sin k-e^{2ik}\Gamma^2\cos k}%
%{(2\sin k+\Gamma) \Gamma\cos k},
%\\
%\psi(0)-{\psi(0)}^\ast&=B\frac{2i(1+\Gamma\sin k)\sin k}%
%{(2\sin k+\Gamma) \Gamma\cos k}.
\end{align}
for the simplest case $\varepsilon_0=\varepsilon_1=0$.
%\begin{align}
%A\pm C^\ast&=\frac{\psi(0)}{1+2\Gamma \sin k +\Gamma^2}
%\times\begin{cases}
%2(1+\Gamma \sin k) +ie^{2ik}\Gamma^2(1+i\Gamma e^{-ik}) \cot k
%\\
%2i\Gamma \cos k +ie^{2ik}\Gamma^2(1+i\Gamma e^{-ik}) \cot k
%\end{cases}
%\end{align}
Note that the component $\psi_\PT^{(\mathrm{L})}(0)$ is real.
If we choose the normalization as $\phi_\PT(0)^{(\mathrm{L})}=1$ we obtain
\begin{align}\label{eq-PTsol}
\phi_\PT^{(\mathrm{L})}(x)=
\begin{cases}
\displaystyle
-\frac{\Gamma}{2\sin k}e^{ik(x+2)}+\left(1 + \frac{\Gamma}{2\sin k}\right) e^{-ikx} & \quad\mbox{for $x\leq-1$},
\\
1 & \quad\mbox{for $x=0$},
\\
\displaystyle
-\frac{\Gamma}{2\sin k}e^{ik(x-2)} +\left(1 + \frac{\Gamma}{2\sin k}\right) e^{-ikx} &\quad\mbox{for $x\geq 1$}
\end{cases}
\end{align}
for $0\leq k\leq \pi$, as our first $\PT$-symmetric solution.

We can instead start from the right-to-left scattering wave Eq.~\eqref{eq-PTasym1}.
%If we start from Eq..~\eqref{eq-PTasym1} of the $\PT$-asymmetric solution instead of Eq.~\eqref{eq-PTasym}, the final $\PT$-symmetrized solution is slightly different from Eq..~\eqref{eq-PTsol}:
Following a similar procedure as above, we obtain
\begin{align}
\psi_\PT^{(\mathrm{R})}(x)=
\begin{cases}
Ce^{ikx}+ (B+D^\ast) e^{-ikx} & \quad\mbox{for $x\leq-1$},
\\
\psi(0) + {\psi(0)}^\ast & \quad\mbox{for $x=0$},
\\
C e^{ikx} + (B^\ast +D) e^{-ikx} &\quad\mbox{for $x\geq 1$}
\end{cases}
\end{align}
for $0\leq k\leq \pi$, where we used the assumption $C\in\mathbb{R}$ in this case.
%\begin{align}
%%\label{eq-PTasym}
%\psi(x)=
%\begin{cases}
%B e^{-ikx} & \quad\mbox{for $x\leq-1$},
%\\
%\psi(0) & \quad\mbox{for $x=0$},
%\\
%C e^{ikx} +D e^{-ikx} &\quad\mbox{for $x\geq 1$},
%\end{cases}
%\end{align}
In the simplest case $\varepsilon_0=\varepsilon_1=0$ we have
\begin{align}
B+D^\ast&=C\frac{e^{-2ik}\Gamma}{2\sin k-\Gamma},
\\
\psi(0)+{\psi(0)}^\ast&=C\frac{2\sin k}{2\sin k-\Gamma}.
%\\
%A-C^\ast&=B\frac{2i\sin k-e^{2ik}\Gamma^2\cos k}%
%{(2\sin k+\Gamma) \Gamma\cos k},
%\\
%\psi(0)-{\psi(0)}^\ast&=B\frac{2i(1+\Gamma\sin k)\sin k}%
%{(2\sin k+\Gamma) \Gamma\cos k}.
\end{align}
After again choosing our normalization such that $\phi_\PT^{(\mathrm{R})}(0)=1$ we obtain
\begin{align}\label{eq-PTsol2}
\phi_\PT^{(\mathrm{R})}(x)=
\begin{cases}
\displaystyle
\left(1 - \frac{\Gamma}{2\sin k}\right)e^{ikx}+\frac{\Gamma}{2\sin k} e^{-ik(x+2)} & \quad\mbox{for $x\leq-1$},
\\
1 & \quad\mbox{for $x=0$},
\\
\displaystyle
\left(1 - \frac{\Gamma}{2\sin k}\right)e^{ikx}+\frac{\Gamma}{2\sin k} e^{-ik(x-2)} &\quad\mbox{for $x\geq 1$}
\end{cases}
\end{align}
for $0\leq k\leq \pi$, as our second $\PT$-symmetric solution, which is indeed obtained simply by flipping the sign of $k$ in the first solution~\eqref{eq-PTsol}:
We therefore conclude that the solution~\eqref{eq-PTsol} holds for the entire first Brillouin zone $-\pi<k\leq\pi$.

\subsection{Jost solutions}\label{app:Jost}

The solutions in the previous subsection~\ref{sec:PT.scattering.2.sym} seem somewhat strange because of the asymmetry with respect to the inversion of $k$.
In this subsection we obtain an alternative $\PT$-symmetric solution %Eq.~\eqref{eq-PTsol} 
by directly finding the Jost solutions of the original Schr\"{o}dinger equation~\eqref{eq-Sch1}--\eqref{eq-Sch3}.
We will find a solution of a more symmetric form, which is indeed a superposition of the solutions~\eqref{eq-PTsol} and~\eqref{eq-PTsol2}.
%, providing an alternate method to that appearing in the main text.
%We can derive the $\PT$-symmetric solution~\eqref{eq-PTsol} using the Jost solutions.
%Let us go back to the Schr\"{o}dinger equation~\eqref{eq-Sch1}--\eqref{eq-Sch3}.
%If we assume from the outset that the wave function is $\PT$-symmetric, such that $\psi(x)=\psi(-x)^\ast$ with $E(k)\in\mathbb{R}$, then Eq.~\eqref{eq-Sch1} becomes equivalent to Eq.~\eqref{eq-Sch3}.
%We thereby map the problem on to the semi-infinite space $x\geq0$ with the boundary condition at $x=0$ given by Eq.~\eqref{eq-Sch2}.
We again restrict ourselves to the simplest case $\varepsilon_0=\varepsilon_1=0$. 

Let us briefly overview the way of constructing a scattering wave solution out of the Jost solutions.
When the potential vanishes far away from the origin, we can assume a solution of the form of a plane wave there.
The solutions thus defined under the boundary conditions~\cite{Newton60,Newton82}
\begin{align}\label{eq-Jost}
f_\pm(x)=\alpha e^{\pm ikx}\quad\mbox{as $x\to\infty$}
\end{align}
with an appropriate constant $\alpha$ are called the Jost solutions.
They, however, do not generally satisfy boundary conditions at the origin.
We therefore take a superposition of the two Jost solutions so that it may satisfy the boundary conditions at the origin, which yields a scattering wave solution.

Since the potential vanishes for $x\geq2$ in the present case, we can use Eq.~\eqref{eq-Jost} in the region $x\geq1$.
Let us now construct a $\PT$-symmetric Jost solutions.
Since $\PT e^{ikx}=e^{ikx}$, we set 
\begin{align}\label{eq-Jost1}
f_\pm(x)=\alpha^\ast e^{\pm ikx}\quad\mbox{as $x\to-\infty$},
\end{align}
which we can use in the region $x\leq -1$.
These Jost solutions, however, do not satisfy the boundary conditions at $x=0$.
Indeed, the Schr\"{o}dinger equation for $x=1$, namely Eq.~\eqref{eq-Sch3}, gives
\begin{align}\label{eq640}
f_\pm(0)&=-f_\pm(2)+(-E(k) -i\Gamma) f_\pm(1)
\nonumber\\
&=\alpha[-e^{\pm 2ik}+(e^{ik}+e^{-ik}-i\Gamma)e^{\pm ik}]
\nonumber\\
&=\alpha(1-i\Gamma e^{\pm ik})
\end{align}
for $\varepsilon_1=0$, while the Schr\"{o}dinger equation for $x=-1$, Eq.~\eqref{eq-Sch1}, gives
\begin{align}\label{eq650}
f_\pm(0)&=-f_\pm(-2)+(-E(k) +i\Gamma) f_\pm(-1)
\nonumber\\
&=\alpha^\ast[-e^{\mp 2ik}+(e^{ik}+e^{-ik}+i\Gamma)e^{\mp ik}]
\nonumber\\
&=\alpha^\ast(1+i\Gamma e^{\mp ik}).
\end{align}
We can make Eqs.~\eqref{eq640} and~\eqref{eq650} continuous at the origin by choosing
\begin{align}
\alpha=1+i\Gamma e^{\mp ik},
\end{align}
which makes $f_\pm(0)=1\pm 2\Gamma\sin k+\Gamma^2$, but the resulting solution $f_\pm(x)$ does not satisfy the Schr\"{o}dinger equation for $x=0$, Eq.~\eqref{eq-Sch2} (even after setting 
$\varepsilon_0=0$ for the present case).

The physical solution that satisfies the Schr\"{o}dinger equation~\eqref{eq-Sch2} must be given by a linear combination of the two Jost solutions:
\begin{align}\label{eq-physsol}
\phi_\PT(x)=A_+f_+(x)+A_-f_-(x)
\end{align}
with two superposing coefficients $A_\pm$, which we set so that $\phi_\PT(x)$ may satisfy Eq.~\eqref{eq-Sch2}.
Let us define the Jost \emph{functions} (not to be confused with the Jost solutions) by
\begin{align}
F_\pm(k)&=-f_\pm(-1)-f_\pm(1)-E(k)f_\pm(0)
\nonumber\\
&=-(1-i\Gamma e^{\pm ik})e^{\mp ik}-(1+i\Gamma e^{\mp ik})e^{\pm ik}
+(e^{ik}+e^{-ik})(1-i\Gamma e^{\pm ik})(1+i\Gamma e^{\mp ik})
\nonumber\\
&=2\Gamma(\Gamma\pm 2\sin k)\cos k.
\end{align}
The Schr\"{o}dinger equation~\eqref{eq-Sch2} is then reduced to
\begin{align}
A_+F_+(k)+A_-F_-(k)=0,
\end{align}
which fixes the ratio between $A_+$ and $A_-$.

%We can confirm that this solution is consistent with Eq.~\eqref{eq-PTsol}.
%Equation~\eqref{eqC60} now reads
%\begin{align}
%F_\pm(k)&=f_\pm(0)\phi_\PT(1)-f_\pm(1)\phi_\PT(0)
%\nonumber\\
%&=\frac{1}{2i\sin k}
%\left[f_\pm(0)\left(F_-(k)f_+(1)-F_+(k)f_-(1)\right)
%-f_\pm(1)\left(F_-(k)f_+(0)-F_+(k)f_-(0)\right)\right]
%\end{align}
%
%\begin{align}
%F_+(k)&=-i(2\sin k+\Gamma),
%\\
%F_-(k)&=-i\Gamma e^{-2ik}.
%\end{align}

By normalizing the function by $\phi_\PT(0)=1$, we can express the final result as follows:
\begin{align}
\phi_\PT(x)=\begin{cases}
\displaystyle
\left(1-\frac{\Gamma}{2\sin k}\right)\frac{1-i\Gamma e^{ik}}{2}e^{ikx}
+\left(1+\frac{\Gamma}{2\sin k}\right)\frac{1-i\Gamma e^{-ik}}{2}e^{-ikx}
&\quad\mbox{for $x\leq-1$},
\\
1 & \quad\mbox{for $x=0$},
\\
\displaystyle
\left(1-\frac{\Gamma}{2\sin k}\right)\frac{1+i\Gamma e^{-ik}}{2}e^{ikx}
+\left(1+\frac{\Gamma}{2\sin k}\right)\frac{1+i\Gamma e^{ik}}{2}e^{-ikx}
&\quad\mbox{for $x\geq1$}.
\end{cases}
\end{align}
This is indeed a linear combination of Eqs.~\eqref{eq-PTsol} and~\eqref{eq-PTsol2} in the form
\begin{align}
\phi_\PT(x)=\frac{1}{2}\left(1+\frac{\Gamma}{2\sin k}\right)\psi_\PT^{(R)}(x)
+\frac{1}{2}\left(1-\frac{\Gamma}{2\sin k}\right)\psi_\PT^{(L)}(x) ;
\end{align}
however, the domain extends over the entire first Brillouin zone $-\pi<k\leq\pi$.

\subsection{$\PT$-current}
\label{sec:PT-current}

Because we have a source and a sink, the discrete states generally do not conserve the particle number and the scattering states do not conserve the current.
We here, however, introduce a current that is conserved for a $\PT$-symmetric scattering state, which we refer to as the $\PT$-current.

The standard current is defined in a one-dimensional continuous space as
\begin{align}
j&=\re\left(\psi(x)^\ast \hat{p}\psi(x)\right)
%\nonumber\\
%&
=\frac{1}{2i}\left(\psi(x)^\ast \frac{d}{dx}\psi(x)-\psi(x)\frac{d}{dx}\psi(x)^\ast\right),
\end{align}
which would normally be independent of $x$, but this does not generally hold true in a $\PT$-symmetric non-Hermitian system.
We here instead introduce the $\PT$-current
\begin{align}\label{eq:PT-current}
j_\PT=\frac{1}{2}\left(\psi(x)^\ast\frac{d}{dx}\psi(-x)-\psi(-x)\frac{d}{dx}\psi(x)^\ast\right).
\end{align}
We can prove that the $\PT$-current is independent of $x$ for an eigenfunction $\psi(x)$ with real eigenvalue $E$ of the Hamiltonian
\begin{align}
H_\PT=-\frac{d}{dx^2}+V_\PT(x)
\end{align}
with $\PT V_\PT(x)=V_\PT(-x)^\ast=V_\PT(x)$.
Computing the $x$ derivative of the $\PT$-current~\eqref{eq:PT-current}, we indeed have
\begin{align}
\frac{d}{dx}j_\PT(x)&=\frac{1}{2i}\left(\psi(x)^\ast\frac{d^2}{dx^2}\psi(-x)-\psi(-x)\frac{d^2}{dx^2}\psi(x)^\ast\right)
\nonumber\\
&=\frac{1}{2i}\left[\psi(x)^\ast\left(V_\PT(-x)-E\right)\psi(-x)-\psi(-x)\left(V_\PT(x)^\ast-E\right)\psi(x)^\ast\right]=0.
\end{align}
Notice that it vanishes identically for a $\PT$-symmetric eigenfunction, because we then have $\psi(x)^\ast=\psi(-x)$ in Eq. (\ref{eq:PT-current}).

\begin{figure}
\centering
\includegraphics[width=0.5\textwidth]{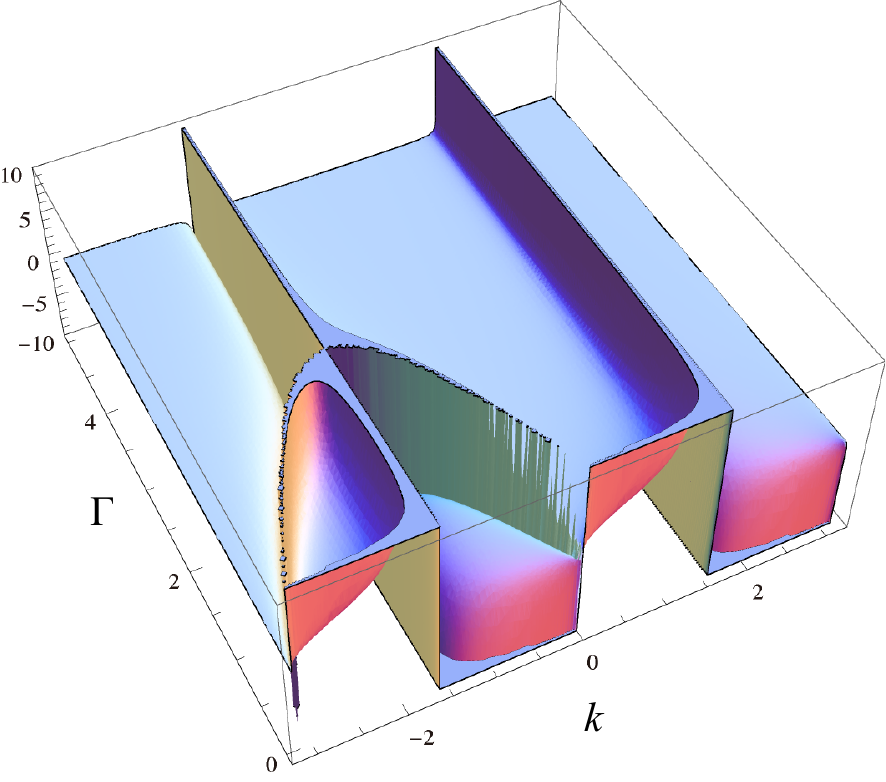}
\caption{The $\PT$-current~\eqref{eq830} without the factor $|B|^2$ in the simplest case $\varepsilon_0=\varepsilon_1=0$.}
\label{fig8}
\end{figure}

In the discretized space of the one-dimensional tight-binding model, the standard current is given by
\begin{align}\label{eq:discretized-current}
j&=\frac{1}{2i}\left[\psi(x)^\ast\left(\psi(x+1)-\psi(x)\right)-\psi(x)\left(\psi(x+1)^\ast-\psi(x)^\ast\right)\right]
\nonumber\\
&=\frac{1}{2i}\left(\psi(x)^\ast\psi(x+1)-\psi(x)\psi(x+1)^\ast\right),
\end{align}
while the $\PT$-current is given by
\begin{align}\label{eq:discretized-PT-current}
j_\PT&=\frac{1}{2}\left(\psi(x)^\ast\psi(-x-1)-\psi(-x)\psi(x+1)^\ast\right).
\end{align}
For a $\PT$-asymmetric left-to-right scattering state of the form~\eqref{eq-PTasym}, the (traditional) current~\eqref{eq:discretized-current} is
\begin{align}
j&=\sin k\times
\begin{cases}
|A|^2+|B|^2 &\quad\mbox{for $x\leq -2$,}\\
|C|^2 & \quad\mbox{for $x\geq 1$,}
\end{cases}
\end{align}
which are generally not equal along the two leads as we showed in Sec.~\ref{sec5}.
The $\PT$-current~\eqref{eq:discretized-PT-current}, on the other hand, is
\begin{align}\label{eq830}
j_\PT&=\sin k\times
\begin{cases}
-iB^\ast C &\quad\mbox{for $x\leq -2$,}\\
iBC^\ast &\quad\mbox{for $x\geq 1$,}
\end{cases}
\\
&=\frac{|B|^2\sin^2k}{(\Gamma+2\sin k)\Gamma\cos k},
\end{align}
which is conserved on both sides of the scattering region.
We plot the $\PT$-current in Fig.~\ref{fig8};
the singularities appearing here correspond to the left-to-right perfect transmission states previously shown in Fig.~\ref{fig7}(a).

%%%%%%%%%%%%%%%%%%%%%
%%%%%%%%%SECTION: CONCLUSION
%%%%%%%%%%%%%%%%%%

\section{Conclusion}
\label{sec:conclusion}

In this paper we have combined two types of non-Hermitian systems, open quantum systems and $\PT$-symmetric systems, in order to study a simple example of a $\PT$-symmetric open quantum system.  This system took the form of a tight-binding model with a $\PT$-symmetric defect potential as shown in Fig. \ref{fig1}, which might be physically realized as an optical lattice array or approximated in a variety of $\PT$ systems with a defect scattering center \cite{PTOptExpt4,PT_WGM,RDM05}.  We used this model to illustrate a number of quite general features of $\PT$-symmetric open quantum systems, including properties of the discrete spectrum as well as the scattering states.

In Sec.~\ref{sec:PT.outgoing} we studied the resonance state in continuum (RIC) as a feature of the discrete spectrum, illustrating that it represents a resonance state appearing directly within the conduction band (scattering continuum) as it crosses from the second Riemann sheet of the complex energy plane into the first sheet.  In Sec.~\ref{sec3B} we showed this state takes the form of an outgoing plane wave from the impurity region into the leads.
As a result, this feature also appears in the scattering continuum and hence we returned to it in Sec.~\ref{sec:PT.scattering} in which we studied the scattering properties of the system; in Sec.~\ref{sec:PT.scattering.RIC} we described that the RIC represents a resonance between the open environment associated with the leads of the system and the $\PT$-symmetric defect potential.  In this sense, the RIC can be viewed as a quite particular feature of $\PT$-symmetric open quantum systems.

We also showed in Sec.~\ref{sec:PT.outgoing.spec.ep1} that the RIC may exit the conduction band as we modify the system parameters by splitting into a bound state and a virtual bound state at the edge of the continuum; we believe that this effect should be experimentally observable, perhaps in a modified version of the experiments presented in Ref.~\cite{PTOptExpt4} or Ref. \cite{PT_WGM}.  We also point out that it has been previously illustrated that a bound state (or virtual bound state) appearing near the edge of the continuum should generally result in an enhancement of the long-time non-exponential decay that is known to appear in open quantum systems~\cite{GPSS13}.  Since both a bound state and a virtual bound state appear near the band edge in the case when the RIC exits the continuum, this may result in an even greater enhancement of the non-exponential effect that could also offer an interesting basis for experimental study.

We illustrated another key difference from Hermitian open quantum systems in that complex-valued solutions are allowed to appear in the first sheet of the complex energy plane in the $\PT$-symmetric case.  These appear as pairs of localized states, one with an amplifying characteristic and the other with an absorbing characteristic, as observed experimentally in Ref.~\cite{PTOptExpt4}.  We also pointed out that some of these states may behave as quasi-bound for large values of the $\PT$ defect parameter 
$\Gamma$; these again might be observable in a system that imitates our defect potential.

We evaluated general scattering properties of $\PT$-symmetric open quantum systems in Sec.~\ref{sec:PT.scattering}, in which we calculated the transmission and reflectance for a $\PT$-asymmetric scattering wave solution in Sec.~\ref{sec:PT.scattering.gen} and verified that these satisfy known symmetry relations~\cite{Mosta14,Ahmed12,GCS12}.  We also studied perfect transmission states in Sec.~\ref{sec:PT.scattering.2.perfect}, with invisible solutions as a subset of these, and illustrated a connection between perfect transmission at the band edges and a delocalization transition of the bound state in the discrete spectrum.

After noting that the eigenvalue $E = - 2 \cos k$ associated with the scattering states is always real, in Sec.~\ref{sec6} we used our model as a mathematical prototype to illustrate the construction of scattering wave solutions that themselves satisfy $\PT$-symmetric boundary conditions (just as the bound state in the discrete spectrum is well known to satisfy such boundary conditions, as we have illustrated in Sec.~\ref{sec:PT.bound}).  In Sec.~\ref{sec:PT-current} we wrote the $\PT$-current for these solutions and pointed out that the previously studied perfect transmission states appeared as a special case.

\section*{Acknowledgements}
We thank Carl M. Bender, Tomio Petrosky, Dvira Segal and Qinghai Wang for helpful comments on topics presented in this work.
S. G. acknowledges support from the Japan Society for the Promotion of Science (Fellowship grant no.~PE12057) as well as a Young Researcher's grant from Osaka Prefecture University.
The research of M. G. was supported in part by her own JSPS fellowship (Fellowship grant no.~PE14011) as well as that of S. G.

%%%%%%%%%%FIGURE prototype:
%%%%%%%%%
%%%%%%%%%
%\begin{figure}
%\hspace*{0.05\textwidth}
% \includegraphics[width=0.4\textwidth]{inline_disc_spec}
%\hfill
% \includegraphics[width=0.4\textwidth]{inline_k_par_plot}
% \hspace*{0.05\textwidth}
%\\
%\vspace*{\baselineskip}
%\hspace*{0.05\textwidth}(a)\hspace*{0.440\textwidth}(b)\hspace*{0.4\textwidth}
%\\
%\vspace*{\baselineskip}
%\caption{(a) Discrete eigenvalue spectrum $z_\pm$ as a function of $\epsd$ and (b) parametric plot of 
%$\im \; k_\pm$ vs. $\re k_\pm$ as $\epsd$ varies as $\epsd \in [-1.2, 1.2].$  Curves for bound states are indicated with a full curve while those for virtual bound states are indicated with dashing.
 %}
 %\label{fig:inline.spec}
 %\end{figure}
%%%%%%%%%
%%%%%%%%%

%%%%%%%%%%%%%%%%%%%%%
%%%%%%%%%APPENDIX
%%%%%%%%%%%%%%%%%%

\begin{appendix}

%%%%%%%%%%%%%%%%%%%%%
%%%%%%%%%APPENDIX: EPs calc
%%%%%%%%%%%%%%%%%%

\section{EP eigenvalue expansion in the case $\varepsilon_0 = \varepsilon_1 = 0$}\label{app:EP.calcs}

Here we briefly detail the eigenvalue expansions obtained in the vicinity of EP2As 
(Eqs.~(\ref{z.A.p.exp}) and~(\ref{z.A.m.exp}))
and EP2Bs (Eqs.~(\ref{z.B.m.exp}) and~(\ref{z.B.p.exp})) for the case $\varepsilon_0 = \varepsilon_1 = 0$, following a variation on the method developed in Ref.~\cite{GRHS12}.  First we find it useful to rewrite the  polynomial equation $P(\lambda) = 0$ from Eq.~(\ref{P.lambda}) directly in terms of the energy eigenvalue $E$.  We accomplish through the substitution $\lambda = -(E + \sqrt{E^2 - 4})$, which yields the equivalent equation $p(E_{\tilde{j}}) = 0$, with
\begin{equation}
  p(E)
    	= \Gamma^2 E^4 
		+  \left( \Gamma^4 - 4 \Gamma^2 - 1 \right) E^2
		+ 4
	.
\label{p.E.0}
\end{equation}
(Note that we have chosen different labeling $\tilde{j}$ for the solutions of this alternative form of the dispersion equation in order to emphasize that there is no consistent labeling that will hold between the sets of solutions as we cross the EP~\cite{EP_Korea}).

The basic idea for our calculation is that we will take advantage of the fact that the derivative of the eigenvalues blow up at the EPs to study the system properties nearby.  First we take a full derivative of the polynomial equation $\textrm{d}p/\textrm{d}\Gamma = 0$ and re-arrange to obtain
\begin{equation}
  \frac{2 \Gamma E^4 + 4 \Gamma \left( \Gamma^2 - 2 \right) E^2}{\partial E / \partial \Gamma}
		+ 2 E \left[  2 \Gamma^2 E^2 + \Gamma^4 - 4 \Gamma^2 - 1 \right]
	= 0
	.
\label{p.E.0.EP.cond}
\end{equation}
Since $\partial E / \partial \Gamma$ diverges, we obtain the a useful relationship between $E = \bar{E}$ and $\Gamma = \bar{\Gamma}$ at the EP by setting the second term on the RHS above to zero, which yields
\begin{equation}
  \bar{E} (\Gamma = \bar{\Gamma})
  	= \pm \frac{\sqrt{1 + 4 \bar{\Gamma}^2 - \bar{\Gamma}^4}}{\sqrt{2} \bar{\Gamma}}
	.
\label{EP.Ebar.gen}
\end{equation}
We can then plug this formula back into the original polynomial dispersion given in Eq.~(\ref{p.E.0}) to find the locations of the EPs in parameter space as
$\Gamma = \pm \bar{\Gamma}_\textrm{A}$ and $\Gamma = \pm \bar{\Gamma}_\textrm{B}$, where
\begin{equation}
  \bar{\Gamma}_\textrm{A}
  	= \sqrt{2} - 1
  						\ \ \ \ \ \ \ \ \ \ \ \ \ \ \ \ \ 
  \bar{\Gamma}_\textrm{B}
  	= 1 + \sqrt{2}
	.
\label{EP.Gbar.A.B}
\end{equation}
We then find the locations of the eigenvalue coalescence points by plugging these values back into Eq.~(\ref{EP.Ebar.gen}) to find 
$E(\bar{\Gamma}_\textrm{A}) = \pm \left| \bar{E}_\textrm{A} \right|$,
$E(- \bar{\Gamma}_\textrm{A}) = \pm \left| \bar{E}_\textrm{A} \right|$,
$E(\bar{\Gamma}_\textrm{B}) = \pm i \left| \bar{E}_\textrm{B} \right|$, and
$E(- \bar{\Gamma}_\textrm{B}) = \pm i \left| \bar{E}_\textrm{B} \right|$, with
\begin{equation}
  \left| \bar{E}_\textrm{A} \right|
  	= \sqrt{ 2 \left( 1 + \sqrt{2} \right)}
  						\ \ \ \ \ \ \ \ \ \ \ \ \ \ \ \ \ 
\left| \bar{E}_\textrm{B} \right|
  	= \sqrt{ 2 \left( \sqrt{2} - 1 \right)}
	.
\label{EP.zbar.A.B}
\end{equation}

Following Ref.~\cite{GRHS12}, we can now write a generic expansion in the vicinity of each the EPs 
%on the $\Gamma > 0$ half of the parameter space 
as
\begin{eqnarray}
  E_{\textrm{A}} (\Gamma)
  &	= & \left| \bar{E}_\textrm{A} \right| + \alpha_{1,2} \sqrt{\Gamma^2 - \bar{\Gamma}_\textrm{A}^2},
						\nonumber  \\
  E_{\textrm{A}} (\Gamma)
  &	= & - \left| \bar{E}_\textrm{A} \right| + \alpha_{3,4} \sqrt{\Gamma^2 - \bar{\Gamma}_\textrm{A}^2}
  	;
\label{z.A.gen.exp}
\end{eqnarray}
and
\begin{eqnarray}
  E_{\textrm{B}} (\Gamma)
  &	= & i \left| \bar{E}_\textrm{B} \right| + \beta_{1,2} \sqrt{\Gamma^2 - \bar{\Gamma}_\textrm{B}^2},
						\nonumber  \\
  E_{\textrm{B}} (\Gamma)
  &	= & - i \left| \bar{E}_\textrm{B} \right| + \beta_{3,4} \sqrt{\Gamma^2 - \bar{\Gamma}_\textrm{B}^2}
	.
\label{z.B.gen.exp}
\end{eqnarray}
%%
%We can of course obtain similar expressions for the $\Gamma < 0$ case 
%(The first line in both of these equations gives the resonances while the second line gives the anti-resonances).  
To find the expansion coefficients $\alpha_{1,2}$, for example, we define 
$\Delta^2 \equiv \Gamma^2 - \bar{\Gamma}_\textrm{A}^2$, plug this into Eq.~(\ref{p.E.0.EP.cond}), and expand in powers of $\Delta$.  Carrying this out for both cases we obtain
\begin{equation}
  \alpha_{1,2} = \alpha_{3,4} = \pm i \frac{1}{2^{1/4} \sqrt{-1 + \sqrt{2}}},
				\ \ \ \ \ \ \ \ \ \ \ \ \ \ \ \ \ \ \ \ 
 \beta_{1,2} = \beta_{3,4} = \pm i \frac{1}{2^{1/4} \sqrt{1 + \sqrt{2}}}
	.
\label{EP.exp.coeffs}
\end{equation}
Putting Eqs.~(\ref{EP.zbar.A.B}) and~(\ref{EP.exp.coeffs}) into Eq.~(\ref{z.A.gen.exp}) and Eq.~(\ref{z.B.gen.exp}), we obtain the expansion associated with the EPs as reported in Sec.~\ref{sec:PT.outgoing.spec}.
%As an example, in Fig.~\ref{fig:PT.1.EP2B} we plot the behavior of the expansions for the two resonance states $z_{\textrm{B},(3,4)}$ in the vicinity of the EP2B located at $\Gamma = \GamB$.

%%%%%%%%%%%%%%%%%%%%%
%%%%%%%%%APPENDIX: complex localized states
%%%%%%%%%%%%%%%%%%

\section{Properties of complex localized states in Region IV for the case $\varepsilon_0 = \varepsilon_1 = 0$}\label{app:iv.calcs}

In this appendix we detail the properties of the complex localized states in Region IV (see Fig.~\ref{fig:PT.0.spec}(b)) for the case $\varepsilon_0 = \varepsilon_1 = 0$, as discussed near the end of Sec.~\ref{sec:PT.outgoing.spec} and Sec.~\ref{sec:PT.outgoing.QBIC}  of the main text.  
We can generally expect the condition $\lambda \gg 1$ to hold throughout this region of the parameter space.
Hence we begin by expanding the solutions of $\lambda_j$, which are reported in Eq.~(\ref{P.lambda.0.solns}), in powers of $1 / \Gamma$ to obtain
\begin{equation}
  \lambda_{1,4}
  	\approx \pm \frac{i}{\Gamma} \left( 1 + \frac{1}{\Gamma^2} \right),
				\ \ \ \ \ \ \ \ \ \ \ \ \ \ \ \ \ \ \ \ 
 \lambda_{2,3}
  	\approx \pm i \left( 1 - \frac{1}{\Gamma^2} \right)
	.
\label{lambda.j.IV.exps}
\end{equation}
We use $E_j = - (\lambda_j + \lambda_j^{-1})$ to obtain expansions for the energy eigenvalues immediately as
\begin{equation}
  E_{1,4}
  	\approx \pm i \left( \Gamma - \frac{2}{\Gamma} \right),
				\ \ \ \ \ \ \ \ \ \ \ \ \ \ \ \ \ \ \ \ 
 E_{2,3}
  	\approx \pm i \frac{2}{\Gamma^2}
	,
\label{E.j.IV.exps}
\end{equation}
as reported in the main text.

To elucidate the asymptotic properties of the wave function for these eigenvalues, we make use of the first and third rows from Eq.~(\ref{outgoing.matrix0}) to write
\begin{equation}
  \frac{ \psi(\mp1) }{\psi(0)}
  =\frac{1}{-\lambda\pm i\Gamma-E(\lambda)}
  %	= \lambda \frac{1 \mp i \Gamma \lambda}{1 + \Gamma^2 \lambda^2},
  =\frac{\lambda}{1\pm i\lambda\Gamma}
		.
\label{psi.imp.amps}
\end{equation}
First, let us evaluate the localization properties for $\psi_1$ associated with the eigenvalue $E_1 \sim i \Gamma$, which appears to be the uncoupled gain site in the limit $\Gamma \rightarrow \infty$.  
The calculation for the wave function $\psi_4$ proceeds along similar lines.  
Applying  $\lambda_1  \Gamma=i(1+1/\Gamma^2)$ we find
\begin{equation}
%  \frac{ \bra -1 | \psi_1 \ket }{ \bra 0 | \psi_1 \ket }
\frac{\psi_1(-1)}{\psi_1(0)}
  	\approx - i \Gamma,
				\ \ \ \ \ \ \ \ \ \ \ \ \ \ \ \ \ \ \ \ 
 \frac{\psi_1(+1)}{ \psi_1(0) }
  	\approx i \frac{1}{2\Gamma}
	.
\label{psi.imp.amps.E1}
\end{equation}
%%
%Clearly we see that for large $\Gamma$, the wave function $\psi_1$ is  heavily localized at the site $-1$ in the impurity region. 
Choosing our normalization such that $\psi_1(-1) = 1$, we  have $\psi_1(0) \approx i/\Gamma$ and $\psi_1(1) \approx -1 / (2\Gamma^2)$.  
%From Eq.~(\ref{outgoing.A.B}) we can find the form of the coefficients $B$ and $C$ in the wave function Eq.~(\ref{outgoing.wave.fcn}) as $B = \Gamma$ and 
%$C = 1 / 2 \Gamma$.  
We then use  $\lambda_1 = e^{i k_1} \approx i / \Gamma$ to write the wave function~\eqref{outgoing.wave.fcn} as
\begin{equation}
  \psi_1 (x) \approx
	\left\{ \begin{array}{ll}
		-i\Gamma \left( \frac{i}{\Gamma} \right)^{|x|}	& \mbox{for $x \le -1$}    	\\
		\frac{i}{\Gamma}						& \mbox{for $x = 0$}   	\\
		\frac{i}{2\Gamma} \left( \frac{i}{\Gamma} \right)^{x}		& \mbox{for $x \ge 1$}
	\end{array}
	\right.  
		,
\label{outgoing.wave.fcn.E1}
\end{equation}
or
\begin{align}
\left|\psi_1(x)\right|^2\sim e^{-|x+1|\log\Gamma},
\end{align}
which shows that the state is localized around the gain site $x = -1$ with the localization length $1/ \log\Gamma$.
A similar calculation for $E_4 \approx - i \Gamma$ shows that the wave function for this eigenvalue is localized around the loss site $x = 1$.
They become sharper and sharper as we increase $\Gamma$.

For the eigenvalues $E_{2,3}$ in Region IV, we can use $\lambda_{2,3} \approx \pm i$ to find
\begin{equation}
  \frac{ \psi_m (\mp1)}{ \psi_m(0)} 
  		\approx \mp i \frac{1}{\Gamma}
\label{psi.imp.amps.E2.E3}
\end{equation}
for both eigenvalues with $m = 2,3$.  
This indicates that the site $0$ is the localization center for the wave functions $\psi_{2,3}$.  
%In this case, we normalize our wave function according to 
%$ \bra 0 | \psi_{2,3} \ket = e^{i \theta_{2,3}}$, with unknown phases fixed by $ \theta_{2,3}$ (we ignored this phase in the previous example as it was superfluous in that case).
%Applying this result in Eq.~(\ref{outgoing.A.B}) gives $B = \pm e^{i \theta_{2,3} / \Gamma}$ and
%$C = \mp e^{i \theta_{2,3} / \Gamma}$ for the $\psi_{2,3}$ wave functions, respectively.  Finally, we can approximate the wave number for these states according to
%%
Let us therefore normalize the wave functions according to $\psi_m(0)=1$.
Because the wave numbers are expanded as
\begin{eqnarray}
  k_{2,3} = - i \log \lambda_{2,3} 
  	& \approx & - i \log \left( \pm i \left( 1 - \frac{1}{\Gamma^2} \right) \right)
		%%
%	= \pm \frac{\pi}{2} - \log \left( 1 - \frac{1}{\Gamma^2} \right) 
								\nonumber \\
	& \approx & \pm \frac{\pi}{2} + i \frac{1}{\Gamma^2}
		,
\label{k2.k3.iv}
\end{eqnarray}
we obtain the wave functions as
\begin{equation}
  \psi_{2,3} (x) \approx 
  		%e^{i \theta_{2,3}} \times
	\left\{ \begin{array}{ll}
  		\mp \frac{1}{\Gamma} e^{\pm i \pi |x| / 2} e^{-|x| / \Gamma^2}	   & \mbox{for $x \le -1$}    	\\
  		1								& \mbox{for $x = 0$}   	\\
		\pm \frac{1}{\Gamma} e^{\pm i \pi x / 2} e^{-x / \Gamma^2}  	& \mbox{for $x \ge 1$}
	\end{array}
	\right.  
		,
\label{outgoing.wave.fcn.E2}
\end{equation}
which shows that the localization length is $\Gamma^2/2$.
In other words, they become broader and broader as we increase $\Gamma$.

\end{appendix}

\end{document}